\documentclass[a4paper,11pt]{article}
\pdfoutput=1
\usepackage{jheppub}
\usepackage{hyperref}
\usepackage{amsmath,amssymb,amsthm,amscd}
\usepackage[all,cmtip]{xy}
\usepackage{xcolor}
\usepackage{blkarray}
\usepackage{float}
\usepackage{pdflscape}

\DeclareFontFamily{U}{wncy}{}
\DeclareFontShape{U}{wncy}{m}{n}{<->wncyr10}{}
\DeclareSymbolFont{mcy}{U}{wncy}{m}{n}
\DeclareMathSymbol{\Sh}{\mathord}{mcy}{"58} 

\usepackage{tikz,textcomp}
\usetikzlibrary{decorations.pathreplacing,calc,fadings,fit,shapes,arrows,positioning,chains,matrix}
\DeclareFontFamily{U}{wncy}{}
    \DeclareFontShape{U}{wncy}{m}{n}{<->wncyr10}{}
    \DeclareSymbolFont{mcy}{U}{wncy}{m}{n}
    \DeclareMathSymbol{\Sh}{\mathord}{mcy}{"58}

\numberwithin{equation}{section}

\DeclareFontFamily{U}{wncy}{}
\DeclareFontShape{U}{wncy}{m}{n}{<->wncyr10}{}
\DeclareSymbolFont{mcy}{U}{wncy}{m}{n}
\DeclareMathSymbol{\Sh}{\mathord}{mcy}{"58} 

\numberwithin{equation}{section}

\title{The discrete Green-Schwarz mechanism in 6D F-Theory and Elliptic Genera of Non-Critical Strings}

\preprint{LMU-ASC 33/22}

\author[a]{Markus Dierigl}
\author[b]{Paul-Konstantin Oehlmann}
\author[c]{Thorsten Schimannek}
\affiliation[a]{Arnold Sommerfeld Center for Theoretical Physics, LMU, Munich, 80333, Germany}  
\affiliation[b]{
Department of Physics, Northeastern University \\
360 Huntington Avenue, Boston, MA 02115, United States
}
\affiliation[c]{Laboratoire de Physique Th\'eorique et Hautes
Energies (LPTHE), UMR 7589 CNRS-Sorbonne Universit\'e,
Campus Pierre et Marie Curie,
4 place Jussieu, F-75005 Paris, France}
\emailAdd{m.dierigl@physik.uni-muenchen.de}
\emailAdd{p.oehlmann@northeastern.edu}
\emailAdd{schimannek@lpthe.jussieu.fr}

\abstract{
We study global anomalies of discrete gauge symmetries in six-dimensional supergravities and their realizations in F-theory.
We explicitly construct a discrete Green-Schwarz mechanism that depends on the choice of a coupling constant and on a certain quadratic refinement in differential cohomology.
By geometrically engineering theories with $G=\mathbb{Z}_3$ gauge symmetry and no tensor multiplets, we observe that a particular choice of the quadratic refinement is singled out in F-theory.
This implies new Swampland constraints on the discrete charge spectra of 6d supergravities.
On the other hand, the discrete Green-Schwarz coupling depends on the geometry of the Calabi-Yau.
We use anomaly inflow to relate this to a 't~Hooft anomaly of the induced global symmetry in the worldsheet theories of non-critical strings.
Using topological symmetry lines, we further relate this anomaly to the modular properties of twisted-twined elliptic genera.
We then argue that the latter are encoded in the A-model topological string partition functions on different torus fibrations that are equipped with a flat torsional B-field.
This allows us to derive a geometric expression for the global discrete anomaly in terms of the height-pairing of a multi-section on a genus one fibered Calabi-Yau.
}

\begin{document}
\maketitle
\flushbottom

\section{Introduction}
\label{sec:intro}

The absence of gauge anomalies is one of the most fundamental consistency constraints on general quantum field theories. Especially gravitational theories in $(2 + 4n)$ dimensions are subject to strong restrictions, imposed by the presence of pure gravitational anomalies. In six-dimensional (6d) $\mathcal{N} = 1$ supergravity theories, the potential anomalies severely limit the multiplicities of hyper, vector, and tensor multiplets. However, these theories also provide the means to cancel part of the anomalies via a generalization of the Green-Schwarz mechanism~\cite{Green:1984bx,GREEN1984117,Sagnotti:1992qw}, for which the (anti-)self-dual tensor fields transform under gauge symmetries and diffeomorphisms, leading to a plethora of seemingly consistent theories.

The Swampland Program~\cite{Vafa:2005ui,Lee:2019skh} analyzes which of these theories actually arise as the low-energy limit of ultra-violet (UV) complete quantum gravity theories such as string theory, defining the Landscape, and which do not, and consequently are in the Swampland~\footnote{For recent reviews of the Swampland Program and pointers to the vast literature see~\cite{Brennan:2017rbf,Palti:2019pca,vanBeest:2021lhn}.}.
One of the richest constructions of UV complete 6d supergravities is provided by F-theory~\cite{Vafa:1996xn,Morrison:1996na,Morrison:1996pp}, a non-perturbative formulation of Type IIB superstring theory.
An advantage of F-theory is, that it encodes the physical properties of the 6d theory in the geometry of the underlying compactification manifold, given by a torus-fibered Calabi-Yau 3-fold. In particular, this includes the realization of anomalies as well as their cancellation. Thus, the absence of anomalies, as required by the consistency of the theory, is equivalent to certain highly non-trivial generic properties of Calabi-Yau manifolds~\cite{Morrison:1996na,Morrison:1996pp,Bershadsky:1996nh,Sadov:1996zm,Aspinwall:2000kf,Grassi:2000we,Grassi:2011hq,Kumar:2009ac,Park:2011ji,Arras:2016evy,Grassi:2018rva}.
However, the set of birational equivalence classes of torus fibered Calabi-Yau threefolds is finite~\cite{Grassi1991,gross1994finiteness} while there are infinite families of apparantly consistent 6d supergravities~\cite{Kumar:2010ru,Taylor:2018khc}.
This suggests the existence of new Swampland constraints of various nature, some of which have been explored in~\cite{Kim:2019vuc,Lee:2019skh}.

The correlation between physical properties, such as anomalies, and geometry is well-explored in the case of continuous symmetries~\cite{Park:2011ji,Krause:2012yh,Morrison:2012ei,Grimm:2013oga}, but is far from being understood in the context of discrete symmetries~\cite{Braun:2014oya,Morrison:2014era,Anderson:2014yva,Klevers:2014bqa,Mayrhofer:2014laa,Cvetic:2015moa,Garcia-Etxebarria:2014qua,Lin:2015qsa,Oehlmann:2016wsb,Kimura:2016crs,Cvetic:2018bni,Anderson:2018heq,Anderson:2019kmx,Cota:2019cjx,Oehlmann:2019ohh,Knapp:2021vkm,Schimannek:2021pau}.
The present work provides some of the missing entries in the F-theory dictionary associated to the cancellation of discrete gauge anomalies, which in turn enables an extension of the Swampland Program to discrete gauge data, shedding light on more general consistency constraints of supergravities that are part of the Landscape.

From the field theory side, we study 6d $\mathcal{N} = 1$ supergravities with a discrete $\mathbb{Z}_n$ gauge group, and in some parts we focus on theories with $n = 3$ and no tensor multiplets.
Discrete gauge anomalies as well as discrete mixed-gravitational anomalies are induced by the presence of $\mathbb{Z}_n$ charged hypermultiplets containing chiral fermions. In the absence of tensor multiplets, the only other field that can contribute to the anomaly cancellation is given by the self-dual 2-form field in the gravity multiplet. 
The geometrical realization within F-theory is given in terms of genus-one fibered Calabi-Yau 3-folds with $n$-section and base $\mathbb{P}^2$ or, equivalently, the associated so-called Jacobian fibration that exhibits a section but is singular~\footnote{For $n\ge 4$ additional geometric realizations of the same F-theory vacua arise, as was studied in~\cite{Knapp:2021vkm,Schimannek:2021pau}.}. Throughout the paper we study the discrete gauge data directly, without relying on un-Higgsing to continuous symmetries.

The analysis of discrete gauge anomalies in a gravitational theory requires the use of modern field theory techniques, see e.g.~\cite{Garcia-Etxebarria:2018ajm,Monnier:2019ytc} for recent reviews. The 6d anomalies are encoded in a 7d invertible topological field, the anomaly theory, which contains information about the phase of the 6d partition function.
For that the 7d manifold is bounded by the 6d spacetime together with the correct gauge field background and tangential structure, such as e.g.\ orientation or Spin structure, which extend into the 7d bulk. The absence of the most general class of global anomalies, often called Dai-Freed anomalies~\cite{Dai:1994kq}, which take into account fluctuations of the spacetime topology, is equivalent to the statement that the above construction does not depend on the extension to seven dimensions.
To ensure that, one needs to evaluate the anomaly theory on a set of generators of deformation classes of closed 7d manifolds with the given structure. These are classified by bordism groups. For 6d theories with $\mathbb{Z}_3$ gauge group there is a single generator of $\Omega^{\text{Spin}}_7(B\mathbb{Z}_3)$ given by $L^7_3$, the 7d lens space with non-tivial $\mathbb{Z}_3$ bundle. We show that theories with the number of charged fermions not divisible by $9$ turn out to be anomalous.

In order to investigate the possibility of discrete anomaly cancellation, one first needs to describe the self-dual tensor as a boundary mode of a 3-form in seven dimensions~\cite{Witten:1996hc, Belov:2006jd,Hsieh:2020jpj}. Moreover, the correct framework of the description of these higher-form fields is given by differential cohomology, in order to account for torsion fluxes. The discrete Green-Schwarz mechanism can then be formulated as a 't Hooft anomaly for the higher-form symmetries of the involved fields and hinges on the realization of a quadratic refinement~\footnote{Here, we refer to the quadratic refinement of the differential cohomology pairing, which on the lens space $L^7_3$ reduces to a quadratic refinement on the torsion pairing on $H^4 (L^7_3;\mathbb{Z})$.}, needed to define a topological term for the 3-form field in the 7d theory. The pure field theory analysis shows that anomalies coming from a fermion multiplicity divisible by $3$ can be cancelled in such a way by appropriate choice of the quadratic refinement.

To compare this to the landscape of theories realized in F-theory, we systematically engineer suitable genus one fibrations by using monad bundles.
All the F-theory models with anomalous fermion sector, however, turn out to exhibit a multiplicity of charged fermions, or equivalently of charged hypermultiplets, that is given by $6 \text{ mod } 9$.
This singles out a particular quadratic refinement and leads us to propose the existence of novel Swampland constraints in the presence of discrete gauge anomalies. 

The special quadratic refinement depends on an additional discrete parameter $\ell \in \{1 \,, 2\}$ that determines the anomaly inflow onto the effective strings coupling to the self-dual tensor field. For the quadratic refinement preferred by F-theory, the discrete 6d anomaly cancellation is independent of this parameter and in principle both values are allowed.
This raises the questions, whether both values are actually realized in the F-theory compactifications and what the physical effect of different values is.

Among the genus one fibered Calabi-Yau manifolds that we have constructed, we observe pairs of geometries that lead to the same massless spectrum in F-theory but that exhibit inequivalent intersection numbers and different Gopakumar-Vafa invariants.
Since the latter carry information about the massive spectrum of the theory, the corresponding theories should be physically distinct.
However, the only parameter that could distinguish them is the value of $\ell$.
This is consistent with the fact that such ``degenerate spectra'' only arise for theories that have a non-vanishing fermion anomaly.
Since F-theory doesn't seem to single out a value, we are then led to the problem of relating the discrete anomaly coefficient $\ell$ to the geometry of the Calabi-Yau.

To solve this problem, we use the fact that the 6d theory also contains non-critical strings that couple to the 2-form fields.
As a result, the discrete gauge anomalies are related to 2d 't~Hooft anomalies of the induced global symmetries on the string worldsheets via anomaly inflow.
The associated 3d anomaly theory, a Dijkgraaf-Witten theory~\cite{cmp/1104180750}, can be represented by an element in group cohomology $[l] \in H^3(B\mathbb{Z}_n,U(1))=\mathbb{Z}_n$ and we argue that the anomaly inflow leads, up to a sign, to the relation
\begin{align}
    [\ell] = [\pm l]\,.
\end{align}
Fixing the sign requires a careful treatment of the Bianchi identity and we leave this for future work.
In the presence of additional tensor fields both $l$ and $\ell$ will be equipped with an index that takes values in the string charge lattice and that we leave implicit here.

A non-vanishing discrete 't~Hooft anomaly manifests itself as a failure of the associated twisted twined elliptic genera to combine into a modular invariant orbifold partition function under gauging.
From another perspective, a global 0-form symmetry is equivalent to the existence of topological symmetry line operators~\cite{Gaiotto:2014kfa}, see also~\cite{Bhardwaj:2017xup,Chang:2018iay}.
The twisting and twining of the partition function can also be realized by inserting corresponding symmetry lines  and the anomaly leads to a non-trivial phase in a crossing relation.
We use this relationship to explicitly derive a general expression for the ``modular anomaly'' of the twisted twined partition function in terms of the element in group cohomology that represents the 't~Hooft anomaly.
This can then be applied to the twisted twined elliptic genera of the non-critical strings.

It remains to relate the modular properties of the twisted twined elliptic genera to the geometry in F-theory.
To this end we use a duality with topological string theory.
Using recent results from~\cite{Schimannek:2021pau}, we first argue that the 6d theory compactified on a 2-torus together with a choice of discrete holonomies $(g,h)$ along the cycles is dual to Type IIA string theory on different torus fibrations $X_g$ that are equipped with a flat torsional B-field labelled by $h$.
The twisted twined elliptic genera are then encoded in the corresponding A-model topological string partition functions.
Using the modular properties of the latter, that have been worked out in~\cite{Schimannek:2019ijf,Cota:2019cjx,Schimannek:2021pau}, we can then derive a geometric expression for the 't~Hooft anomaly in terms of the height-pairing of a multi-section on the genus one fibered Calabi-Yau.
Applying this to our family of Calabi-Yau manifolds, we find that there are indeed different anomalies for the pairs of inequivalent theories that exhibit the same massless spectra and therefore the ``degeneracy'' is lifted.

\begin{figure}[t!]
\centering
\begin{tikzpicture}[
scale=.8, every node/.style={scale=0.8},
container/.style = {draw, rectangle, dashed, align=center,rounded corners=0.5em,fill=white!20},
fieldth/.style = {draw, rectangle, align=center, rounded corners=0.5em,fill=white!20}
]

\node [fieldth, minimum width=5cm, minimum height=3cm, align=center ] (6D)   {
\underline{6d $\mathcal{N}=1$ SUGRA on $M_6$}\\[.8em]
{\begin{minipage}{5cm}\begin{itemize}\setlength\itemsep{.0em}
\item $G=\mathbb{Z}_3,\,T=0$
\item From F-theory on $X_0$
\item Discrete GS-coupling $\ell$
\end{itemize}\end{minipage}  
}
};  

\node [ minimum width=1cm, minimum height=5cm, align=center, below= of 6D  ] (Empt1)   {
};  

 \node[fieldth, minimum width=5cm, minimum height=3cm, align=center, below=of Empt1 ] (2D)   {
\underline{2d Non-Crit. String on $\Sigma$}\\[.8em]
{\begin{minipage}{5cm}\begin{itemize}\setlength\itemsep{.0em}
\item From D3's on $C \subset \mathbb{P}^2$
\item Induced global $G_F=\mathbb{Z}_3 $
\item 't Hooft Anomaly $l$
\end{itemize}\end{minipage}  
}
};  

 \node[fieldth, minimum width=1cm, minimum height=3cm, align=center, right=of Empt1 ] (Elliptic)   {
\underline{Elliptic Genera $Z^{(r,s)}_C$}\\[.8em]
{\begin{minipage}{6.3cm}\begin{itemize}\setlength\itemsep{.0em}
\item $(r,s)$-Twisted/twined
\item Encoded in topological string\\partition function on $X_{r,\text{n.c.}s}$
\end{itemize}\end{minipage}  
}
};  

 \node[  minimum width=1cm, minimum height=5cm, align=center, right=of Elliptic ] (Empt2)   { 
};  

\node [fieldth, minimum width=5.3cm, minimum height=3cm, align=center, above= of Empt2  ] (7D)   {
\underline{7d Anomaly Theory on $N_7$}\\[.8em]
{\begin{minipage}{6.5cm}\begin{itemize}\setlength\itemsep{.0em}
\item  $\eta$-inv. + CS-like theory
 \item F-theory fixes quad. refinement\\
 $\rightarrow\,n_{\pm1} \in \{0,6\} \text{ mod }9 $  
\end{itemize}\end{minipage}  
}
};  

 \node[fieldth, minimum width=5cm, minimum height=3cm, align=center, below=of Empt2 ] (3D)   {
\underline{3d Anomaly Theory on $N_3$}\\[.8em]
{\begin{minipage}{5.5cm}\begin{itemize}\setlength\itemsep{.0em}
\item Dijkgraaf-Witten theory\\$[l]\in H^3(B\mathbb{Z}_3,U(1))=\mathbb{Z}_3$
\item Enters crossing relation of\\topological symmetry lines
\end{itemize}\end{minipage}  
}
};

\draw[<-, very thick, align=center] ([shift={(0,0)}]7D.west) to node[above] {\begin{tabular}{c} Extension\\$M_6=\partial N_7$\end{tabular} }(6D.east);

\draw[<-, very thick, align=center] (3D.west) to node[below] {\begin{tabular}{c} Extension\\$\Sigma=\partial N_3$\end{tabular}  }(2D.east);

\draw[<->, very thick] (2D.north) to node[left] {\begin{tabular}{l}Anomaly\\ \, Inflow\\$[\ell]=[\pm l]$ \end{tabular} }(6D.south);

\draw[->, very thick] (6D.south east) to node[right,xshift=0.3cm] {\begin{tabular}{l}
4d Instantons from Strings on $S^1_F\times S^1_M$\\w/ discrete holonomies $(r,s)$\end{tabular} }([shift={(-1,0)}]Elliptic.north);

\draw[<-, very thick] ([shift={(-.5,0)}]Elliptic.south) to node[left,yshift=-0.0cm] { \begin{tabular}{c} $T^2$-Partition\\function  \end{tabular}   }(2D.north east);

\draw[<->, very thick] (3D.north west) to node[right,yshift=-0.0cm] { \begin{tabular}{c} Modular\\anomaly \end{tabular}   }([shift={(.5,0)}]Elliptic.south);

\end{tikzpicture}
\caption{The relationship between the different theories and objects that enter our analysis.}
\label{fig:overview}
\end{figure}
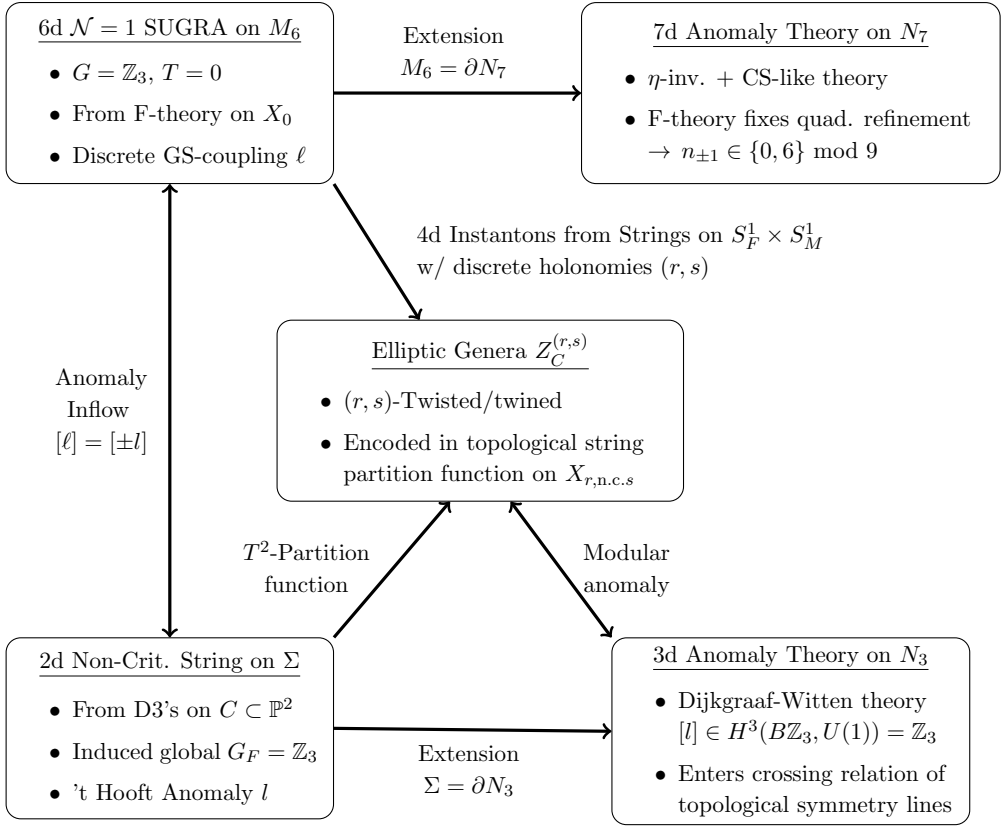

An overview of the different theories and objects that are involved in our analysis is provided in Figure~\ref{fig:overview} and the paper is organized as follows. In Section \ref{sec:fthdisc} we describe the geometric realization of discrete gauge symmetries in F-theory, involving genus-one fibrations with multi-sections and their associated singular Jacobian fibrations. We also discuss the relationship between discrete holonomies and Type IIA compactifications on different torus fibrations with flat torsional B-fields and construct a set of genus one fibrations that leads to F-theory vacua with $G=\mathbb{Z}_3$. In Section \ref{sec:6danomalies} we discuss the discrete Dai-Freed anomalies from a purely field theoretical viewpoint and study the possibilities of anomaly cancellation via a discrete version of the Green-Schwarz mechanism and its implication on the anomaly-inflow onto the effective strings in the setup. 
In Section~\ref{sec:2danomalies} we study the relationship between a discrete 't~Hooft anomaly in 2d CFTs and the modular properties of the twisted twined partition functions. We use topological symmetry lines to derive a general expression for the anomalous phase under modular transformations in terms of the anomaly. 
In Section~\ref{sec:egan} we then relate the twisted twined elliptic genera of non-critical strings to the A-model topological string partition functions on different torus fibered Calabi-Yau manifolds with flat torsional B-fields.
Using the modular properties of the latter, we derive a geometric expression for the discrete 't~Hooft anomaly.

 Some technical details are relegated to appendices.
 In Appendix~\ref{app:MonadBundles} we elaborate on our construction of genus one fibrations with a 3-section using monad bundles.
 In Appendix~\ref{app:gravanom} we review the anomaly theories associated to pure gravitational anomalies in six-dimensional supergravities.
 In Appendix~\ref{app:modjacobi} we collect basic definitions related to modular and Jacobi forms.
Appendix~\ref{app:AbelianAnomalies} provides an overview of some earlier results on anomalies of six-dimensional supergravities.
Finally, in Appendix~\ref{sec:gvinvariants}, we compare the Gopakumar-Vafa (GV) invariants, including their recently proposed $\mathbb{Z}_3$ refinements, for a pair of geometries that leads to identical massless spectra in F-theory and show that the excitation spectra of the self-dual strings are different.

\section{F-theory vacua with discrete gauge symmetries}
\label{sec:fthdisc}
We will assume familiarity with the basics of the ``F-theory dictionary'' and refer to~\cite{Weigand:2018rez,Cvetic:2018bni} for reviews.
However, the geometric realizations of discrete symmetries in F-theory and the corresponding dualities to M-theory and Type IIA string theory have usually been discussed from the perspective of Higgs transition~\cite{Braun:2014oya, Morrison:2014era,Mayrhofer:2014haa,Mayrhofer:2014laa,Lin:2015qsa,Cvetic:2015moa}, see also~\cite{Bhardwaj:2015oru}, where a $\mathbb{Z}_n$ symmetry is a remnant of a partially broken $U(1)$.
For our purpose it will prove beneficial to approach the topic of discrete symmetries intrinsically, without invoking an un-Higgsing.
This will also allow us to connect the discussion to recent insights on the role of non-commutative resolutions of singular torus fibrations in the duality to Type IIA compactifications discovered in~\cite{Schimannek:2021pau}, which will be important later on.
To begin, let us first discuss the effect of discrete holonomies in circle compactifications of 6d supergravity.

\subsection{Discrete symmetries and discrete holonomies}
Before discussing how discrete symmetries are geometrically engineered in F-theory, let us start by considering a six-dimensional $\mathcal{N}=1$ supergravity with gauge group
\begin{align}
    G_{6d}=\mathbb{Z}_n\,,
\end{align}
and for simplicity assume that $n$ is prime.
In addition to the Planck scale, the continuous parameters that enter the theory are the vacuum expectation values of the real scalar fields in $T$ tensor multiplets that parametrize the tensor branch, as well as the complex scalar fields in $H$ hypermultiplets that parametrize the Higgs branch.

After compactifying on a circle, the five-dimensional gauge symmetry is given by
\begin{align}
    G_{5d}=\mathbb{Z}_n\times U(1)_{KK}\times U(1)^T\,.
    \label{eqn:5dgauge}
\end{align}
From the six-dimensional perspective, the additional factors arise from the Kaluza-Klein photon associated to the metric and from the dual vectors of the $T$ anti-self-dual tensor fields.
The vacuum expectation values of the real scalar fields in the vector multiplets parametrize the  $T+1$-dimensional Coulomb branch of the theory.

While there is no continuous Coulomb branch associated to the $\mathbb{Z}_n$ factor of the gauge group, one can nevertheless turn on a discrete holonomy.
After fixing a generator $g\in\mathbb{Z}_n$ this amounts to a choice,
\begin{align}
    \xi_{\mathbb{Z}_n}=r\,,\quad r\in 0,\ldots,n-1\,.
\end{align}
such that, denoting the circle direction by $x^6$, a shift $x^6\mapsto x^6+2\pi$ is equivalent to acting with $g^r$ in the corresponding five-dimensional theory.
Note that this is nothing but a choice of $\mathbb{Z}_n$-bundle along the circle.

Let us denote the $\mathbb{Z}_n\times U(1)_{KK}$ charges of states in the theory with $\xi_{\mathbb{Z}_n}=0$ by $(q_{\mathbb{Z}_n},q_{KK})$.
To understand the physical effect of the discrete holonomy, recall that shifts along the circle generate $U(1)_{KK}$, such that a general coordinate dependent translation $x^6\mapsto x^6+\chi(x_{1,\ldots,5})$ amounts to a gauge transformation
\begin{align}
        A_{KK}\mapsto A_{KK}+d\chi\,,\quad \phi\mapsto e^{iq\chi}\phi\,,
\end{align}
where $A_{KK}$ is the associated 1-form potential and $\phi$ is any field of KK-charge $q$.
After turning on a non-trivial holonomy $\xi_{\mathbb{Z}_n}=r\ne 0$, the KK-charges are effectively shifted,
\begin{align}
    q_{KK}\mapsto q_{KK}+\frac{r}{n}\cdot q_{\mathbb{Z}_n}\,.
\end{align}
As a result, the $\mathbb{Z}_n$ and $U(1)_{KK}$ gauge symmetries combine into a new symmetry $U(1)_M$ that is encoded in the short exact sequence
\begin{align}
    1\rightarrow\mathbb{Z}_n\rightarrow U(1)_M\xrightarrow[]{\cdot n} U(1)_{KK}\rightarrow 1\,,
    \label{eqn:u1seq}
\end{align}
where the generator $g\in\mathbb{Z}_n$ maps to $e^{2\pi i r/n}\in U(1)_M$, and the corresponding charges are
\begin{align}
q_{U(1)_M}=n\cdot q_{KK}+r\cdot  q_{\mathbb{Z}_n}\,.
\end{align}
Since we have assumed that $n$ is prime, the discrete symmetry is now fully embedded into $U(1)_M$ such that the gauge group becomes
\begin{align}
    G_{5d,k}=U(1)_M\times U(1)^T\,.
\end{align}

After compactifying on an additional circle to four dimensions, there is another Kaluza-Klein gauge symmetry $U(1)_{M'}$ as well as a second choice of a discrete holonomy $s=0,\ldots,n-1$.
For $r\ne 0$, since the $\mathbb{Z}_n$ is fully embedded inside $U(1)_M$, this holonomy just amounts to a large $U(1)_M$ gauge transformation.
If on the other hand $r=0$, a non-trivial choice $s\ne 0$ is again absorbed by the second Kaluza-Klein $U(1)_{M'}$, analogous to the discussion in the previous paragraph.

\subsection{The geometry of discrete holonomies}
\label{sec:geodhol}
Let us now discuss how the six-dimensional supergravity and the circle compatifications with discrete holonomies are geometrically engineered in F-theory, M-theory and Type IIA string theory.
Recall that in F-theory, the axio-dilaton profile of a non-perturbative Type IIB compactification on a surface $B$ is encoded geometrically in a torus fibered Calabi-Yau threefold $\pi:X\rightarrow B$.
If there is no discrete gauge symmetry, the circle compactification of this theory is dual to M-theory on $X$ while the 2-torus compactification is dual to Type IIA string theory on $X$.
In the presence of a discrete gauge symmetry, this picture needs to be refined in order to account for the discrete holonomies.

The possibility of discrete holonomies is obvious from the six-dimensional perspective but it has the non-trivial geometric implication that multiple torus fibrations over a given base $B$ can encode the same axio-dilaton profile.
While each of these torus fibrations therefore corresponds to the same F-theory vacuum, they lead to different compactifications in M-theory and geometrically realize the different choices of discrete holonomies.

In the following, we will denote the torus fibration associated to the holonomy $r=0,\ldots,n-1$ by $X_r$.
Note that since the holonomies $r$ and $-r$ are related by choosing the $\mathbb{Z}_n$ generator $g^{-1}$ instead of $g$, which in particular preserves the charges of all hypermultiplets, the corresponding theories are physically equivalent and one expects $X_r=X_{-r}$.
The five dimensional gauge symmetry~\eqref{eqn:5dgauge} in the absence of a discrete holonomy is only preserved when compactifying M-theory on $X_0$.

In fact, another consequence is that this set of geometries is equipped with a group law.
For any finite Abelian group $G$ it is clear that the set of $G$-bundles on a circle is itself equipped with a $G$ group structure.
Given two bundles that correspond to the respective twists $g,g'\in G$, the product of the two bundles is just the bundle with twist $g\cdot g'$.
The holonomies themselves can therefore be interpreted as elements $[r]\in G$ and the group of holonomies is isomorphic to the six-dimensional discrete symmetry in F-theory.

The group of discrete holonomies is geometrically realized as the \textit{Weil-Châtelet group} $WC(J/B)$ associated to the so-called Jacobian fibration $J=X_0$.
The elements correspond to pairs $(Y,\sigma)$, where $Y$ is a torus fibration that encodes the same axio-dilaton profile as $J$ and $\sigma$ is an action
\begin{align}
    \sigma:\,Y\times J\rightarrow Y\,.
\end{align}
We will not go into the subtleties related to the precise definition of this group and instead refer to~\cite{silverman2009arithmetic,joseph1994advanced} for details.
Restricting to elements of $WC(J/B)$ that correspond, roughly speaking, to geometries without multiple fibers leads to the \textit{Tate-Shafarevich group} $\Sh(J/B)$.
The physics associated to multiple fibers is still not fully understood, although some aspects have been discussed in~\cite{deBoer:2001wca,Bhardwaj:2015oru,Anderson:2018heq,Anderson:2019kmx,Oehlmann:2019ohh,Kohl:2021rxy}.
We will therefore only consider the situation when $\Sh(J/B)=WC(J/B)$ and to highlight this we will in the following refer to the Tate-Shafarevich (TS) group instead of the Weil-Châtelet group.

A non-trivial TS-group arises for example when $X$ is a smooth genus one fibered Calabi-Yau threefold that does not have a section but only an $n$-section, meaning a divisor that intersects the generic fiber $n$ times.
The associated Jacobian fibration $J=X_0$ can be constructed by replacing the fibers of $X$ with the corresponding Weierstra{\ss} models.
Assuming e.g.~that $B=\mathbb{P}^2$ and that the fibration is sufficiently generic, one can then show that $\Sh(J/B)=\mathbb{Z}_n$ and choose $X_1=X$~\cite{oggshafa}.

After compactifying on an additional circle, the theory becomes dual to Type IIA string theory on the torus fibered Calabi-Yau threefold $X_r$.
From the six-dimensional perspective, as discussed before, we now have the freedom to choose a pair of discrete holonomies $r,s\in\mathbb{Z}_n$ along the cycles of the torus.
On the other hand, in M-theory this choice corresponds to a compactification on $X_r\times S^1$ together with discrete holonomy $s$ along the circle.

In Type IIA string theory, the additional degree of freedom $[s]\in\mathbb{Z}_n$ was recently found to correspond to a choice of flat fractional B-field on $X_r$~\cite{Schimannek:2021pau}.
From a geometrical perspective, it was argued that $X_r$ is singular whenever $n'=\text{gcd}(r,n)\ne 1$, given our assumption that $n$ is prime this implies $r=0$, and that the B-field leads to a so-called \textit{non-commutative resolution} $X_{r,\text{n.c.}s}$.
Moreover, the group of holonomies along the second circle can be identified with $\text{Tors}\,H^3(\widehat{X},\mathbb{Z})$, where $\widehat{X}$ is a non-K\"ahler small analytic resolution of $X$, see also~\cite{wip}.
Again, we will not go into the rich details of this aspect and rather refer to~\cite{Schimannek:2021pau} and~\cite{wip} for more information.

The relationship between the different compactifications and choices of discrete holo\-nomies are summarized in Table~\ref{tab:FMA}.
\begin{table}[t!]
\centering
\begin{tabular}{|c|ccc|}\hline
$d$&F-theory&M-theory&Type IIA\\\hline
$6$&$X$&&\\\hline
$5$&$\begin{array}{c}X\times S^1\\\mathbb{Z}_n\text{-hol. }r\end{array}$&$X_r$&\\\hline
$4$&$\begin{array}{c}X\times T^2\\\mathbb{Z}_n\text{-hol. }(r,s)\end{array}$&$\begin{array}{c}X_r\times S^1\\\mathbb{Z}_n\text{-hol. }s\end{array}$&$X_{r,\text{n.c.}s}$\\\hline
\end{tabular}
\caption{The duality between F-theory, M-theory and Type IIA string theory with $\mathbb{Z}_n$-holonomies.}
\label{tab:FMA}
\end{table}

\subsection{F-theory vacua with $G=\mathbb{Z}_3$ and $T=0$}
To initiate the study of discrete anomalies in F-theory, we will focus on compactifications that lead to six-dimensional $\mathcal{N}=1$ effective supergravities with gauge group $G=\mathbb{Z}_3$ and no tensor multiplets, i.e. $T=0$, such that the only anti-symmetric two-form field that participates in the generalized Green-Schwarz mechanism is the generic self-dual two-form.

Geometrically, this leads us to consider smooth Calabi-Yau threefolds $X_1$ that are genus one fibered over $B=\mathbb{P}^2$ and that do not admit a section but only a 3-section.
The associated Jacobian fibration $X_0=J(X_1/B)$, that can be obtained by replacing each fiber in $X_1$ with the corresponding Weierstra{\ss} model, exhibits terminal nodal singularities~\cite{caldararu2001derived} but encodes the same F-theory vacuum.
The corresponding Tate-Shafarevich group $\Sh(X_0/B)=\mathbb{Z}_3$ takes the form~\cite{Cvetic:2015moa}
\begin{align}
    \Sh(X_0/B)=\{(X_0,i),(X_1,\sigma),(X_1,\sigma^{-1})\}\,,
\end{align}
where the details of the actions $i,\sigma,\sigma^{-1}$ of $X_0$ on itself and on $X_1$ are not important to us.
As discussed above, this can physically be thought of as the group of discrete holonomies and is isomorphic to the discrete gauge symmetry itself.

The 3-section of $X_1$ determines a degree $3$ line bundle on each fiber and the sections embed the fiber as a hypersurface into $\mathbb{P}^2$~\cite{Braun:2014oya}.
As a consequence, the Calabi-Yau $X_1$ itself can be realized as an anti-canonical hypersurface in a $\mathbb{P}^2$ bundle over the base $B$.
Assuming that the cohomological Brauer group of $B$ vanishes, which is the case for $B=\mathbb{P}^2$, such bundles can always be obtained as a projectivization $\mathbb{P}(V)$ of a rank $3$ vector bundle $V$ on $B$~\footnote{A non-trivial element of the Brauer group can be used to define twisted bundles, which lead to additional $\mathbb{P}^n$ bundles that are not projectivizations of ordinary rank $n+1$ vector bundles.}.
To construct all smooth Calabi-Yau threefolds that are genus one fibered with a 3-section over $B$, one would therefore in principle have to construct all rank 3 vector bundles $V$ on $B$ for which a generic anti-canonical hypersurface in $\mathbb{P}(V)$ is smooth.
However, according to Bertini's theorem, this is equivalent to determining whether the anti-canonical bundle $\mathbb{P}(V)$ is basepoint-free, which is in general a hard problem even for $B=\mathbb{P}^2$.

Instead of attempting a full classification, we therefore focus on particular monad bundles $V$ that fit into a short exact sequence
\begin{align}
    0\rightarrow V\rightarrow W\xrightarrow[]{f}U \rightarrow 0\,,
\end{align}
where we further assume that $W$ and $U$ are sums of line bundles on $\mathbb{P}^2$.
We then find that the anti-canonical hypersurface in $\mathbb{P}(V)$ can equivalently be described as a complete intersection in a toric $\mathbb{P}^5$ bundle over $\mathbb{P}^2$ and the technology developed in~\cite{Batyrev:1994pg} can be applied.
As a result, we obtain a set of 13 inequivalent Calabi-Yau threefolds with $h^{1,1}=2$ that are genus one fibered over $\mathbb{P}^2$ and exhibit a $3$-section.
The details of our construction are discussed in~Appendix~\ref{app:MonadBundles} and the data of the Calabi-Yaus is summarized in Table~\ref{tab:z3list}. 

For the readers convenience, we reproduce the resulting numbers $n_q$ of hypermultiplets with $\mathbb{Z}_3$ charge $q$ in F-theory, as well as the corresponding triple intersection numbers
\begin{align}
    c_{ijk}= D_i \cdot D_j \cdot D_k, \text{ with } i,j,k=1,2 \, .  
\end{align}   
of the Calabi-Yau manifolds in Table~\ref{tab:z3listred}.
Here $D_1$ and $D_2$ respectively denote the class of a 3-section and the vertical divisor associated to the hyperplane class of the base.
We have already included the value
\begin{align}
    l=\frac12(c_{112}+3)\text{ mod }3\,,
\end{align}
for the $\mathbb{Z}_3$ 't~Hooft anomaly, applying the expression~\eqref{eqn:thooftheightpairing} in terms of the geometric height-pairing that will be derived in Section~\ref{sec:egan}.
In each case the massless F-theory spectrum satisfies the 6d irreducible
 gravitational anomaly cancellation condition
\begin{align}
    n_{\pm1} +  n_0 = 273 \, ,
\end{align}
as expected.
We can summarize our main results in terms of the properties of this table:

\begin{table}[t!]
\begin{center}
\begin{tabular}{|c|c|c|cc|c|cc|c|c|c|c|c|c|}\hline
\# & 1 & 2 & 3 & 4 & 5 & 6 & 7 &8 &9 & 10 & 11 & 12 & 13 \\  \hline
$n_{\pm 1}$& 177 & 180 & 186 & 186 & 189 & 195 & 195 & 198 & 204 & 207 & 213 & 216 & 225 \\  
$n_{0}$& 96 & 93 & 87 & 87 & 84 & 78 & 78 & 75 & 69 & 66 & 60 & 57 & 48 \\ \hline
$c_{111}$& 14 & 3 & 5 & 11 & 0 & 2 & 8 & 15 & 5 & 12 & 2 & 9 & 6 \\
$c_{112}$& 7 & 3 & 5 & 7 & 3 & 5 & 7 & 9 & 7 & 9 & 7 & 9 & 9 \\ \hline
$l$ & 2  & 0 & 1&2&0&1&2&0&2&0&2&0&0  \\ \hline
\end{tabular}
\caption{F-theory multiplicities $n_{\pm q}$ of hypermultiplets with $\mathbb{Z}_3$ charge $\pm q$ and triple intersection numbers $c_{ijk}$ associated to 13 inequivalent Calabi-Yau threefolds that are genus one fibered with a 3-section over $\mathbb{P}^2$.}
	\label{tab:z3listred}
 \end{center}
\end{table}

\begin{enumerate}
\item \textbf{Multiplicity of  charged hypermultiplets:} The multiplicities of $\mathbb{Z}_3$ charged hypermultiples $n_{\pm1}$ are always divisible by three and the spectra therefore satisfy the (refined)
Monnier-Moore \cite{Monnier:2018nfs} discrete anomaly cancellation conditions (see Appendix~\ref{app:AbelianAnomalies} for a summary). However, we observe an even stronger condition.
All of the multiplicities satisfy the congruency relation
\begin{align}
n_{\pm 1} \in \{0,6\} \text{ mod }9 \,.
\label{eq:chargemulti}
\end{align}
In other words, none of the spectra exhibit $n_{\pm1} = 3$ mod $9$!
We will explain this in Section~\ref{sec:6danomalies} as a  highly non-trivial consequence of a particular choice of a quadratic refinement that F-theory uses in the generalized Green-Schwarz mechanism for global anomalies.
This suggests that six-dimensional theories with $T=0,\,G=\mathbb{Z}_3$ and $n_{\pm1} = 3$ mod $9$ are in the Swampland.

\item {\bf Degenerate spectra}: There are two pairs of models that lead to the same six-dimensional massless spectra in F-theory, respectively corresponding to the entries $3,4$ and $6,7$ in Table~\ref{tab:z3listred}.
At a geometric level, the two Calabi-Yau manifolds in each pair have the same Hodge numbers but inequivalent triple intersection numbers.
In both cases the discrete fermion anomaly does not vanish, i.e. $n_{\pm 1}=6\text{ mod }9$, and needs to be cancelled.
As will be discussed in Section~\ref{sec:6danomalies}, the discrete Green-Schwarz mechanism then leads to a topological term in the six-dimensional theory and a corresponding coupling constant $\ell\in\{1,2\}$, which is not fixed by the number of charged fermions.
Up to an overall sign, this can be identified under anomaly inflow with the 't~Hooft anomaly $[l]\in H^3(B\mathbb{Z}_3,U(1))=\mathbb{Z}_3$ of the induced global $\mathbb{Z}_3$ symmetry of the worldsheet theories of a single self-dual string.
The value of $l$, that will be derived in Sections~\ref{sec:2danomalies} and~\ref{sec:egan}, distinguishes the two theories in each pair.
In Appendix~\ref{sec:gvinvariants}, using recent results from~\cite{Schimannek:2021pau}, we show that the associated Jacobian fibrations $X_0$ encode different $\mathbb{Z}_3$-refined Gopakumar-Vafa invariants and therefore exhibit different excitation spectra for the self-dual strings.
\end{enumerate}

In the rest of the paper we will carefully work out the details of these statements.
We start by studying the cancellation of global $\mathbb{Z}_3$ anomalies in six dimensions.

\section{Discrete anomalies in 6d and their cancellation} 
\label{sec:6danomalies}

In this section we discuss the presence of anomalies for discrete gauge symmetries in six dimensions. Since these are non-perturbative anomalies in a gravitational theory one needs to take into account topology changes of the spacetime manifold in connection with the associated Dai-Freed anomalies \cite{Dai:1994kq, Witten:2015aba, Yonekura:2016wuc, Garcia-Etxebarria:2018ajm, Hsieh:2018ifc}. We then discuss a minimal mechanism to cancel such anomalies which is reminiscent of a discrete version of the usual Green-Schwarz mechanism for six-dimensional theories discussed in \cite{Green:1984bx, Sagnotti:1992qw}, and relies on the description of (anti-)self-dual tensor fields as boundary modes in a seven-dimensional theory \cite{Witten:1996hc, Belov:2006jd, Hsieh:2020jpj}. We will conclude the section with several remarks pointing out some interesting directions for further investigations. We will further focus on the case of a single self-dual 2-form in 6d, since it simplifies the discussion and describes the low-energy dynamics of F-theory models with a $\mathbb{P}^2$ base.
 
 \subsection{Discrete 6d anomalies from an anomaly theory}

The modern formulation of anomalies \cite{Freed:2014iua} describes the phase of the partition function of the theory of interest as the boundary value of an invertible topological field theory, the so-called anomaly theory $\mathcal{A}$, in one higher dimension. In our case of six-dimensional theories described by the partition function $Z$, one finds for a 7-manifold $N$ with boundary given by the 6-manifold $\partial N = M$ that
\begin{align}
Z[M] = e^{2 \pi i \int_N \mathcal{A}} \enspace \big|Z[M]\big| \,.
\end{align}
Of course, in order to evaluate the partition function, one further needs to specify backgrounds for the various symmetries as well as other structures, e.g. a Spin structure, on the manifold $M$ which extend to $N$. The absence of anomalies \cite{Dai:1994kq, Witten:2015aba, Yonekura:2016wuc, Garcia-Etxebarria:2018ajm, Hsieh:2018ifc, Debray:2022wcd} is then equivalent with the statement that this procedure does not depend on the extension from $M$ to $N$. Thus, gluing two different extensions along their common boundary, forming a closed 7-manifold with the imposed structure and field backgrounds, one finds that the absence of anomalies can be reformulated as
\begin{align}
e^{2 \pi i \int_Y \mathcal{A}} = 1 \quad \rightarrow \quad \int_Y \mathcal{A} = 0 \text{ mod } \mathbb{Z} \,,
\label{eq:anomalyfree}
\end{align}
for all closed 7-manifolds $Y$ with the necessary structure. These are classified by the bordism groups
\begin{align}
\Omega^{\xi}_7 (BG) \,,
\end{align}
where $\xi$ denotes a tangential structure on $Y$ and $BG$ indicates that one also incorporates maps into the classifying space of the symmetry group $G$, parametrizing the gauge field backgrounds.  

The trivial element in the bordism group can be described as a boundary of an 8-manifold, i.e.~is null-bordant. Via the connection between the anomaly theory $\mathcal{A}$ and various index theorems, this piece captures the perturbative anomalies of the six-dimensional theories. In our case these are the pure gravitational anomalies. The non-trivial elements, however, encode potential non-perturbative global anomalies. To verify their presence one has to find a complete set of generators of $\Omega^{\xi}_7 (BG)$ and then evaluate the anomaly theory $\mathcal{A}$ on them to show that \eqref{eq:anomalyfree} is violated.

In our case with gauge group $G=\mathbb{Z}_3$, we are interested in seven-dimensional Spin manifolds equipped with a $\mathbb{Z}_3$ bundle\footnote{In general the 6d supergravity theories should require a twisted String structure, which we believe should be related to our discussion as mentioned in Section \ref{subsec:furtherdiranom}.} which are described by
\begin{align}
\Omega^{\text{Spin}}_7 (B\mathbb{Z}_3) = \mathbb{Z}_9 \,.
\end{align}
This can be derived for example using the Atiyah-Hirzebruch spectral sequence, see e.g.\ \cite{Garcia-Etxebarria:2018ajm}. Consequently, there is a single generator, which in our case is given by the seven-dimensional lens space
\begin{align}
L^7_3 = S^7 / \mathbb{Z}_3 \,,
\end{align}
which can be viewed as the asymptotic boundary of $\mathbb{C}^4$ with complex coordinates $z_i$ divided by the $\mathbb{Z}_3$ action
\begin{align}
(z_1, z_2, z_3, z_4) \rightarrow (\omega \, z_1, \omega \, z_2, \omega \, z_3, \omega \, z_4) \quad \text{with} \quad \omega = e^{2 \pi i/3} \,.
\end{align}
Importantly, there is a non-trivial $\mathbb{Z}_3$ bundle on $L^7_3$ given by the fibration
\begin{align}
\begin{split}
\mathbb{Z}_3 \hookrightarrow & \, \, S^7 \\
& \, \downarrow \\
& \, L^7_3
\end{split}
\end{align}
This means that there will be a non-trivial discrete $\mathbb{Z}_3$ transition function acting on charged states once one traverses the torsion 1-cycle of $L^7_3$. This is associated to an element in $H^1 (L^7_3; \mathbb{Z}_3)$, which specifies the discrete gauge bundle\footnote{As usual the gauge bundles are classified by homotopy classes of maps from spacetime into the classifying space which here is given by $B\mathbb{Z}_3$. Since $B \mathbb{Z}_3 = K(\mathbb{Z}_3,1)$ the Eilenberg-Mac Lane space we see that the  $\mathbb{Z}_3$ bundles on $N$ are classified by cohomology classes $H^1(N; \mathbb{Z}_3)$. See also the discussion below.}.

The anomaly theory of a 6d Weyl fermion with charge $q \in \{0,1,2\}$ under the $\mathbb{Z}_3$ gauge symmetry is given by the $\eta$-invariant for the Dirac operator, see e.g.\ \cite{Hsieh:2020jpj},
\begin{align}
\eta^\text{D}_q [L^7_3] = - \frac{1}{3} \sum_{\ell = 1}^{2} \frac{e^{-2 \pi i \ell q/3}}{\big( 2 \sin (\pi \ell/3) \big)^4} \,.
\end{align}
Giving the values
\begin{align}
\eta^{\text{D}}_0 [L^7_3] = - \tfrac{2}{27} \,, \quad \eta^{\text{D}}_1 [L^7_3] = \eta^{\text{D}}_2 [L^7_3] = \tfrac{1}{27} \,.
\end{align}
The fact that the $\eta$-invariant does not vanish even for $q=0$ indicates the presence of a perturbative gravitational anomaly. Indeed, since $\Omega^{\text{Spin}}_7 (\text{pt}) = 0$ forgetting about the $\mathbb{Z}_3$ bundle we can describe $L^7_3$ as the boundary of a 8d Spin manifold $W$. Using the connection of the $\eta$-invariant to various index theorems, see e.g.\ \cite{Hsieh:2020jpj}, we find
\begin{align}
\eta^{\text{D}}_0 [L^7_3] = - \int_W \hat{A} (R) \text{ mod } \mathbb{Z} \,.
\label{eq:gravanom}
\end{align}
Here, $\hat{A}$ is the A-hat genus, whose 8-form part reads
\begin{align}
\hat{A}_8 = \tfrac{1}{5760} (7 p_1^2 - 4 p_2)
\end{align}
with Riemann curvature $\tfrac{R}{2\pi} = \mathcal{R}$ entering the Pontryagin classes
\begin{align}
p_1 = - \tfrac{1}{2}  \text{tr} \mathcal{R}^2 \,, \quad p_2 = \tfrac{1}{8} \Big( \big( \text{tr} \mathcal{R}^2 \big)^2 - 2 \, \text{tr} \mathcal{R}^4 \Big) \,.
\end{align}
Plugging this into \eqref{eq:gravanom} one finds
\begin{align}
- \int_W \hat{A} (R) = - \tfrac{1}{5760} \int_W \Big( \text{tr} \mathcal{R}^4 + \tfrac{5}{4} \big( \text{tr} \mathcal{R}^2 \big)^2 \Big) \,,
\end{align}
which, as necessary, reproduces the gravitational contribution of a Weyl fermion or hypermultiplet to the anomaly polynomial. We can therefore extract the global anomaly involving the discrete $\mathbb{Z}_3$ symmetry by subtracting the pure gravitational part, which for the full 6d theory is cancelled anyway by the anomaly contribution of other fields, by considering the bordism invariant
\begin{align}
\big( \eta^{\text{D}}_q - \eta^{\text{D}}_0 \big) [L^7_3] = \tfrac{1}{9} \,, \quad \text{with} \quad q \in \{ 1, 2 \} \,.
\end{align}
Since for the 6d theories of interest the only charged degrees of freedom are contained in hypermultiplets, the anomaly contribution comes exclusively from the presence of charged Weyl fermions. For a theory with $n_1$ hypermultiplets of charge $q = 1$ and $n_{-1}$ hypermultiplets of charge $q = 2 \sim -1$ the part of the 7d anomaly theory capturing global anomalies involving the $\mathbb{Z}_3$ gauge symmetry is given by
\begin{align}
\mathcal{A}_{\mathbb{Z}_3}^F = n_1 \, \eta^{\text{D}}_1 + n_{-1} \, \eta^{\text{D}}_2 - (n_1 + n_{-1}) \, \eta^{\text{D}}_0 \,,
\end{align}
which, evaluated on $L^7_3$, produces
\begin{align}
\int_{L^7_3} \mathcal{A}_{\mathbb{Z}_3}^F = \tfrac{1}{9} \, (n_1 + n_{-1}) \,.
\end{align}
This indicates that the theory is anomalous unless the number of charged hypermultiplets is divisible by nine.
As we review in Appendix~\ref{app:AbelianAnomalies}, this agrees with the findings of Monnier and Moore~\cite{Monnier:2018nfs}.
However, since we have seen that F-theory produces models with $n_{\pm 1} = (n_1 + n_{-1})$ not divisible by nine, this cannot be the only contribution to the anomaly.

\subsection{Anomaly cancellation and quadratic refinements}

In general there are various ways to get rid of the discrete gauge anomalies above. For example it could happen that the fundamental gauge group is not $\mathbb{Z}_3$ but $\mathbb{Z}_9$ with fermions charged in $3 \mathbb{Z}$ mod $9$. In this case the discrete anomaly would vanish, see e.g.\ \cite{Ibanez:1991hv, Hsieh:2018ifc, Garcia-Etxebarria:2018ajm}. In our models where we have an explicit high-energy completion in terms of the underlying F-theory model, we do not believe that such a gauge enhancement occurs since the $\mathbb{Z}_3$ has a geometric realization. 

Another possibility to cancel anomalies is to add additional topological sectors whose (higher-form symmetry) anomalies are related to the gauge symmetry background, a topological version of the Green-Schwarz mechanism \cite{Garcia-Etxebarria:2017crf}. While the topological sectors do not modify the local dynamics of the theory, the completeness hypothesis still demands the presence of possibly extended objects coupling to the fields in the topological theory. This would result in a variety of different UV completions that all differ by their spectrum of extended objects and topological sectors. The consistency of such a `discrete landscape' is not clear yet, see also \cite{Debray:2021vob}.

This is why we focus on a minimal way to cancel the discrete anomalies which does not modify the gauge group or additional topological sectors of the theory and is closely related to a discrete version of the usual Green-Schwarz mechanism \cite{Green:1984bx, Sagnotti:1992qw}. For that we modify the transformation properties of the self-dual 2-form in six dimensions depending on the discrete $\mathbb{Z}_3$ symmetry background, a similar mechanism in ten dimension was discussed in \cite{Debray:2021vob}. In order to do so we first discuss the realization of the (anti-)self-dual tensor fields as boundary modes of 3-form fields in seven dimensions closely following the presentation of \cite{Hsieh:2020jpj}. 

\subsection*{Self-dual fields and quadratic refinements}

As was discussed in \cite{Witten:1996hc, Belov:2006jd, Hsieh:2020jpj} it is useful to describe chiral degrees of freedom as boundary modes of non-chiral fields in one higher dimension. In our case we want to describe the self-dual 2-form on $M$ as a boundary mode\footnote{Or rather a mode localized close to the boundary.} of a 3-form field $C$ in seven dimensions, i.e.~on $N$ with $\partial N = M$. The way to describe (higher-form) fields is via differential cohomology $\check{H}^\bullet(N)$, as it was introduced in \cite{10.1007/BFb0075216}, see also \cite{Freed:2006yc, Hsieh:2020jpj}.
Roughly, the information in the associated differential character $\check{F}_C$, an element of $\check{H}^4(N)$, is the fieldstrength $\mathcal{F}_C$, the information about (torsion) fluxes $F_C$, as well as the holonomies of $C$. Moreover, one chooses Dirichlet boundary conditions 
\begin{align}
\check{F}_C \big|_{\partial N = M} = 0 \,.
\label{eq:boundaryC}
\end{align}
The naive Euclidean action one would write down for $C$ is given by, see \cite{Hsieh:2020jpj}, 
\begin{align}
S \sim 2 \pi \int_N \Big( \tfrac{1}{2g^2} \mathcal{F}_C \wedge \ast \mathcal{F}_C - i \tfrac{\kappa}{2} C \wedge \mathcal{F}_C - i C \wedge \mathcal{F}_X \Big) \,,
\end{align}
with the gauge invariant fieldstrength $\mathcal{F}_C$ interpreted as an element in de~Rham cohomology.
Here, $X$ denotes the coupling to a background 3-form gauge field $X$ also part of a differential character $\check{X} \in \check{H}^4(N)$, which will play an important role later.
Next, one needs to lift this action to differential cohomology.

While the first term, encoding the kinematics of $C$, is unproblematic, the two topological terms need to be treated with care. Using the product
\begin{align}
\star: \quad \check{H}^4 (N) \times \check{H}^4 (N) \rightarrow \check{H}^8 (N) \,,
\end{align}
one can define a differential cohomology pairing $(\check{F}_X,\check{F}_C)$ by evaluating the holonomy of $\check{F}_X \star \check{F}_C$ on $N$. With this one can also describe the second term given now by $\tfrac{1}{2} (\check{F}_C, \check{F}_C)$. However, the multiplication by $\tfrac{1}{2}$ is not well-defined and instead needs to be described by a quadratic refinement\footnote{Since the differential character contain integer cohomology information, such as the flux described in $H^4(N,\mathbb{Z})$, one cannot simply multiply with rational numbers such as $\tfrac{1}{2}$ above.}, which we will denote by $\mathcal{Q}$ and can be understood as a map
\begin{align}
\mathcal{Q}: \quad \check{H}^4 (N) \rightarrow U(1) \,.
\end{align}
The relation between the differential cohomology pairing and $\mathcal{Q}$ is given by
\begin{align}
(\check{F}_X, \check{F}_C) = \mathcal{Q} (\check{F}_X + \check{F}_C) - \mathcal{Q} (\check{F}_X) - \mathcal{Q} (\check{F}_C) + \mathcal{Q} (0) \,.
\label{eq:qrtor}
\end{align}
With these generalizations the Euclidean action of $C$ on $N$ on the level of differential cohomology is given by \cite{Hsieh:2020jpj}
\begin{align}
S = \int_N \Big( \tfrac{2 \pi}{2g^2} \mathcal{F}_C \wedge \ast \mathcal{F}_C \Big) - 2 \pi i \kappa \, \mathcal{Q} (\check{F}_C) - 2 \pi i \, (\check{F}_C, \check{F}_X) \,.
\label{eq:diffcohoaction}
\end{align}
Together with the boundary conditions \eqref{eq:boundaryC} the equations of motion shows that there is a localized boundary mode 
\begin{align}
C = d ( e^{m r} ) \wedge B \,,
\end{align}
where $r < 0$ denotes a coordinate perpendicular to the boundary $M$, that can be identified with the $B$ field and is self-dual for the choice
\begin{align}
\kappa = -1 \,,
\end{align}
where the minus sign is due to the fact that we work in Euclidean signature.

In order to determine the anomaly contribution we can neglect the kinetic term, by sending the coupling $g \rightarrow \infty$.
Evaluating the path integral carefully, see \cite{Hsieh:2020jpj}, one finds that the anomaly of the self-dual tensor field $B$ above is given by
\begin{align}
\mathcal{A}^B = \kappa \big( \mathcal{A}^B_{\text{grav}} - \mathcal{Q} (\check{F}_X)\big) \,.
\end{align}
The first term denotes the gravitational anomaly which for six dimensions is given by
\begin{align}
\mathcal{A}^B_{\text{grav}} = 28 \, \eta^\text{D}_0 \,,
\label{eq:anomdual}
\end{align}
i.e.~the same as 28 uncharged chiral fermions. Notably, there is a second term that depends on the background $\check{F}_X$, this term will be playing a crucial role in the cancellation of the discrete gauge anomalies. Note also that with the contribution \eqref{eq:anomdual} the irreducible gravitational anomaly is cancelled, see Appendix \ref{app:gravanom}.

\subsection*{Quadratic refinement and anomaly cancellation}

We see that the background field enters the anomaly theory of the self-dual tensor. By analyzing the boundary behavior one finds that $\check{F}_X$ acts as background field for the natural 2-form symmetry of $B$.
Since $B$ is self-dual, its electric and magnetic 2-form symmetries are identified. The modern interpretation of a Green-Schwarz mechanism as a 't Hooft anomaly for the higher-form symmetries of the involved $p$-form fields \cite{Hsieh:2020jpj, Lee:2022spd} can now be applied directly to our setup. For that we relate the background $\check{F}_X$ to the discrete gauge data of the underlying background.

First, even for vanishing gauge background the perturbative gravitational anomaly is not yet fully cancelled. This is due to the contribution of the gravitino to the reducible gravitational anomaly, which contributes 
\begin{align}
I_8 \supset - \tfrac{1}{4} p_1^2 =  - \tfrac{1}{16} \big( \text{tr} \mathcal{R}^2 \big)^2 \,,
\end{align}
to the anomaly polynomial, see Appendix \ref{app:gravanom}. To compensate this reducible part of the anomaly we demand that
\begin{align}
- \kappa \int_N \mathcal{Q} (0) = \tfrac{1}{4} \int_W p_1^2 \,, 
\label{eq:gravred}
\end{align}
for $N = \partial W$ and $N$ interpreted as a seven-dimensional Spin manifold. With this the perturbative gravitational anomalies are cancelled completely.\footnote{Note that this shows that the 6d supergravity theories with F-theory origin implement a different quadratic refinement compared to \cite{Hsieh:2020jpj} for which $\mathcal{Q}(0) = - \mathcal{A}^B_{\text{grav}}$.}

The remaining global anomalies involving the discrete gauge data needs to be cancelled by 
\begin{align}
- \kappa \, \widetilde{\mathcal{Q}}(\check{F}_X) = - \kappa \big(\mathcal{Q} (\check{F}_X) - \mathcal{Q} (0) \big) \,,
\label{eq:anomcontr}
\end{align}
which is non-zero only for non-trivial gauge background $\check{F}_X$. We describe how this works in detail for gauge group $\mathbb{Z}_3$ next.

\subsection*{Cancellation of $\mathbb{Z}_3$ gauge anomalies}

Let us apply the general approach above to our supergravity theory in six dimensions with $\mathbb{Z}_3$ gauge symmetry. We have seen that the only non-trivial bordism generator is given by $L^7_3$ with a non-trivial $\mathbb{Z}_3$ gauge bundle. The integer cohomology of the lens space is given by
\begin{align}
H^n (L^7_3; \mathbb{Z}) = \begin{cases} \mathbb{Z} \,, \quad \text{for } n \in \{0 \,, 7 \} \,, \\ \mathbb{Z}_3 \,, \quad \text{for } n \in \{2 \,, 4 \,, 6 \} \,, \\
0 \,, \quad \text{otherwise} \,. \end{cases}
\end{align}
The quadratic refinement $\mathcal{Q}$ is with respect to the torsion pairing
\begin{align}
H^4 (L^7_3; \mathbb{Z}) \times H^4 (L^7_3; \mathbb{Z}) \rightarrow \mathbb{R} / \mathbb{Z} \sim U(1) \,.
\end{align}
Denoting by $x$ the generator of $H^4 (L^7_3; \mathbb{Z})$, see also \cite{Hsieh:2020jpj}, one has for $a,b \in \mathbb{Z} \text{ mod }3$
\begin{align}
(a \, x , b \, x) = \tfrac{2}{3} \, a b \text{ mod } \mathbb{Z} \,.
\label{eq:torspair}
\end{align}
This class can be lifted to a differential character $\check{F}_x$ with trivial field strength, but non-trivial holonomy $A_x$ and topological fluxes $F_x$
\begin{align}
[F_x] = x \in H^4 (L^7_3; \mathbb{Z}) \,, \quad [F_x] = \text{Bock} [A_x] \,,
\label{eq:diflift}
\end{align}
where $[A_x]$ is the gauge invariant information of the 3-form holonomy, i.e.~an element of $H^3 \big(L^7_3; \text{U}(1)\big)$, and $\text{Bock}$ denotes the Bockstein homomorphism associated to the short exact sequence
\begin{align}
0 \longrightarrow \mathbb{Z} \longrightarrow \mathbb{R} \longrightarrow U(1) \sim \mathbb{R}/\mathbb{Z} \longrightarrow 0 \,.
\label{eq:Bockseq}
\end{align}
Demanding that the quadratic refinement vanishes for the trivial element in $H^4 (L^7_3; \mathbb{Z})$ and using the property \eqref{eq:qrtor}, one finds three possibilities for $\widetilde{\mathcal{Q}}$:
\begin{align}
\renewcommand{\arraystretch}{1.3}
\begin{tabular}{c | c | c}
$a$ & $1$ & $2$ \\ \hline \hline
$\widetilde{\mathcal{Q}}^0 (a \, \check{F}_x)$ & $\tfrac{1}{3}$ & $\tfrac{1}{3}$ \\ \hline 
$\widetilde{\mathcal{Q}}^1 (a \, \check{F}_x)$ & $\tfrac{2}{3}$ & $0$ \\ \hline 
$\widetilde{\mathcal{Q}}^2 (a \, \check{F}_x)$ & $0$ & $\tfrac{2}{3}$ \\ \hline  
\end{tabular}
\renewcommand{\arraystretch}{1}
\label{eq:qrvalues}
\end{align}
These can be written in terms of the torsion pairing of the lens space \eqref{eq:torspair} and its fractional first Pontryagin class $\tfrac{1}{4} p_1 (L^7_3) = x$ as
\begin{align}
\widetilde{\mathcal{Q}}^k (a \, \check{F}_x) = - \big( a \, x, a \, x + \tfrac{k}{4} \, p_1 \big) = - \tfrac{2}{3} a (a+k) \,.
\end{align}
We are left with the identification of $\check{F}_X$ in terms of the $\mathbb{Z}_3$ gauge background.

The $\mathbb{Z}_3$ bundles on $L^7_3$ are classified by homotopy classes of maps $L^7_3$ into the classifying space $B\mathbb{Z}_3$.
Since $B \mathbb{Z}_3$ is given in terms of the Eilenberg-MacLane space $K(\mathbb{Z}_3,1)$, we find that $\mathbb{Z}_3$ bundles are classified by homotopy classes of maps from spacetime $N$ into $K(\mathbb{Z}_3,1)$, i.e.~by $H^1 (N; \mathbb{Z}_3)$, which for $L^7_3$ is given by
\begin{align}
H^1 (L^7_3; \mathbb{Z}_3) = \mathbb{Z}_3 \,.
\end{align}
We will call the generator $A$. Using the short exact sequence
\begin{align}
0 \longrightarrow \mathbb{Z} \longrightarrow \mathbb{Z} \longrightarrow \mathbb{Z}_3 \longrightarrow 0 \,,
\end{align}
we obtain another Bockstein homomorphism
\begin{align}
\beta: \quad H^1 (L^7_3 ; \mathbb{Z}_3) \rightarrow H^2 (L^7_3; \mathbb{Z}) = \mathbb{Z}_3 \,,
\end{align}
from which we can build a class in $H^4 (L^7_3; \mathbb{Z}) = \mathbb{Z}_3$ using the cup product
\begin{align}
\beta(A) \cup \beta(A) \in H^4 (L^7_3; \mathbb{Z}) \,.
\end{align}
Alternatively, this can be seen as the pull-back of the non-trivial group cohomology class in $H^4 (B\mathbb{Z}_3; \mathbb{Z})$ under the classifying map. Thus, choosing the generator appropriately we have
\begin{align}
\beta (A) \cup \beta (A) = x \,,
\end{align}
which can be lifted to the differential character $\check{F}_x$, see \eqref{eq:diflift}. Since $\beta$ is a homomorphism and $x$ is quadratic in $A$ one has
\begin{align}
\beta (2A) \cup \beta (2A) = 4 \big( \beta (A) \cup \beta (A) \big) \sim \beta (A) \cup \beta (A) = x \,,
\end{align}
since it is measured mod $3$. This means that $\check{F}_x$ only distinguishes between the presence of a non-trivial $\mathbb{Z}_3$ bundle and its absence but maps both $A$ as well as $2A$ to the generating element $x \in H^4 (L^7_3;\mathbb{Z})$. We now define the gauge background $\check{F}_X$ as\footnote{Since we are working mod $3$ this is the same as $\check{F}_X = \pm \check{F}_x$.}
\begin{align}
\check{F}_X = \ell \, \check{F}_x \,, \quad \ell \in \{ 1 \,, 2\} \,.
\end{align}
This means we can read off the contribution of the non-trivial background $\check{F}_X$ to the anomaly theory by combining \eqref{eq:anomcontr} and \eqref{eq:qrvalues}. The full anomaly theory for global anomalies involving the $\mathbb{Z}_3$ gauge sector in the presence of a single self-dual tensor field reads
\begin{align}
\mathcal{A}_{\mathbb{Z}_3} = \mathcal{A}^F_{\mathbb{Z}_3} + \widetilde{\mathcal{Q}}^k (\check{F}_X) = \mathcal{A}^F_{\mathbb{Z}_3} + \widetilde{\mathcal{Q}}^k (\ell \check{F}_x) \,.
\end{align}
For the F-theory vacua with non-trivial $\mathbb{Z}_3$ anomaly, see \eqref{eq:chargemulti}, we found
\begin{align}
\mathcal{A}^F_{\mathbb{Z}_3} = \tfrac{2}{3} \text{ mod } 3 \,,
\end{align}
which singles out the quadratic refinement $\widetilde{\mathcal{Q}}^0$, which evaluates to $\tfrac{1}{3}$ for the non-trivial bordism generator $L^7_3$.

It therefore appears that F-theory vacua with discrete $\mathbb{Z}_3$ gauge factor single out a specific quadratic refinement in the discrete version of the Green-Schwarz mechanism, which can be written as the negative of the torsion pairing
\begin{align}
\widetilde{\mathcal{Q}}^0 (\ell \check{F}_x ) = - \ell^2 (x, x) = \tfrac{1}{3} \ell^2 \,.
\end{align}
We see that this does not fix the cancellation term completely, and one is left with $\ell \in \{1 \,, 2\}$, both of which cancel the discrete anomalies in six dimensions. Luckily $\ell$ does have a physical consequence, in that it determines different anomaly inflow onto the self-dual string and thus can be extracted from the localized degrees of freedom on the string worldsheet. This can be seen form the action \eqref{eq:diffcohoaction}, by considering the boundary contribution of the last term, producing a term schematically of the form
\begin{align}
\int_{\partial N} B \cup F_X = \ell \int_M B \cup x = \ell \int_M B \cup \beta(A) \cup \beta(A) \,.
\label{eqn:BAAterm}
\end{align}
We will extract this information in the next section via the topological string partition function and find that indeed both possibilities for $\ell$ are realized in F-theory vacua\footnote{This also suggests that $\widetilde{\mathcal{Q}}^0$ is singled out in F-theory, since it is the only quadratic refinement that produces non-trivial contributions to the discrete anomaly for both values of $\ell$.}.

\subsection{Anomaly inflow on non-critical strings}
\label{subsec:anominflow}

The six-dimensional $\mathbb{Z}_3$ gauge symmetry induces a global $\mathbb{Z}_3$ symmetry on the worldvolume theories of non-critical strings.
For continuous symmetries, the relationship between the six-dimensional local gauge anomalies and the corresponding two-dimensional 't~Hooft anomalies has been worked out in~\cite{Shimizu:2016lbw}.

The usual Green-Schwarz coupling takes the form $B\wedge I_4$ and the corresponding local anomaly exactly cancels the contribution of the anomaly polynomial, which factorizes as
\begin{align}
    I_8\simeq \frac12 I_4\wedge I_4\,.
\end{align}
Up to contributions related to the gravitational anomaly, the anomaly inflow then essentially identifies $I_4$ with the anomaly polynomial of the two-dimensional theory on the string worldsheet~\cite{Shimizu:2016lbw}.
While an analogous calculation of the inflow for the global discrete anomaly requires a careful treatment of the corresponding Bianchi identity, with potential subtleties related to the String structure being discussed in the next subsection, this result for local anomalies admits a very natural generalization.

As will be discussed further in Section~\ref{sec:2danomalies}, the 't~Hooft anomaly associated to a two-dimensional global $\mathbb{Z}_n$ symmetry is classified by an element~\cite{cmp/1104180750}
\begin{align}
    [l]\in H^3(B\mathbb{Z}_n,U(1))=\mathbb{Z}_n\,.
\end{align}
and if $n$ is odd one can use the Atiyah-Hirzebruch spectral sequence to identify
\begin{align}
\label{eq:2DAnomaly}
H_3(B\mathbb{Z}_n,\mathbb{Z})\simeq\Omega^{\text{Spin}}_3(B\mathbb{Z}_n)\,.
\end{align}
Since $H^3(B\mathbb{Z}_n,U(1))\simeq \text{Hom}(H_3(B\mathbb{Z}_n,\mathbb{Z}),U(1))$, the element $[l]$ can be directly identified with the anomaly theory itself.
Using the universal coefficient theorem one can also show that
\begin{align}
    H^3(B\mathbb{Z}_n,U(1))\simeq H^4(B\mathbb{Z}_n,\mathbb{Z})\,,
\end{align}
and the pullback under the inclusion $L^7_3\subset L^\infty_3=B\mathbb{Z}_3$ further identifies
\begin{align}
    H^4(L^7_3,\mathbb{Z})\simeq H^4(B\mathbb{Z}_n,\mathbb{Z})\,.
\end{align}

Considering the six-dimensional discrete Green-Schwarz coupling in~\eqref{eqn:BAAterm}, these isomorphisms identify the image  $\ell\,\beta(A)\cup\beta(A)$ in $H^3(B\mathbb{Z}_3,U(1))$ with the two-dimensional $\mathbb{Z}_3$ 't~Hooft anomaly $[l]$, such that up to a choice of generator
\begin{align}
    [l]= [\pm\ell]\in\mathbb{Z}_3\,.
\end{align}

Note that in the presence of additional 2-form fields, both $\ell$ and $B$ should carry additional indices that respectively take values in the string charge lattice and its dual.
Since we focus on the case where there is only a single self-dual 2-form we can suppress this index.

\subsection{Discrete anomaly cancellation and String structures?}
\label{subsec:furtherdiranom}

The mechanism above made extensive use of the construction of self-dual fields as boundary modes in a 7d theory. The mathematical framework is that of differential cohomology. While this captures the gauge invariant information contained in the various higher-form fields it does not explicitly capture the modification of the Bianchi identity due to the Green-Schwarz coupling, describing a higher-group structure. However, this mixture between the higher-form symmetries of the 2-form fields and the gauge background could be captured within the framework of (twisted) String structures, see e.g.\ \cite{Sati:2009ic}.

For the usual String structure one demands that 
\begin{align}
\tfrac{1}{2} p_1 = 0 \,.
\end{align}
Since this is a structure of the 6d theory it should also extend over the seven-dimensional manifold defined for the anomaly theory, similar to the Spin structure above. But this means that for closed 7-manifolds one needs to restrict to manifolds with vanishing characteristic class $\tfrac{1}{2} p_1$.
However, for the generator $L^7_3$ above one has
\begin{align}
\tfrac{1}{2} p_1 = 2 x \neq 0 \in H^4 (L^7_4; \mathbb{Z}) \,.
\end{align}
Similar to $x$ this has a lift to a differential character, which we denote as $\tfrac{1}{2} \check{p}_1 = 2 \check{F}_x$.

For twisted String structures one allows the characteristic classes of the tangent bundle to be compensated by characteristic classes of a gauge background. Thus, the presence of a non-trivial $\mathbb{Z}_3$ background might be exactly what we need in the situation at hand. Indeed we see that
\begin{align}
\tfrac{1}{2} \check{p}_1 \pm \check{F}_X = 2 \check{F}_x + \check{F}_x = 0 \in \check{H}^4 (L^7_4) \,,
\end{align}
for $\pm$ sign for $\ell = 1$ and $\ell = 2$, respectively. It would further be interesting to explore what role the cancellation of the remaining (reducible) gravitational anomaly in \eqref{eq:gravred} implies for the potential definition of such a discretely twisted String structure.

This shifts via (fractional) characteristic classes also point towards the use of shifted differential cohomology as in \cite{Monnier:2018nfs}, see also \cite{Monnier:2017oqd}, which also pointed out the importance of a definition of a trivialization of a Wu structure in the intermediate steps of the anomaly cancellation. We believe that the proper definition of the quadratic refinement form first principles and its connection to twisted String structures and shifted differential cohomology will shed some light on the topological intricacies of Green-Schwarz anomaly cancellation in general, but we postpone a detailed study to future work. 
 
\section{Discrete anomalies in 2d theories}
\label{sec:2danomalies}
In order to connect the six-dimensional discrete anomaly to the geometry of the compactification space in F-theory, we are going to use the modular properties of elliptic genera of non-critical strings.
Before doing that in Section~\ref{sec:egan}, we will study the general properties of the partition functions associated to two-dimensional conformal field theories with a global $\mathbb{Z}_n$ symmetry that exhibits a 't Hooft anomaly.

\subsection{Discrete 't Hooft anomalies as a failure of modularity}
\label{sec:CFTtHooft}
Let us consider a general two-dimensional conformal field theory on a torus with a global finite Abelian symmetry $G$ and assume that all local anomalies are cancelled.
We can then define partition functions $Z^{(g,h)}$, where $g,h\in G$ respectively correspond to twists along the time and space direction.
This can be equivalently expressed as
\begin{align}
	Z^{(g,h)}=\text{Tr}_{\mathcal{H}_g}\left(\ldots \cdot h\right)\,,
\end{align}
where $\mathcal{H}_g$ is the Hilbert space of $g$-twisted states and the insertion of $h$ is sometimes called a \textit{twining}.

If $G$ does not suffer from global 't~Hooft anomalies, it can be gauged.
This leads to an orbifold of the original CFT and the $SL(2,\mathbb{Z})$ invariant partition function is given by
\begin{align}
	Z_O=\frac{1}{|G|}\sum\limits_{g,h\in G}Z^{(g,h)}\,.
\end{align}
However, it was argued in~\cite{Numasawa:2017crf,Chang:2018iay,Kikuchi:2019ytf} that if the 't Hooft anomaly does not vanish this manifests itself as a non-vanishing phase in the relation
\begin{align} 
	T^n Z^{(g,1)}=\gamma_g^\alpha Z^{(g,1)}\,,
        \label{eqn:tnphase}
\end{align}
where $T$ acts as $\tau\rightarrow\tau+1$, $n$ is the order of $g$ and the phase $\gamma_n^\alpha$ depends on a class in $H^3(BG,U(1))$ that represents the anomaly.
Using $S Z^{(g,1)}=Z^{(1,g)}$ with $S$ acting as $\tau\rightarrow -1/\tau$, this is equivalent to the failure of invariance of $Z^{(1,g)}$ under
\begin{align}
	U_n^{-1}\equiv S T^n S:\quad\tau\mapsto\frac{\tau}{1-n\tau}\,.
	\label{eqn:utrafo}
\end{align}
We will explicitly derive this result for a general finite group $G$, as well as the precise form of the phase, using topological symmetry line operators below in Section~\ref{sec:toplines}.

Let us first specialize to the case  $G=\mathbb{Z}_n$, such that the $(r,s)$ twisted fields satisfy boundary conditions
\begin{align}
	\phi(x_1+2\pi,x_2)=\mu_n^{r}\phi(x_1,x_2)\,,\quad \phi(x_1,x_2+2\pi)=\mu_n^{s}\phi(x_1,x_2)\,,
\end{align}
with $\mu_n=e^{2\pi i /n}$ and the partition functions take the form
\begin{align}
	Z^{(r,s)}=\text{Tr}_{\mathcal{H}_r}\left(\ldots\mu_n^{s J_{\mathbb{Z}_n}}\right)\,,
\end{align}
where we use $J_{\mathbb{Z}_n}$ to denote the $\mathbb{Z}_n$ charge operator.
The twisted boundary conditions transform under large diffeomorphisms $\gamma\in SL(2,\mathbb{Z})$ as
\begin{align}
\label{eq:ModBoundary}
	\gamma=\left(\begin{array}{cc}a&b\\c&d\end{array}\right):\,(r,s)\mapsto (ar+cs,br+ds)\,,
\end{align}
and are in general only preserved by the subgroup $\Gamma(n)\subset SL(2,\mathbb{Z})$~\cite{ZUBER1986127}, which in particular contains the $U_n$-transformation~\eqref{eqn:utrafo}.
However, note that the boundary conditions $(r,0)$ and $(0,s)$ are actually preserved by the respective larger subgroups $\Gamma^1(n),\Gamma_1(n)\subset \Gamma(n)$.
The 't Hooft anomaly corresponds to a cocycle
\begin{align}
	[l]\in H^3\left(B\mathbb{Z}_n,U(1)\right)=\mathbb{Z}_n\,,
\end{align}
and, as we derive below in Section~\ref{sec:toplines}, this is measured by the phases that $Z^{(1,0)}$ and $Z^{(0,1)}$ respectively acquire under $T^n$- and $U_n$-transformations~\eqref{eqn:utrafo},
\begin{align}
	T^n:\,Z^{(1,0)}\mapsto \mu_n^{l} Z^{(1,0)}\,,\quad U_n:\,Z^{(0,1)}\mapsto \mu_n^{-l} Z^{(0,1)}\,.
	\label{eqn:0nthooft}
\end{align}

\subsection{Modular anomalies from topological line operators}
\label{sec:toplines}
We now use the topological line operators implementing the discrete global 0-form symmetry on the worldsheet in a 2-dimensional CFT~\cite{Gaiotto:2014kfa,Bhardwaj:2017xup} to derive the relation of the 't Hooft anomaly to the modular properties of the twisted sector partition functions. 
Note that topological line operators have also been used in~\cite{Chang:2018iay} to deduce the result for $G=\mathbb{Z}_3$ but the corresponding argument does not easily generalize to arbitrary groups.
We will instead develop an iterative algorithm that leads to a closed expression which is valid for any finite group $G$.

As already mentioned above, the 't~Hooft anomaly of a $G$ symmetry corresponds to an element in group cohomology~\cite{cmp/1104180750}
\begin{align}
    [\alpha]\in H^3(BG,U(1))\,.
    \label{eqn:ancoh}
\end{align}
For a discrete group, an element $[\alpha]\in H^k(BG,U(1))$ can be represented by a map $\alpha:\,G^k\rightarrow U(1)$ that is annihilated by the coboundary operator
\begin{align}
	\begin{split}
	\delta\alpha(g_1,\ldots,g_{k+1})=&\alpha(g_1,\ldots,g_k)^{(-1)^{k+1}}\alpha(g_2,\ldots,g_{k+1})\\
	&\times\prod\limits_{i=1}^k\alpha(g_1,\ldots,g_ig_{i+1},\ldots,g_{k+1})^{(-1)^i}\,.
	\end{split}
	\label{eqn:coboundary}
\end{align}
Physically, the cohomology class~\eqref{eqn:ancoh} can be interpreted from two rather different perspectives.
On the one hand, assuming for the moment that $G=\mathbb{Z}_n$ with $n$ odd, we can use $H_3(B\mathbb{Z}_n,\mathbb{Z})=\Omega^{\text{Spin}}_3(B\mathbb{Z}_n)$ as well as
\begin{align}
	H^3(BG,U(1))=\text{Hom}\left(H_3(BG,\mathbb{Z}),U(1)\right)\,,
\end{align}
to think of it as a bordism invariant that defines the 3-dimensional invertible topological field theory which encodes the global discrete anomaly.
On the other hand, it corresponds to a crossing relation of the topological line operators that generate the action of the symmetry.

We will briefly introduce the very basics of symmetry line operators that are relevant for our argument but refer to~\cite{Gaiotto:2014kfa,etingof2016tensor,Bhardwaj:2017xup,Chang:2018iay} for details. 
The topological operators are oriented lines $\mathcal{L}_g$ that are labeled by group elements $g\in G$.
Contracting a loop $\mathcal{L}_g$ around a local operator $\mathcal{O}_p$ that transforms in some unitary representation $U$ of $G$ amounts to acting on the operator with $U_g$.
On a cylinder, a line that wraps the spacelike circle can be pushed on the boundary and then acts on the corresponding Hilbert space.
On the other hand, a line that is oriented along the timelike direction should rather be interpreted as a defect that changes the boundary conditions and therefore changes the Hilbert space to the Hilbert space of $g$-twisted states.

\begin{figure}[h!]
	\centering
\begin{tikzpicture}[scale=.8]
	\begin{scope}[xshift=0cm,yshift=0cm]
	\begin{scope}[xshift=0cm,yshift=0cm]
	\node[align=center] at (3,-3) {\includegraphics[width=.24\linewidth]{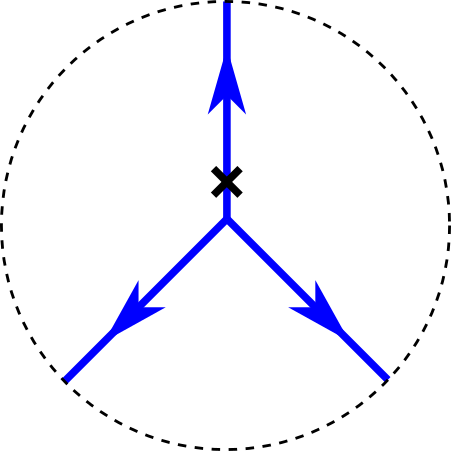}};
	\node[] at (1.6,-3.3) {\color{blue}$g_1$};
	\node[] at (4.4,-3.3) {\color{blue}$g_2$};
	\node[] at (3.7,-1.8) {\color{blue}$g_3$};
	\end{scope}
	\begin{scope}[xshift=6cm,yshift=0cm]
	\node[align=center] at (3,-3) {\includegraphics[width=.24\linewidth]{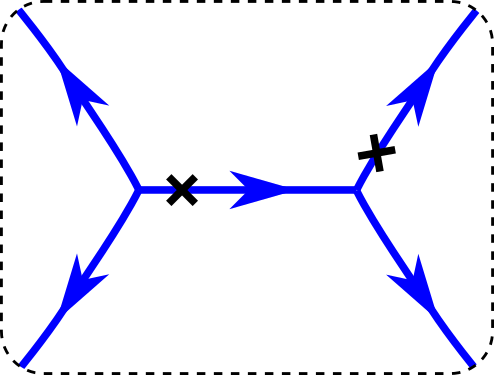}};
	\node[] at (2,-2) {\color{blue}$g_1$};
	\node[] at (2,-4) {\color{blue}$g_2$};
	\node[] at (4,-4) {\color{blue}$g_3$};
	\node[] at (4,-2) {\color{blue}$g_4$};
	\end{scope}
	\begin{scope}[xshift=13.5cm,yshift=0cm]
	\node[align=center] at (3,-3) {\includegraphics[width=.24\linewidth]{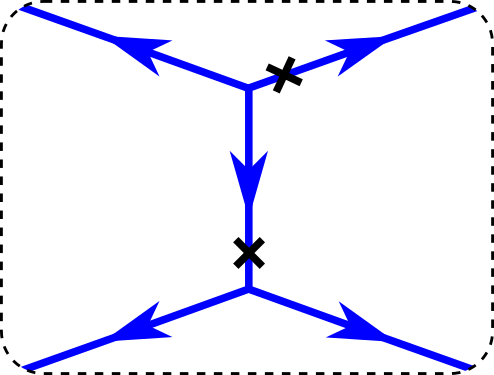}};
	\node[] at (2,-2.2) {\color{blue}$g_1$};
	\node[] at (2,-3.8) {\color{blue}$g_2$};
	\node[] at (4,-3.8) {\color{blue}$g_3$};
	\node[] at (4,-2.2) {\color{blue}$g_4$};
	\end{scope}
	\draw[->,black,thick] (11.5,-3) to (14,-3);
	\node[] at (12.75,-2.5) {$\alpha(g_1,g_2,g_3)$};
	\node[] at (3,-6.0) {$\mathcal{L}_{g_1}\otimes\mathcal{L}_{g_2}\rightarrow\mathcal{L}_{g_1g_2}$};
	\node[] at (9,-5.5) {$(\mathcal{L}_{g_1}\otimes\mathcal{L}_{g_2})\otimes\mathcal{L}_{g_3}\rightarrow\mathcal{L}_{g_1g_2g_3}$};
	\node[] at (16.5,-5.5) {$\mathcal{L}_{g_1}\otimes(\mathcal{L}_{g_2}\otimes\mathcal{L}_{g_3})\rightarrow\mathcal{L}_{g_1g_2g_3}$};
	\end{scope}
\end{tikzpicture}
\caption{The fusion of symmetry lines can be depicted as a vertex with fixed ordering of the outgoing legs while the 't Hooft anomaly manifests itself as a non-trivial phase in a crossing relation.}
	\label{fig:vert}
\end{figure}
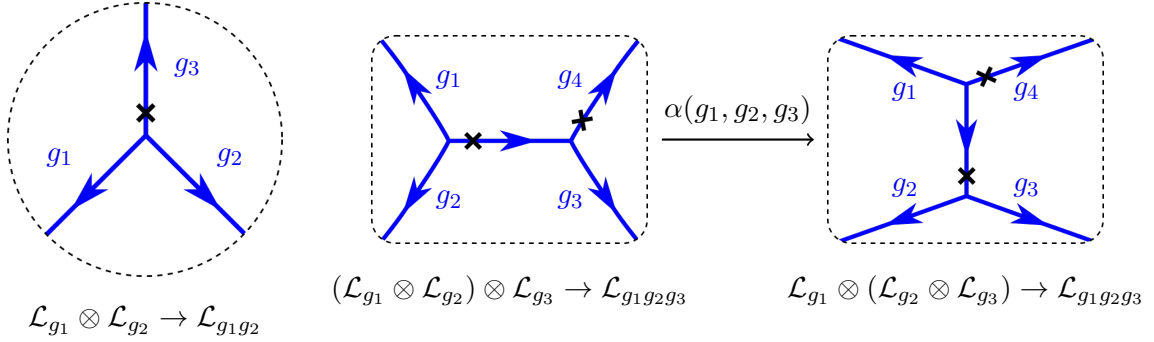

We require all lines that are not loops  to start and end on trivalent vertices $V_{\mathcal{L}_{g_1},\mathcal{L}_{g_2},\mathcal{L}_{g_3}}$ with $g_1g_2g_3=1$ in the convention that all lines are outgoing.
Each such vertex corresponds to a fusion of symmetry lines $\mathcal{L}_{g_1}\otimes\mathcal{L}_{g_2}\rightarrow\mathcal{L}_{g_1g_2}$ and, in the presence of a 't Hooft anomaly, it is important to keep track of the order of lines at each vertex.
Following~\cite{Chang:2018iay}, we assume counterclockwise ordering and mark the last leg as illustrated in Figure~\ref{fig:vert}.
Since we want to work on a torus we will assume that the metric is flat.
The lines can then be continuously deformed as long as the order of lines at each vertex is preserved. 
Moreover, a line $\mathcal{L}_g$ can always be replaced by a line $\mathcal{L}_{g^{-1}}$ with opposite orientation.

The 't Hooft anomaly is encoded in the crossing kernel $\alpha:\,G^3\rightarrow U(1)$, that changes the order of fusion
\begin{align}
(\mathcal{L}_{g_1}\otimes\mathcal{L}_{g_2})\otimes\mathcal{L}_{g_3}\xrightarrow[]{\alpha(g_1,g_2,g_3)}\mathcal{L}_{g_1}\otimes(\mathcal{L}_{g_2}\otimes\mathcal{L}_{g_3})\,,
\end{align}
as also illustrated in Figure~\ref{fig:vert}.
The requirement that $\alpha$ is annihilated by the coboundary operator~\eqref{eqn:coboundary} follows from the so-called pentagon identity, a sequence of five non-trivial crossings that leads back to the original diagram, see~\cite{Bhardwaj:2017xup}.
Different representatives $\alpha,\alpha'$ of the same anomaly are related by a local operator redefinition, that acts on the Hilbert spaces at the vertices $V_{\mathcal{L}_{g_1},\mathcal{L}_{g_2},\mathcal{L}_{g_3}}$ by multiplication with $\delta\epsilon(g_1,g_2,g_3)$ for some $\epsilon:\,G^2\rightarrow U(1)$.
We can use part of this freedom to assume that $\alpha(g_1,g_2,g_3)$ vanishes if one of the elements $g_1,g_2,g_3$ is the identity.
From a mathematical perspective, the line operators form a \textit{pointed fusion category} $\text{Vec}^\alpha_G$~\cite{etingof2016tensor}.

Before we begin with our actual argument let us note two useful properties that can be easily derived from the crossing relations:
\begin{enumerate}
	\item Clockwise rotation of the labels at a vertex $V_{\mathcal{L}_{g_1},\mathcal{L}_{g_2},\mathcal{L}_{g_3}}$ introduces a phase $\alpha(g_1,g_2,g_3)$.
		This follows from setting $g_4=1$ in the crossing relation in Figure~\ref{fig:vert}.
	\item A counterclockwise loop $\mathcal{L}_g$ can be replaced by an overall phase $\alpha(g,g^{-1},g)$.
		This can be derived by setting $g_1=g_2^{-1}=g_3=g_4^{-1}=g$ in Figure~\ref{fig:vert} and connecting the legs associated to $g_1,g_2$ as well as those associated to $g_3,g_4$.
\end{enumerate}
\begin{figure}[h!]
	\centering
\begin{tikzpicture}[scale=.8]
	\begin{scope}[xshift=1.0cm,yshift=0cm]
		\draw[->,black,thick] (11.5,-4) to[in=30,out=-30,looseness=.5] (11.5,-8);
	\draw[fill=white,draw=none] (11.5,-5.55) rectangle ++(0.9,-0.9);
		\node[align=center] at (13.3,-5.6) {Gauge\\transformation};
	\begin{scope}[xshift=-1.5cm,yshift=0cm]
	\node[align=center] at (3,-3) {\includegraphics[width=.24\linewidth]{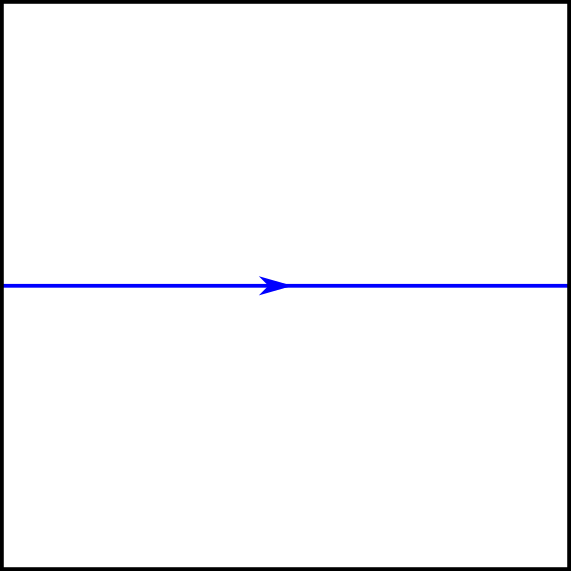}};
	\node at (.95,-.95) {$1$};
	\draw[draw=black,thick] (.75,-.75) rectangle ++(0.4,-0.4);
		\node[align=center] at (.5,-3) {\tiny$0$};
		\node[align=center] at (5.5,-3) {\tiny$0$};
		\node[align=center] at (3,-2.5) {\color{blue}$g$};
	\draw[->,black,thick] (6,-3) to (7.5,-3);
	\node[align=center] at (6.75,-2.5) {$T^n$};
	\end{scope}
	\begin{scope}[xshift=6.0cm,yshift=0cm]
	\draw[draw=black,thick] (.75,-.75) rectangle ++(0.4,-0.4);
	\node at (.95,-.95) {$2$};
	\node[align=center] at (3,-3) {\includegraphics[width=.24\linewidth]{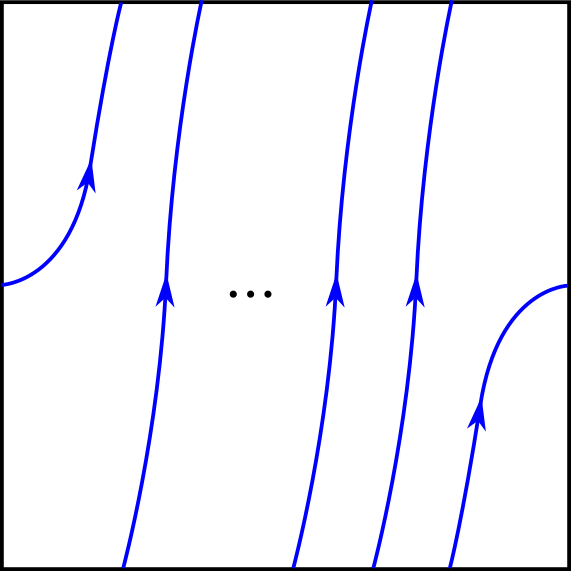}};
	\node[align=center] at (.5,-3) {\tiny$0$};
	\node[align=center] at (5.5,-3) {\tiny$0$};
	\node[align=center] at (1.7,-.5) {\tiny$1$};
	\node[align=center] at (1.7,-5.5) {\tiny$1$};
	\node[align=center] at (2.3,-.5) {\tiny$2$};
	\node[align=center] at (2.9,-5.5) {\tiny$n-2$};
	\node[align=center] at (3.7,-0.5) {\tiny$n-1$};
	\node[align=center] at (3.7,-5.5) {\tiny$n-1$};
	\node[align=center] at (4.3,-0.5) {\tiny$n$};
	\node[align=center] at (4.3,-5.5) {\tiny$n$};
	\node[align=center] at (1.2,-3.2) {\color{blue}$g$};
	\node[align=center] at (4.6,-3.2) {\color{blue}$g$};
	\end{scope}
	\begin{scope}[xshift=6.0cm,yshift=-6cm]
	\draw[draw=black,thick] (.75,-.75) rectangle ++(0.4,-0.4);
	\node at (.95,-.95) {$3$};
	\node[align=center] at (3,-3) {\includegraphics[width=.24\linewidth]{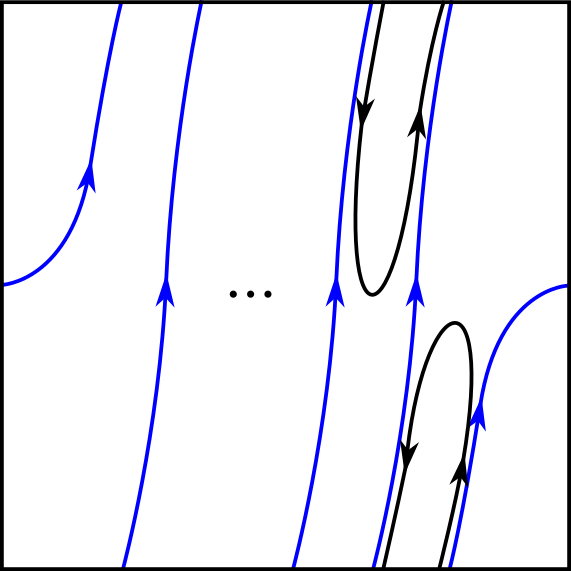}};
	\node[align=center] at (.5,-3) {\tiny$0$};
	\node[align=center] at (5.5,-3) {\tiny$0$};
	\node[align=center] at (1.7,-.5) {\tiny$1$};
	\node[align=center] at (1.7,-5.5) {\tiny$1$};
	\node[align=center] at (2.3,-.5) {\tiny$2$};
	\node[align=center] at (2.9,-5.5) {\tiny$n-2$};
	\node[align=center] at (3.7,-0.5) {\tiny$n-1$};
	\node[align=center] at (3.7,-5.5) {\tiny$n-1$};
	\node[align=center] at (4.3,-0.5) {\tiny$n$};
	\node[align=center] at (4.3,-5.5) {\tiny$n$};
	\node[align=center] at (3.9,-1.5) {$g$};
	\node[align=center] at (4.2,-4.5) {$g$};
	\end{scope}
	\begin{scope}[xshift=-1.5cm,yshift=-6cm]
	\draw[->,black,thick] (7.5,-3) to (6,-3);
	\node[align=center] at (6.75,-2.5) {Fusion};
	\draw[draw=black,thick] (.75,-.75) rectangle ++(0.4,-0.4);
	\node at (.95,-.95) {$4$};
	\node[align=center] at (3,-3) {\includegraphics[width=.24\linewidth]{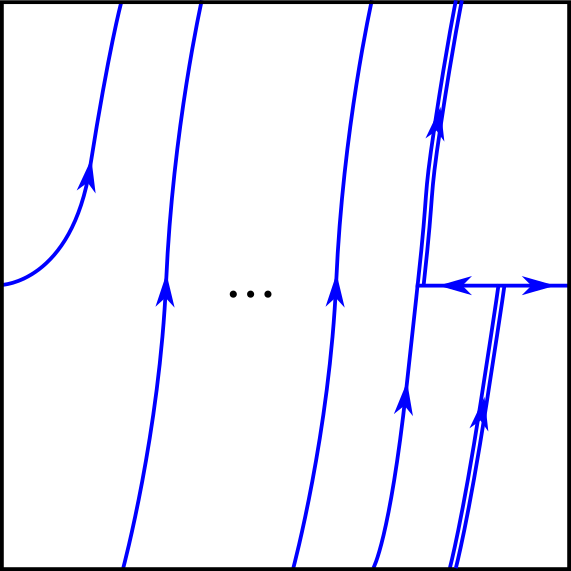}};
	\node[align=center] at (.5,-3) {\tiny$0$};
	\node[align=center] at (5.5,-3) {\tiny$0$};
	\node[align=center] at (1.7,-.5) {\tiny$1$};
	\node[align=center] at (1.7,-5.5) {\tiny$1$};
	\node[align=center] at (2.3,-.5) {\tiny$2$};
	\node[align=center] at (2.9,-5.5) {\tiny$n-3$};
	\node[align=center] at (3.7,-0.5) {\tiny$n-2$};
	\node[align=center] at (3.7,-5.5) {\tiny$n-2$};
	\node[align=center] at (4.3,-0.5) {\tiny$n$};
	\node[align=center] at (4.3,-5.5) {\tiny$n$};
	\node[align=center] at (1.2,-3.2) {\color{blue}$g$};
	\node[align=center] at (4.6,-1.7) {\color{blue}$g^2$};
	\node[align=center] at (4.9,-4.1) {\color{blue}$g^2$};
	\end{scope}
	\end{scope}
	\draw[->,black,thick] (2.5,-12) to (2.5,-19.5);
	\begin{scope}[xshift=0,yshift=-12.5cm]
	\draw[draw=black,fill=lightgray!30,very thick] (0,0) rectangle ++(\linewidth,-6.3);
	\draw[draw=black,fill=white,thick] (0,0) rectangle ++(0.4,-0.4);
		\node at (.20,-.20) {$5$};
		\node[] at (7.5,-.5) {Iterate $n-2$ times};
	\begin{scope}[xshift=-.5cm,yshift=-.3cm]
	    \node [shape=rectangle, fill=white, minimum width=3.6cm, minimum height=3.6cm, anchor=center] at (3,-3) {};
	\draw[draw=black,thick] (.75,-.75) rectangle ++(0.4,-0.4);
		\node at (.95,-.95) {$a$};
		\node[align=center] at (3,-3) {\includegraphics[width=.24\linewidth]{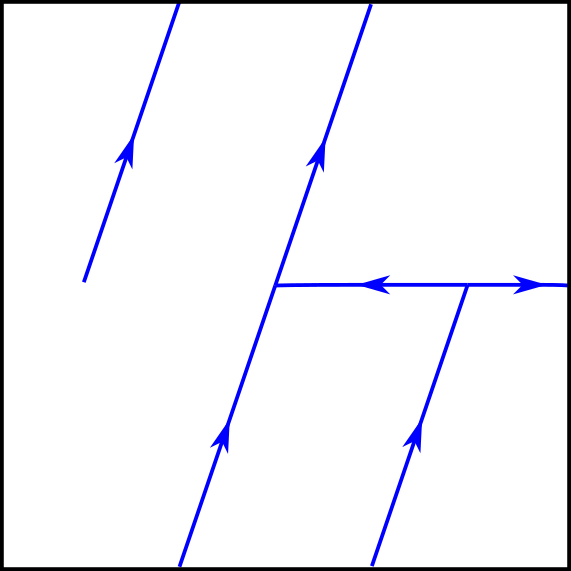}};
	\node[align=center] at (2,-2) {$g$};
	\node[align=center] at (3.9,-2) {$g^{a+1}$};
	\node[align=center] at (4.6,-4.3) {$g^{a+1}$};
	\node[align=center] at (2.8,-4.3) {$g$};
	\node[align=center] at (3.5,-3.4) {$g^a$};
	\node[align=center] at (4.8,-3.4) {$g$};
	\node[align=center] at (2.2,-.5) {\tiny $n-1$};
	\node[align=center] at (2.2,-5.5) {\tiny $n-1$};
	\node[align=center] at (3.7,-.5) {\tiny $n$};
	\node[align=center] at (3.7,-5.5) {\tiny $n$};
	\end{scope}
	\begin{scope}[xshift=4.5cm,yshift=-.3cm]
	    \node [shape=rectangle, fill=white, minimum width=3.6cm, minimum height=3.6cm, anchor=center] at (3,-3) {};
	\draw[draw=black,thick] (.75,-.75) rectangle ++(0.4,-0.4);
		\node at (.95,-.95) {$b$};
	\node[align=center] at (3,-3) {\includegraphics[width=.24\linewidth]{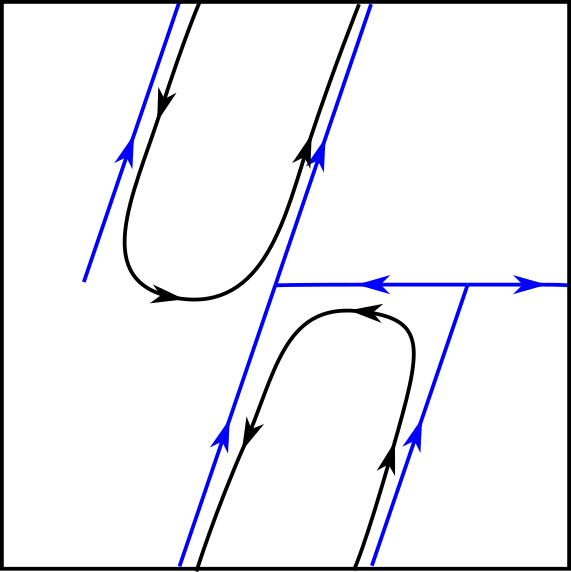}};
	\node[align=center] at (2.3,-2.5) {$g$};
	\node[align=center] at (3.3,-4.5) {$g$};
	\end{scope}
	\begin{scope}[xshift=9.5cm,yshift=-.3cm]
	    \node [shape=rectangle, fill=white, minimum width=3.6cm, minimum height=3.6cm, anchor=center] at (3,-3) {};
	\draw[draw=black,thick] (.75,-.75) rectangle ++(0.4,-0.4);
		\node at (.95,-.95) {$c$};
	\node[align=center] at (3,-3) {\includegraphics[width=.24\linewidth]{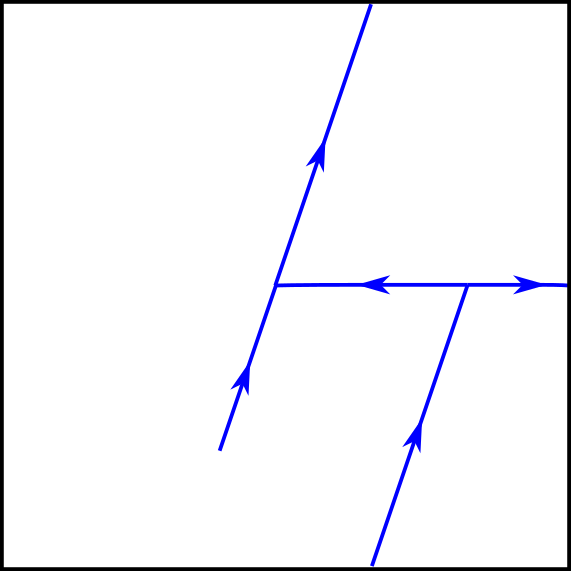}};
	\node[align=center] at (3.9,-2) {$g^{a+2}$};
	\node[align=center] at (4.6,-4.3) {$g^{a+2}$};
	\node[align=center] at (2.8,-4.3) {$g$};
	\node[align=center] at (3.5,-3.4) {$g^{a+1}$};
	\node[align=center] at (4.8,-3.4) {$g$};
	\node[align=center] at (3.7,-.5) {\tiny $n$};
	\node[align=center] at (3.7,-5.5) {\tiny $n$};
	\end{scope}
	\end{scope}
	\begin{scope}[xshift=-0.5cm,yshift=-19cm]
	\node[align=center] at (3,-3) {\includegraphics[width=.24\linewidth]{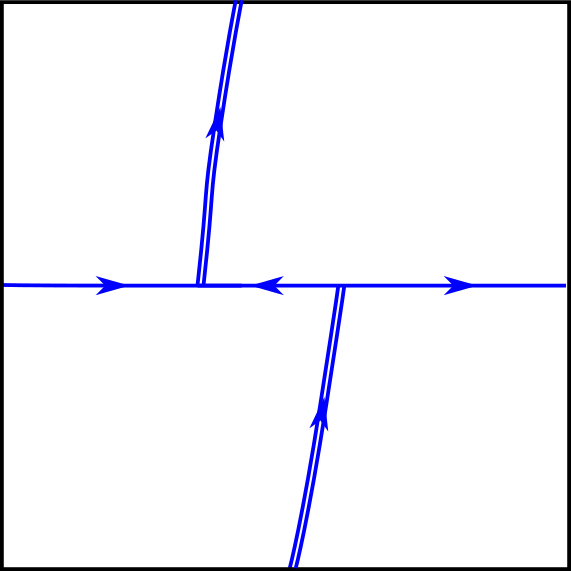}};
	\node at (.95,-.95) {$6$};
	\draw[draw=black,thick] (.75,-.75) rectangle ++(0.4,-0.4);
		\node[align=center] at (.5,-3) {\tiny$0$};
		\node[align=center] at (5.5,-3) {\tiny$0$};
		\node[align=center] at (1.5,-2.5) {\color{blue}$g$};
		\node[align=center] at (4.5,-2.5) {\color{blue}$g$};
		\node[align=center] at (2.9,-1.9) {\color{blue}$g^n$};
		\node[align=center] at (3.7,-4.3) {\color{blue}$g^n$};
		\node[align=center] at (2.7,-3.5) {\color{blue}$g^{n-1}$};
	\draw[->,black,thick] (6,-3) to (7.5,-3);
	\node[align=center] at (6.75,-2.5) {$=$};
	\end{scope}
	\begin{scope}[xshift=7.0cm,yshift=-19cm]
	\node[align=center] at (3,-3) {\includegraphics[width=.24\linewidth]{figures/f1.png}};
	\node at (.95,-.95) {$7$};
	\draw[draw=black,thick] (.75,-.75) rectangle ++(0.4,-0.4);
		\node[align=center] at (.5,-3) {\tiny$0$};
		\node[align=center] at (5.5,-3) {\tiny$0$};
		\node[align=center] at (3,-2.5) {\color{blue}$g$};
		\node[align=center] at (7.0,-3) {$\times\,\gamma^\alpha_g$};
	\end{scope}
\end{tikzpicture}
\caption{
	In $(1)$ we start with a defect line on a torus that generates the twisted Hilbert space $\mathcal{H}_g$.
Acting with $T^n$ leads to configuration $(2)$. A sequence of gauge transformation relates this to the original configuration. The first step is illustrated in $(3)$ and $(4)$.
Iterating the transformation in step $(5)$ $n-2$ times produces $(6)$ which is equivalent to the original defect line in $(7)$.}
	\label{fig:lines1}
\end{figure}

Let us assume that $g\in G$ is an element of order $n$. The corresponding defect line that leads to the twisted Hilbert space $\mathcal{H}_g$ is illustrated in step $(1)$ of Figure~\ref{fig:lines1}.
Our goal is to show that the partition function of the $g$-twisted sector transforms in the presence of a 't~Hooft anomaly $[\alpha]\in H^3(BG,U(1))$ as~\eqref{eqn:tnphase} and to derive an explicit expression for the phase $\gamma^\alpha_g$ that only depends on the cohomology class.

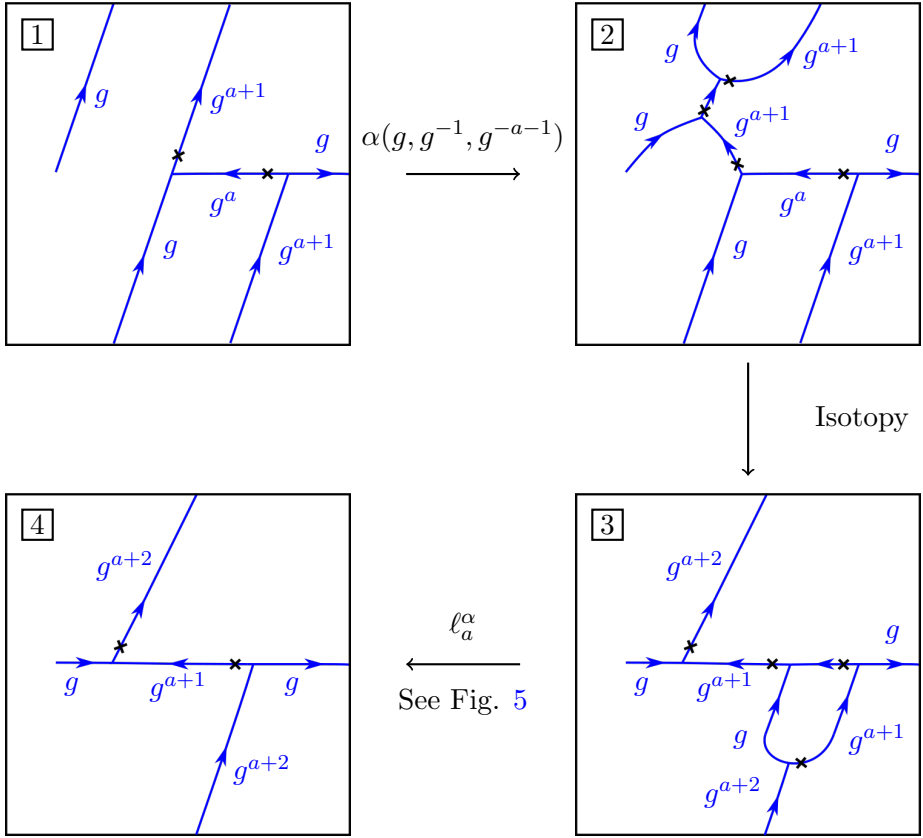
\begin{figure}[h!]
	\centering
\begin{tikzpicture}[scale=1]
	\begin{scope}[xshift=1.0cm,yshift=0cm]

	\begin{scope}[xshift=-1.5cm,yshift=0cm]
	\node[align=center] at (3,-3) {\includegraphics[width=.3\linewidth]{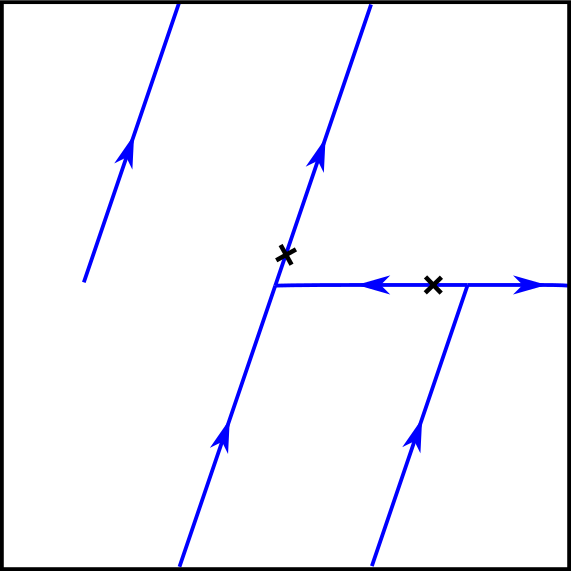}};
	\begin{scope}[xshift=.2cm,yshift=-.2cm]
	\node at (.95,-.95) {$1$};
	\draw[draw=black,thick] (.75,-.75) rectangle ++(0.4,-0.4);
	\end{scope}
	\node[align=center] at (2,-2) {\color{blue}$g$};
	\node[align=center] at (3.8,-2) {\color{blue}$g^{a+1}$};
	\node[align=center] at (2.9,-4) {\color{blue}$g$};
	\node[align=center] at (4.7,-4) {\color{blue}$g^{a+1}$};
	\node[align=center] at (3.6,-3.4) {\color{blue}$g^a$};
	\node[align=center] at (4.9,-2.6) {\color{blue}$g$};
	\draw[->,black,thick] (6,-3) to (7.5,-3);
		\node[align=center] at (6.75,-2.5) {$\alpha(g,g^{-1},g^{-a-1})$};
	\end{scope}

	\begin{scope}[xshift=6.0cm,yshift=0cm]
	\begin{scope}[xshift=.2cm,yshift=-.2cm]
	\node at (.95,-.95) {$2$};
	\draw[draw=black,thick] (.75,-.75) rectangle ++(0.4,-0.4);
	\end{scope}
	\node[align=center] at (3,-3) {\includegraphics[width=.3\linewidth]{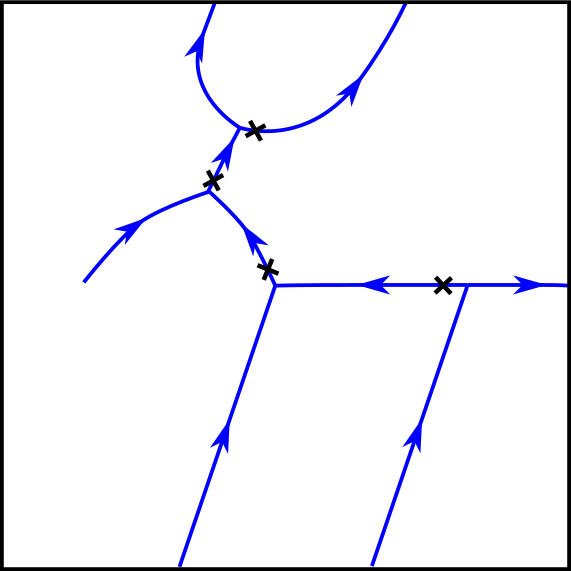}};
	\node[align=center] at (2.9,-4) {\color{blue}$g$};
	\node[align=center] at (4.7,-4) {\color{blue}$g^{a+1}$};
	\node[align=center] at (3.6,-3.4) {\color{blue}$g^a$};
	\node[align=center] at (4.9,-2.6) {\color{blue}$g$};
	\node[align=center] at (1.6,-2.3) {\color{blue}$g$};
	\node[align=center] at (3.2,-2.3) {\color{blue}$g^{a+1}$};
	\node[align=center] at (2.0,-1.4) {\color{blue}$g$};
	\node[align=center] at (4.1,-1.4) {\color{blue}$g^{a+1}$};
	\draw[->,black,thick] (3,-5.5) to (3,-7.0);
	\node[align=center] at (4.5,-6.25) {Isotopy};
	\end{scope}

	\begin{scope}[xshift=6.0cm,yshift=-6.5cm]
	\begin{scope}[xshift=.2cm,yshift=-.2cm]
	\node at (.95,-.95) {$3$};
	\draw[draw=black,thick] (.75,-.75) rectangle ++(0.4,-0.4);
	\end{scope}
	\node[align=center] at (3,-3) {\includegraphics[width=.3\linewidth]{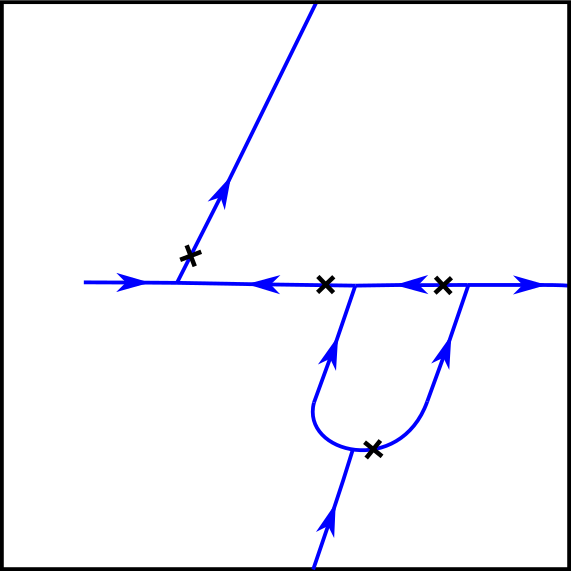}};
	\node[align=center] at (2.9,-4) {\color{blue}$g$};
	\node[align=center] at (4.7,-4) {\color{blue}$g^{a+1}$};
	\node[align=center] at (2.7,-3.3) {\color{blue}$g^{a+1}$};
	\node[align=center] at (4.9,-2.6) {\color{blue}$g$};
	\node[align=center] at (1.6,-3.3) {\color{blue}$g$};
	\node[align=center] at (2.3,-1.7) {\color{blue}$g^{a+2}$};
	\node[align=center] at (2.8,-4.7) {\color{blue}$g^{a+2}$};
	\end{scope}

	\begin{scope}[xshift=-1.5cm,yshift=-6.5cm]
	\draw[->,black,thick] (7.5,-3) to (6,-3);
	\node[align=center] at (6.75,-2.5) {$\ell^\alpha_a$};
	\node[align=center] at (6.75,-3.5) {See Fig. \ref{fig:triloop}};
	\begin{scope}[xshift=.2cm,yshift=-.2cm]
	\node at (.95,-.95) {$4$};
	\draw[draw=black,thick] (.75,-.75) rectangle ++(0.4,-0.4);
	\end{scope}
	\node[align=center] at (3,-3) {\includegraphics[width=.3\linewidth]{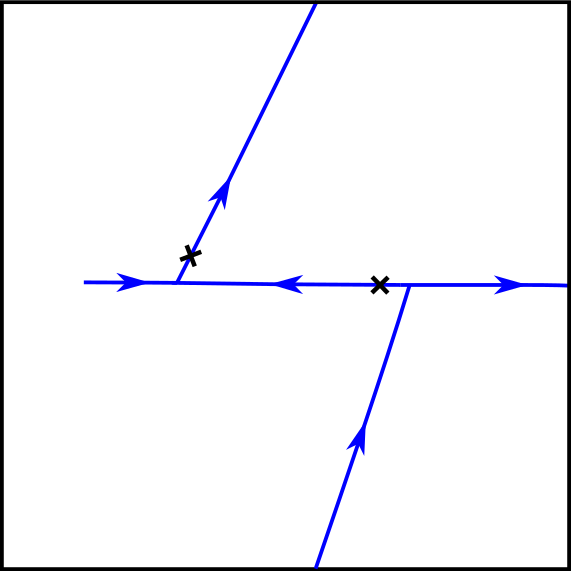}};
	\node[align=center] at (1.6,-3.3) {\color{blue}$g$};
	\node[align=center] at (2.3,-1.7) {\color{blue}$g^{a+2}$};
	\node[align=center] at (3.0,-3.3) {\color{blue}$g^{a+1}$};
	\node[align=center] at (4.5,-3.3) {\color{blue}$g$};
	\node[align=center] at (4.1,-4.4) {\color{blue}$g^{a+2}$};
	\end{scope}
	\end{scope}
\end{tikzpicture}
\caption{The iteration step produces an overall phase $\alpha(g,g^{-1},g^{-a-1})\ell^\alpha_a$, where the phase $\ell^\alpha_a$ arises from contracting a trivalent loop and is defined in Figure~\ref{fig:triloop}.}
\label{fig:linetrafo1}
\end{figure}

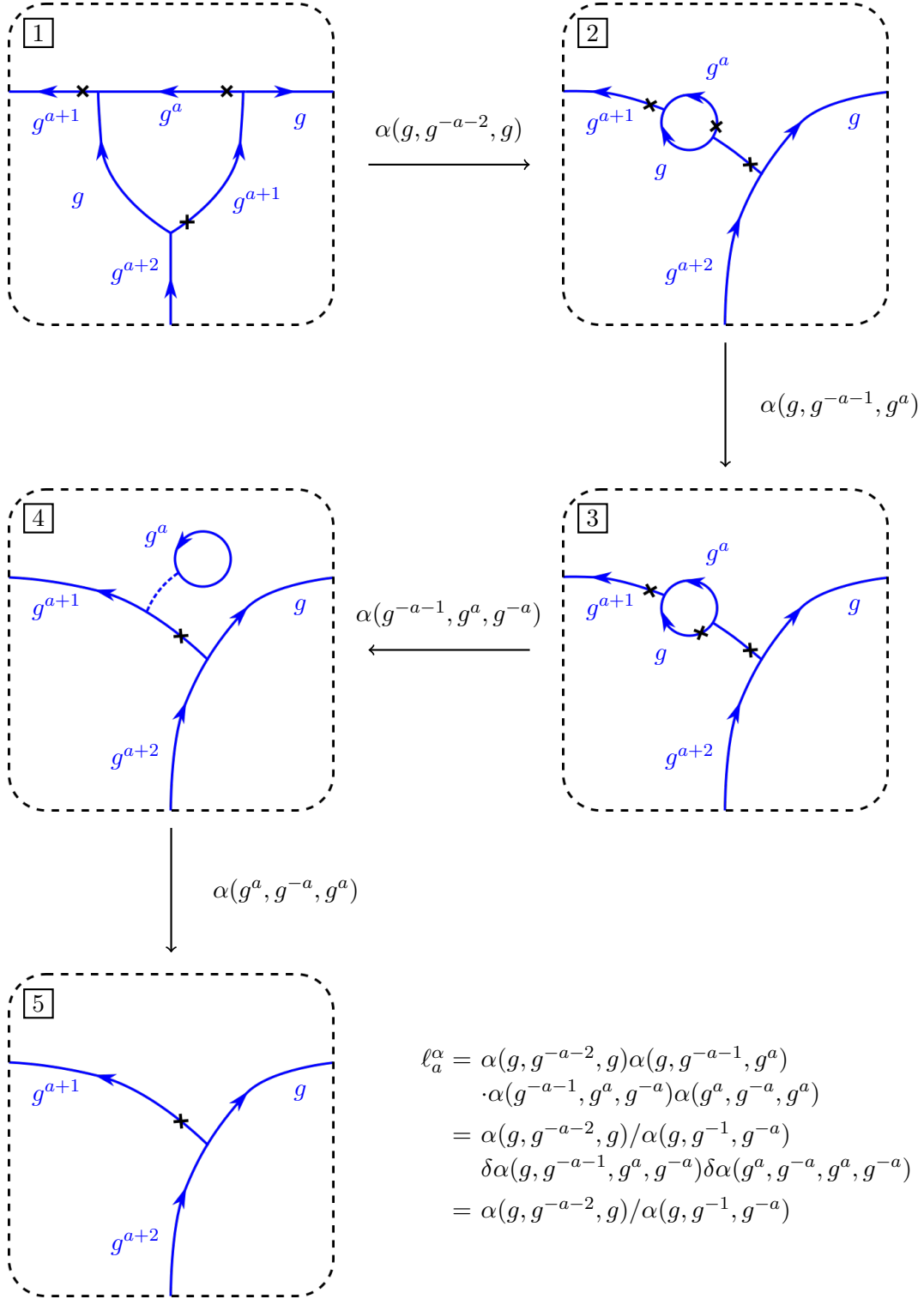
\begin{figure}[h!]
	\centering
\begin{tikzpicture}[scale=1]
	\begin{scope}[xshift=1.0cm,yshift=0cm]

	\begin{scope}[xshift=-1.5cm,yshift=0cm,scale=1.1]
	\node[align=center] at (3,-3) {\includegraphics[width=.33\linewidth]{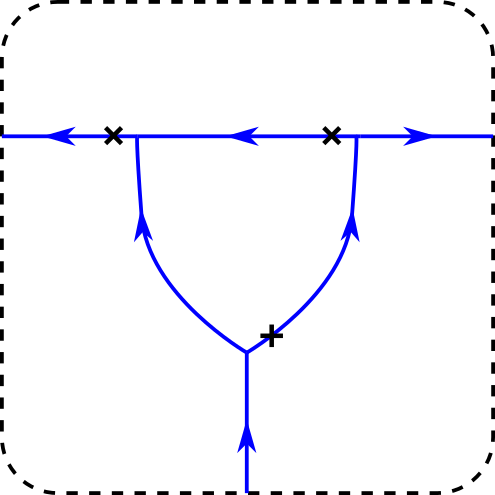}};
	\begin{scope}[xshift=.2cm,yshift=-.2cm]
	\node at (.95,-.95) {$1$};
	\draw[draw=black,thick] (.75,-.75) rectangle ++(0.4,-0.4);
	\end{scope}
	\node[] at (1.4,-2.4) {\color{blue}$g^{a+1}$};
	\node[] at (4.8,-2.4) {\color{blue}$g$};
	\node[] at (2.5,-4.5) {\color{blue}$g^{a+2}$};
	\node[] at (1.7,-3.5) {\color{blue}$g$};
	\node[] at (4.2,-3.5) {\color{blue}$g^{a+1}$};
	\node[] at (3,-2.3) {\color{blue}$g^a$};
	\draw[->,black,thick] (5.75,-3) to (8.0,-3);
	\node[align=center] at (6.875,-2.5) {$\alpha(g,g^{-a-2},g)$};
	\end{scope}

	\begin{scope}[xshift=7.0cm,yshift=0cm,scale=1.1]
	\begin{scope}[xshift=.2cm,yshift=-.2cm]
	\node at (.95,-.95) {$2$};
	\draw[draw=black,thick] (.75,-.75) rectangle ++(0.4,-0.4);
	\end{scope}
	\node[] at (1.4,-2.4) {\color{blue}$g^{a+1}$};
	\node[] at (4.8,-2.4) {\color{blue}$g$};
	\node[] at (2.5,-4.5) {\color{blue}$g^{a+2}$};
	\node[] at (2.1,-3.1) {\color{blue}$g$};
	\node[] at (2.9,-1.65) {\color{blue}$g^{a}$};
	\node[align=center] at (3,-3) {\includegraphics[width=.33\linewidth]{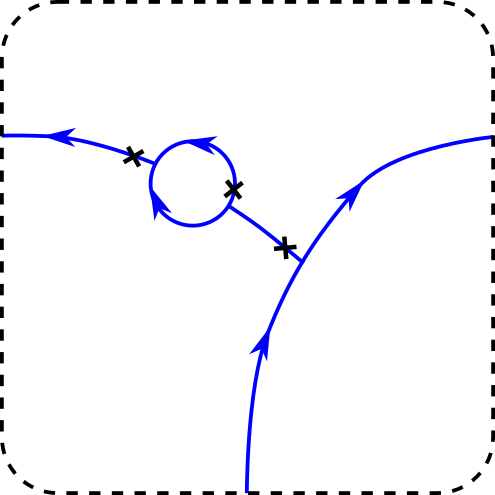}};
	\draw[->,black,thick] (3,-5.5) to (3,-7.25);
	\node[align=center] at (4.6,-6.375) {$\alpha(g,g^{-a-1},g^a)$};
	\end{scope}

	\begin{scope}[xshift=7.0cm,yshift=-7.5cm,scale=1.1]
	\begin{scope}[xshift=.2cm,yshift=-.2cm]
	\node at (.95,-.95) {$3$};
	\draw[draw=black,thick] (.75,-.75) rectangle ++(0.4,-0.4);
	\end{scope}
	\node[] at (1.4,-2.4) {\color{blue}$g^{a+1}$};
	\node[] at (4.8,-2.4) {\color{blue}$g$};
	\node[] at (2.5,-4.5) {\color{blue}$g^{a+2}$};
	\node[] at (2.1,-3.1) {\color{blue}$g$};
	\node[] at (2.9,-1.65) {\color{blue}$g^{a}$};
	\node[align=center] at (3,-3) {\includegraphics[width=.33\linewidth]{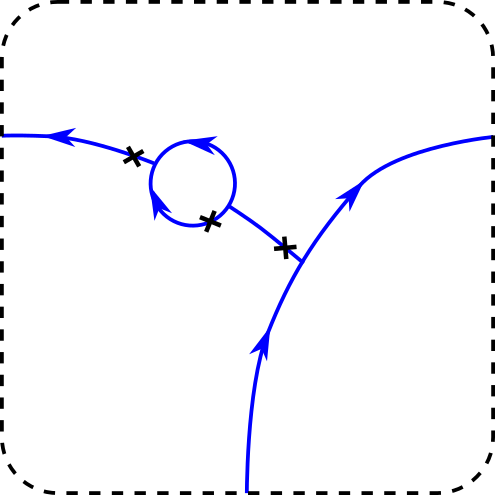}};
	\end{scope}

	\begin{scope}[xshift=-1.5cm,yshift=-7.5cm,scale=1.1]
	\draw[->,black,thick] (8.0,-3) to (5.75,-3);
	\node[align=center] at (6.875,-2.5) {$\alpha(g^{-a-1},g^{a},g^{-a})$};
	\begin{scope}[xshift=.2cm,yshift=-.2cm]
	\node at (.95,-.95) {$4$};
	\draw[draw=black,thick] (.75,-.75) rectangle ++(0.4,-0.4);
	\end{scope}
	\node[] at (1.4,-2.4) {\color{blue}$g^{a+1}$};
	\node[] at (4.8,-2.4) {\color{blue}$g$};
	\node[] at (2.5,-4.5) {\color{blue}$g^{a+2}$};
	\node[] at (2.8,-1.4) {\color{blue}$g^{a}$};
	\node[align=center] at (3,-3) {\includegraphics[width=.33\linewidth]{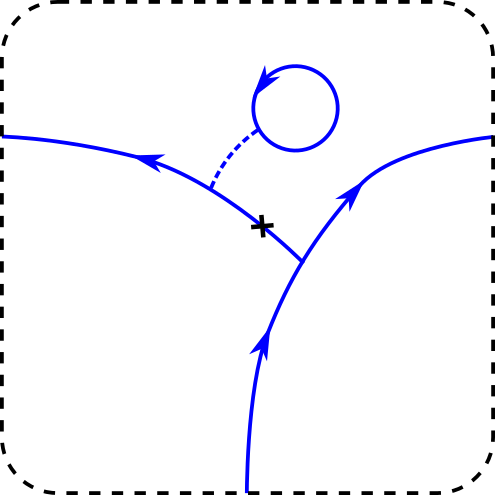}};
	\draw[->,black,thick] (3,-5.5) to (3,-7.25);
		\node[align=center] at (4.6,-6.375) {$\alpha(g^a,g^{-a},g^a)$};
	\end{scope}

	\begin{scope}[xshift=-1.5cm,yshift=-15cm,scale=1.1]
	\begin{scope}[xshift=.2cm,yshift=-.2cm]
	\node at (.95,-.95) {$5$};
	\draw[draw=black,thick] (.75,-.75) rectangle ++(0.4,-0.4);
	\end{scope}
	\node[] at (1.4,-2.4) {\color{blue}$g^{a+1}$};
	\node[] at (4.8,-2.4) {\color{blue}$g$};
	\node[] at (2.5,-4.5) {\color{blue}$g^{a+2}$};
	\node[align=center] at (3,-3) {\includegraphics[width=.33\linewidth]{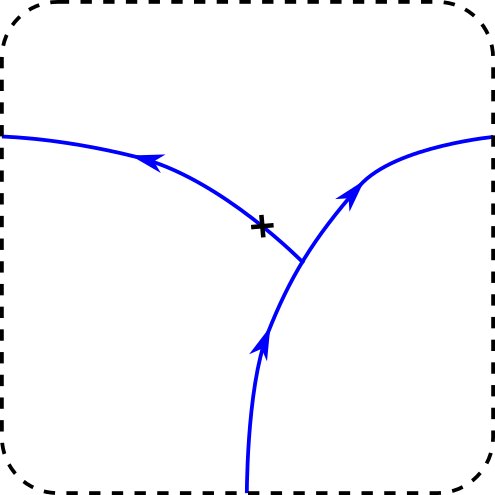}};
	\end{scope}
	\end{scope}

	\begin{scope}[xshift=6.0cm,yshift=-15cm,scale=1.1]
		\node[] at (4.0,-3) {
			$\begin{array}{rl}
			\ell^\alpha_a=&\alpha(g,g^{-a-2},g)\alpha(g,g^{-a-1},g^a)\\
				&\cdot\alpha(g^{-a-1},g^a,g^{-a})\alpha(g^a,g^{-a},g^a)\\[.3em]
			=&\alpha(g,g^{-a-2},g)/\alpha(g,g^{-1},g^{-a})\\
				&\delta\alpha(g,g^{-a-1},g^a,g^{-a})\delta\alpha(g^a,g^{-a},g^a,g^{-a})\\[.3em]
			=&\alpha(g,g^{-a-2},g)/\alpha(g,g^{-1},g^{-a})
			\end{array}$
			};
	\end{scope}
\end{tikzpicture}
\caption{Contracting the trivalent loop shown in step $(1)$ can be performed in four crossings. This leads to an overall phase $\ell^\alpha_a=\alpha(g,g^{-a-2},g)/\alpha(g,g^{-1},g^{-a})$.}
\label{fig:triloop}
\end{figure}

The general strategy to derive the phase is outlined in steps $(2)-(7)$ of Figure~\ref{fig:lines1} and essentially boils down to iterative application of step $(5)$.
However, care has to be taken in evaluating the phases that are picked up from crossing relations at each step.
Iterative applications of the transformations in Figure~\ref{fig:linetrafo1} and taking into account the phases produced by the trivalent loops in Figure~\ref{fig:triloop} leads to the result 
\begin{align}
	\begin{split}
		\gamma_g^\alpha=&\prod\limits_{a=0}^{n-2}\frac{\alpha(g,g^{-1},g^{-a-1})}{\alpha(g,g^{-1},g^{-a})}\alpha(g,g^{-a-2},g)\\
		=&\alpha(g,g^{-1},g^{1-n})\prod\limits_{a=0}^{n-2}\alpha(g,g^{-a-2},g)=\prod\limits_{a=0}^{n-1}\alpha(g,g^a,g)\,.
		\label{eqn:gamma}
	\end{split}
\end{align}
Note that in the second step we have contracted the telescoping product that arises from the quotient.
It is easy to show that this is invariant under changes of $\alpha$ by coboundaries $\delta\epsilon$ of maps $\epsilon:\,G^2\rightarrow U(1)$,
\begin{align}
	\begin{split}	\gamma_g^{\delta\epsilon}=&\prod\limits_{a=0}^{n-1}\frac{\epsilon(g,g^{a+1})\epsilon(g^a,g)}{\epsilon(g,g^a)\epsilon(g^{a+1},g)}=\frac{\epsilon(g,g^n)\epsilon(1,g)}{\epsilon(g,1)\epsilon(g^n,g)}=1\,.
\end{split}
\end{align}

Let us now specialize this again to $G=\mathbb{Z}_n$ with $H^3(\mathbb{Z}_n,U(1))=\mathbb{Z}_n$.
Representing elements of $\mathbb{Z}_n$ additively by integers $a=0,\ldots,n-1$, a generator of cocycles is
\begin{align}
    \alpha(a,b,c)=e^{2\pi i a (b+c-\langle b+c\rangle)/n^2}\,,
\end{align}
where $\langle b+c\rangle$ denotes the mod $n$ reduction of $b+c$~\cite{deWildPropitius:1995cf}.
Representing an element $g\in\mathbb{Z}_n$ of order $m$ by $r\in\{0,\ldots,n-1\}$ and using the telescoping nature of the product we find
\begin{align}
\gamma_r^\alpha=\prod\limits_{a=0}^{m-1}\alpha(r,\langle a\cdot r\rangle,r)=e^{2\pi i \frac{m r^2}{n^2}}\,,
\end{align}
and therefore in particular $\gamma_1^{l\cdot \alpha}=\mu_n^l$.

\section{Elliptic genera and anomalies}
\label{sec:egan}
We now apply the general discussion of the relation between the modular anomaly and the 't~Hooft anomaly in 2d CFTs from Section~\ref{sec:2danomalies} to the twisted twined elliptic genera of non-critical string in six-dimensional F-theory vacua.
Using the duality with topological string amplitudes and the corresponding modular properties derived in~\cite{Cota:2019cjx,Schimannek:2021pau}, this will allow us to obtain a generic expression for the anomaly in terms of the geometric height-pairing of a multi-section on the corresponding Calabi-Yau.

\subsection{Elliptic genera and the topological string partition function}
Let us first consider a generic six-dimensional F-theory vacuum with or without discrete gauge symmetry.
To illustrate the effect of a continuous gauge symmetry, we assume that $G=U(1)\times\ldots$ but this is not essential.

From the Type IIB perspective, a non-critical string arises from a D3-brane that wraps a curve $C$ in the base $B$ of the torus fibration $X$.
The corresponding worldsheet theories are two-dimensional $(0,4)$-SCFTs and the elliptic genera can be defined as traces
\begin{align}
	Z_C(\tau,m,\lambda)=\text{Tr}_R(-1)^F q^{H_0}\bar{q}^{\bar{H}_0}y^{J_3^L}\xi^{J}\,,\quad q=e^{2\pi i \tau},\,y=e^{2\pi i \lambda},\,\xi=e^{2\pi i m}\,.
	\label{eqn:01twist}
\end{align}
Here $J_3^L$ corresponds to the Cartan charge associated to a global $SU(2)_L$ symmetry.
After compactifying the theory on a circle to five dimensions, this becomes part of the $SO(4)=SU(2)_L\times SU(2)_R$ little group.
We use $J$ to represent the Cartan generator associated to the additional global $U(1)$ symmetry that is induced by the six-dimensional $U(1)$ gauge symmetry.
For details on the worldsheet theories, as well as references to the vast surrounding literature, we refer e.g. to~\cite{Haghighat:2013gba,Haghighat:2015ega,Lawrie:2016axq,DelZotto:2018tcj}.

The modular properties of the elliptic genus encode the two-dimensional local 't Hooft anomalies.
Under $U_1:\,\tau\rightarrow \tau/(\tau+1)$ it transforms as
\begin{align}
	Z_C\left(\frac{\tau}{\tau+1},\frac{m}{\tau+1},\frac{\lambda}{\tau+1}\right)=e^{2\pi i a}e^{2\pi i \frac{f(m,\lambda)}{\tau+1}}Z_C(\tau,m,\lambda)\,,
	\label{eqn:streg}
\end{align}
where $a=\frac12 c_1(B)\cdot C$ encodes the gravitational anomaly and $f(m,\lambda)$ is an equivariant integral over the anomaly polynomial $\mathcal{A}_4$~\cite{Bobev:2015kza,Benjamin:2016fhe,DelZotto:2016pvm,DelZotto:2017mee}
\begin{align}
	f(m,\lambda)=\int\limits_{\text{eq.}}\mathcal{A}_4\,.
\end{align}
The relevant terms of this are given by~\cite{DelZotto:2017mee,Weigand:2017gwb,DelZotto:2018tcj,Lee:2018urn}
\begin{align}
	\mathcal{A}_4=&-\frac12 C^2\cdot c_2\left(SU(2)_L\right)-\frac14\left(C\cdot c_1(B)\right)\cdot p_1(X)-\frac12(C\cdot b)\cdot c_1\left(U(1)\right)^2\,.
	\label{eqn:a4poly}
\end{align}
Here the $U(1)$ anomaly coefficient $b$ is a certain height pairing associated to an additional section on the Calabi-Yau~\cite{Park:2011ji} and will be further discussed in Section~\ref{sec:anmod}.
Evaluating the integral amounts to the replacement~\cite{DelZotto:2016pvm}
\begin{align}
	c_2\left(SU(2)_L\right)\rightarrow -\lambda^2\,,\quad c_1\left(U(1)\right)^2\rightarrow -m^2\,,\quad p_1(X)\rightarrow 2\lambda^2\,.
\end{align}
The transformation behaviour of the elliptic genus in~\eqref{eqn:streg} is then precisely that of a weight $0$ lattice Jacobi form with two elliptic parameters and corresponding indices
\begin{align}
	r_\lambda(C)=\frac12C(C-c_1(B))\,,\quad r_m=\frac12 C\cdot b\,,
\end{align}
Some basic properties of Jacobi forms are summarized in Appendix~\ref{app:modjacobi}.
While $r_\lambda(C)$ is another contribution from the gravitational anomaly, $r_m(C)$ is the $U(1)$ 't~Hooft anomaly.
The generalized Green-Schwarz mechanism~\cite{GREEN1984117,Sagnotti:1992qw} relates these two-dimensional 't Hooft anomalies to the six-dimensional gauge anomalies via anomaly inflow~\cite{Kim:2016foj,Shimizu:2016lbw}.

After compactifying further down to four dimensions, F-theory on $X_0\times T^2$ becomes dual to Type IIA string theory on $X_0$~\footnote{In the presence of discrete symmetries we denote by $X_0$ the associated Weierstra{\ss} fibration. To study the topological string A-model, which is independent of the complex structure moduli, we can further replace any singular geometry by its smooth deformation.}.
The strings wrapping the torus lead to instanton corrections and as a result the elliptic genera are encoded in the A-model topological string partition function~\cite{Klemm:1996hh,Haghighat:2013gba,Haghighat:2014vxa,Huang:2015sta,DelZotto:2017mee,Lee:2018spm,Lee:2018urn,Cota:2019cjx}.
More precisely, the topological string partition function on a torus fibration $X$ admits an expansion
\begin{align}
	Z_{\text{top.}}[X](t,\tau,m,\lambda)=Z_0[X](\tau,m,\lambda)\cdot\left[1+\sum\limits_{C\in H_2(B,\mathbb{Z})} Z_{C}[X](\tau,m,\lambda) e^{2\pi i t\cdot C}\right]\,,
	\label{eqn:topstringex}
\end{align}
and the elliptic genera~\eqref{eqn:streg} directly correspond to the coefficients on $X_0$,
\begin{align}
    Z_{C}[X_0](\tau,m,\lambda)=Z_C(\tau,m,\lambda)\,,
\end{align}
while the overall factor of $Z_0[X_0](\tau,m,\lambda)$ arises due to field theoretic corrections and encodes the massless spectrum of the six-dimensional theory.

In~\eqref{eqn:topstringex} the arguments are complexified K\"ahler parameters of the Calabi-Yau and for $X_0$ the complexified volume of the generic fiber is $\tau$ while $t$ represents the complexified volumes of curves in the base.
On the other hand, $m$ represents the complexified volumes of rational components of reducible fibers which can be absent on a generic torus fibration.
From the topological string perspective, the modular properties -- and therefore the local anomalies -- can be derived from the automorphic properties of the topological string partition function under certain generic monodromies in the stringy K\"ahler moduli space~\cite{Candelas:1994hw,Huang:2015sta,Schimannek:2019ijf,Cota:2019cjx}.
This turns out to generalize to the global discrete anomalies and we will come back to this point in Section~\ref{sec:anmod}.

\subsection{Twisting and twining from geometry and B-fields}
\label{sec:twfg1}
Let us now include discrete symmetries and consider a six-dimensional F-theory vacuum with gauge group $\mathbb{Z}_n\subset G$.
This induces a corresponding global discrete symmetry in the worldsheet SCFTs of the non-critical strings.
We can therefore generalize the trace~\eqref{eqn:01twist} and consider the set of \textit{twisted twined elliptic genera}
\begin{align}
    Z_C^{(r,s)}(\tau,\lambda)=\text{Tr}_{\mathcal{H}_r}(-1)^F q^{H_0}\bar{q}^{\bar{H}_0}y^{J_3^L}\mu_n^{s J_{\mathbb{Z}_n}}\,,
    \label{eqn:tteg}
\end{align}
where we denote the $\mathbb{Z}_n$ charge operator by $J_{\mathbb{Z}_n}$ and $\mu_n=e^{2\pi i /n}$.

Using our analysis from Section~\ref{sec:fthdisc}, we can again relate these to topological string amplitudes.
As discussed in Section~\ref{sec:geodhol} there is a unique Weierstra{\ss} fibration $X_0$ associated to the F-theory vacuum, and F-theory on $X_0\times T^2$, without discrete holonomies, is dual to Type IIA on $X_0$.
Moreover, the elliptic genus can be interpreted as the torus partition functions of a non-critical string that is wrapped on this geometric $T^2$.
The twisting and twining can therefore be induced by turning on non-trivial discrete holonomies.

Let us first consider the twined elliptic genera with $(r,s)=(0,s)$.
As discussed in~\cite{Schimannek:2021pau}, the Jacobian fibration $X_0$ is singular but one can turn on a fractional $B$-field, labelled by a choice $s\in\mathbb{Z}_n$, that stabilizes the singularities and leads to a regular closed string worldsheet theory.
In the open string sector, this leads to a so-called non-commutative resolution of $X_0$ and in both sectors we denote the corresponding background by $X_{0,\text{n.c.}s}$.
Following our analysis in Section~\ref{sec:fthdisc}, we expect F-theory on $X_0\times T^2$ with discrete holonomy $(r,s)=(0,s)$ to be dual to Type IIA on $X_{0,\text{n.c.}s}$.

At the level of the topological string partition function, the effect of the $B$-field is to multiply each contribution by $\mu_n^{msq}$, where $q$ is the $\mathbb{Z}_n$ charge and $m$ takes into account multi-coverings of a curve~\cite{Schimannek:2021pau}.
When $C$ is primitive, there are no such multi-covering contributions and $m=1$.
The phase then precisely matches the twining of the elliptic genus in~\eqref{eqn:tteg}.
However, the discrete holonomy induces the same twining on the worldsheets of bound states of multiple strings.
We therefore conclude that
\begin{align}
	Z_C[X_{0,\text{n.c.}s}](t,\tau,\lambda)=Z_{C}^{(0,s)}(\tau,\lambda)\,,
	\label{eqn:topstringex1}
\end{align}
in terms of the expansion~\eqref{eqn:topstringex} of the topological string partition function on $X_{0,\text{n.c.}s}$.

This relation essentially generalizes to arbitrary choices $(r,s)$, but there is a subtlety related to the normalization of the K\"ahler parameters. This can already be seen when considering $(r,s)=(1,0)$.
Let us again denote the geometry that corresponds to an element $r$ of the Tate-Shafarevich group $\Sh(X_0)=\mathbb{Z}_n$ by $X_r$.
As discovered in~\cite{Schimannek:2021pau}, the topological string partition function on $X_1=X_{1,\text{n.c.}0}$ is related to that on $X_{0,\text{n.c.}1}$ by a Fricke involution
\begin{align}
    F_n:\,\tau\rightarrow -\frac{1}{n\tau}\,.
    \label{eqn:frrel}
\end{align}
However, from the two-dimensional perspective discussed in Section~\ref{sec:2danomalies}, the twisted twined elliptic genera corresponding to $(r,s)=(0,1),(1,0)$ are related by a simple S-transformation
\begin{align}
    S:\,\tau\rightarrow -\frac{1}{\tau}\,.
\end{align}

The reason for this difference is, that the K\"ahler parameters are normalized such that the exponents in the expansion of the topological string partition function are integral.
In particular, the K\"ahler parameter $\tau$ associated to $X_1$ is choosen to be $1/n$ times the complexified volume of the generic fiber~\cite{Cota:2019cjx}.
Taking this into account, we find that the twisted elliptic genera are encoded in the topological string partition function on $X_r$ as
\begin{align}
	Z_C[X_{r,\text{n.c.}0}]\left(t,\frac{\text{gcd}(r,n)}{n}\tau,\lambda\right)=Z_{C}^{(r,0)}(\tau,\lambda)\,.
	\label{eqn:topstringex2}
\end{align}
If $n$ is prime, all of the twisted elliptic genera $Z_C^{(r,s)}$ with $r\ne 0$ can then be obtained from $Z_C^{(1,0)}$ by $SL(2,\mathbb{Z})$ transformations.

For F-theory vauca that have $\mathbb{Z}_n$ symmetries with non-prime values of $n$, the remaining partition functions can also be calculated.
The Tate-Shafarevich group $\Sh(X_0)$ then contains additional singular genus one fibrations and one has to consider all of the smooth deformations of the associated (partial) non-commutative resolutions. The relevant networks of geometries for examples with $n=2,\ldots,5$, as well as the corresponding vector valued modular transformations of the topological string partition functions, have been worked out in~\cite{Schimannek:2021pau}.
However, the relation to the twisted twined elliptic genera is slightly more involved, due to the mixing of symmetries encoded in~\eqref{eqn:u1seq}, and we leave a detailed discussion for future work.

\subsection{The discrete 't Hooft anomaly as a height-pairing}
Using the modular properties of the topological string amplitudes, we are now in a position to relate the worldsheet $\mathbb{Z}_n$ 't Hooft anomaly, and by anomaly inflow the six-dimensional global discrete anomaly, to the geometry of the genus one fibration.
Here we will first summarize the result and then in Section~\ref{sec:anmod} show that all of the anomalies, global and local, are directly encoded in certain auto-equivalences of the category of topological B-branes on $X_1$.
Recall that some basic definitions related to modular and Jacobi forms are summarized in Appendix~\ref{app:modjacobi}.

It was found in~\cite{Cota:2019cjx}, generalizing results for elliptic fibrations from~\cite{Haghighat:2013gba,Huang:2015sta}, that the base degree $\beta$ topological string partition function on a smooth genus one fibration $X_1$ with an $n$-section takes the form 
\begin{align}
	Z_C[X_1](\tau,\lambda)=Z^{(1,0)}_C(n\tau,\lambda)=\frac{\Delta_{2n}(\tau)^{\frac{l_C}{n}}}{\eta(n\tau)^{12c_1(B)\cdot C}}\cdot \tilde{\phi}_C(\tau,\lambda)\,,
\end{align}
where $\Delta_{2n}(\tau)\in M_{2n}(\Gamma_1(n))$ is a particular cusp form and $\tilde{\phi}_C(\tau,\lambda)$ is some meromorphic $\Gamma_1(n)$ Jacobi form.
The weight and index of $\tilde{\phi}_C(\tau,\lambda)$ are such that $Z_C[X_1]$ itself transforms like a Jacobi form of weight $0$ and index $r_\lambda(C)$ but its precise form will not be relevant to us.
At least for low degrees $C$, it can usually be fixed using known boundary data or Gopakumar-Vafa invariants.

To determine the exponent $l_C/n$ of $\Delta_{2n}$, one first needs to choose some $n$-section $E_0^{(n)}$ on $X_1$, which, as will be explained in the following section, also defines the K\"ahler parameter $\tau$, and introduce the height pairing
\begin{align}
    D=-\pi^{-1}\pi_*\left(E_0^{(n)}\cdot E_0^{(n)}\right)\,.
    \label{eqn:highpair}
\end{align}
The fractional part of the exponent is then determined by the congruence relation~\cite{Cota:2019cjx}
\begin{align}
    l_C=\frac12\left[n^2c_1(B)-D\right]\cdot C\text{ mod }n\,, 
 \label{eqn:thooftheightpairing}
\end{align}
where we implicitly identify the vertical divisor $D\in H_4(X_1)$ with its image in $H_2(B)$.
Using the transformation properties discussed in Appendix~\ref{app:modjacobi} it follows that 
\begin{align}
\begin{split}
    T^n:\,Z_C^{(1,0)}(\tau,\lambda)\mapsto\,& \mu_n^{-k_C}Z_C^{(1,0)}(\tau,\lambda)\,,\quad k_C=\frac12 D\cdot C\,.
	\end{split}
\end{align}

The phase $\mu_n^{-k_C}$ receives two contributions.
The first contribution of $e^{\pi i n c_1(B)\cdot C}$ comes from the factor of $\eta(\tau)^{-12c_1(B)\cdot C}$, where $-c_1(B)\cdot C/2$ corresponds to the vacuum energy on the non-critical string worldsheet~\cite{haghighat2016}.
This already appears for the ordinary elliptic genera associated to smooth elliptic fibrations and measures the gravitational anomaly, as can also be seen by comparing with~\eqref{eqn:a4poly}.
On the other hand, the remaining contribution of $\mu_n^{l_C}$ comes precisely from the $\mathbb{Z}_n$ 't Hooft anomaly on the worldsheet that appears in~\eqref{eqn:0nthooft} of Section~\ref{sec:CFTtHooft}.
We thus find that the anomaly theory is
\begin{align}
	[l_C]\in H^3(B\mathbb{Z}_n,U(1))\,.
\end{align}

\subsection{Anomalies from monodromies}
\label{sec:anmod}
In the previous subsection we have derived a geometric expression for the discrete 't~Hooft anomaly by using the modular Ansatz for the topological string partition function on a smooth genus one fibration $X_1$.
Since the derivation of the precise form of this Ansatz in~\cite{Cota:2019cjx} involved the use of Higgs transitions in F-theory, let us now discuss how the anomalies can be directly related to the intersection theory on the smooth genus one fibration $X_1$ by using auto-equivalences on the category of topological B-branes.

For local anomalies associated to general continuous gauge groups, this has been carried out in~\cite{Schimannek:2019ijf,Cota:2019cjx}, giving an alternative derivation of earlier results that were obtained using duality with M-theory~\cite{Ferrara:1996wv,Kumar:2009ac,Grimm:2010ks,Bonetti:2011mw,Grimm:2013oga}.
We therefore focus on the case where the six-dimensional gauge symmetry is
\begin{align}
    G=\mathbb{Z}_n\times U(1)\,.
\end{align}
The $U(1)$ gauge symmetry is included to illustrate the effect of the corresponding local anomaly but does not affect results for the global anomaly. 

We can then assume that $X_1$ is a smooth genus one fibered Calabi-Yau threefold with an $n$-section over a base $B$.
The presence of the $U(1)$ gauge symmetry implies that the rank of the Mordell-Weil group $\text{MW}(X_0)$ of the Jacobian fibration is $r=1$
and, by a generalization of the Shioda-Tate-Wazir formula~\cite{Braun:2014oya}, $X_1$ exhibits two independent $n$-sections $E_0^{(n)},E_1^{(n)}$.
In addition, there are also vertical divisors $D_l=\pi^{-1}\tilde{D}_l,\,l=1,\ldots,b_2(B)$ which are preimages of divisors $\tilde{D}_l\in H^2(B,\mathbb{Z})$.
To simplify the exposition, we assume that the Mori cones are simplicial, with a basis $\tilde{C}^l,\,l=1,\ldots,b_2(B)$ on $B$, and choose the divisors $\tilde{D}_l$ to be dual.
We also introduce the curves $C^l=E_0^{(n)}\cdot \pi^{-1}\tilde{C}^l$ on $X_1$, such that
\begin{align}
    C^i\cdot D_j=n\cdot \delta^i_j\,.
\end{align}

Following~\cite{Cota:2019cjx,Knapp:2021vkm}, to construct a good modular parametrization of the K\"ahler form we first need to introduce the shifted zero $n$-section
\begin{align}
    \widetilde{E}_0^{(n)}=E_0^{(n)}+\frac{D}{2n}\,,\quad D=-\pi^{-1}\pi_*\left(E_0^{(n)}\cdot E_0^{(n)}\right)\,,
    \label{eqn:shiftede0}
\end{align}
where $D$ is the height pairing that also appeared in~\eqref{eqn:highpair}.
For any vertical divisor $D_l$ these satisfy the relations
\begin{align}
    \widetilde{E}_0^{(n)}\cdot\widetilde{E}_0^{(n)}\cdot D_l=0 \quad\Leftrightarrow\quad E_0^{(n)}\cdot E_0^{(n)}\cdot D_l=-E_0^{(n)}\cdot D\cdot D_l\,,
\end{align}
which generalizes corresponding identities for rational sections $E_0^{(1)}$ on elliptic fibrations where $D=c_1(B)$.
We also need the image of $E_1^{(n)}$ under an $n$-section analogue $\sigma^{(n)}$ of the Shioda map
\begin{align}
    \widetilde{E}_1^{(n)}=\sigma^{(n)}(E_1^{(n)})\equiv E_1^{(n)}-E_0^{(n)}-\frac{1}{n}\left[\pi^{-1}\pi_*\left(E_1^{(n)}\cdot E_0^{(n)}\right)+D\right]\,,
\end{align}
which is defined such that $\widetilde{E}_1^{(1)}$ is orthogonal to all curves $C^l$.
This satisfies $\pi_*(\widetilde{E}_1^{(n)}\cdot \widetilde{E}_0^{(n)})=0$ and we introduce the corresponding height pairing
\begin{align}
    b=-\pi^{-1}\pi_*\left(\widetilde{E}_1^{(n)}\cdot \widetilde{E}_1^{(n)}\right)\,.
\end{align}
A suitable parametrization of the K\"ahler form on $X_1$ is then given by
\begin{align}
    \omega=\tau\cdot\widetilde{E}_0^{(n)}+m\cdot\widetilde{E}_1^{(n)}+\sum\limits_{i=1}^{b_2(B)}t^i\cdot D_i\,.
    \label{eqn:kaehler}
\end{align}
We denote the complexified volume associated to a curve $C$ in the base by $t_C=\omega\cdot C$.
Note that the choice of $E_0^{(n)},E_1^{(n)}$ just reflects the freedom of choosing different generators for the gauge symmetry.

The strategy is now to interpret the complexified K\"ahler parameters as central charges of $2$-branes and to lift the actions
\begin{align}
    T:\,\tau\rightarrow \tau+1\,,\quad U:\,\tau\rightarrow\frac{\tau}{n\tau+1}\,,
\end{align}
to auto-equivalences on the category of topological B-branes.
This can be done for every torus fibered Calabi-Yau with an $n$-section and leads to generic expressions for the transformation of the complete set of K\"ahler parameters.
We refer to~\cite{Schimannek:2019ijf,Cota:2019cjx} for the technical details and here just state the result
\begin{align}
\begin{split}
    U:&\quad\left\{\begin{array}{c}
    \tau\mapsto\frac{\tau}{n\tau+1}\,,\quad m\mapsto\frac{m}{n\tau+1}\\[.2em]
    t_C\mapsto t_C+\frac12 c_1(B)\cdot C-\frac{nm^2}{n\tau+1}\frac12  b\cdot C+\mathcal{O}(Q)
    \end{array}\right.\,,\\[.3em]
    T:&\quad\tau\mapsto\tau+1\,,\quad m\mapsto m\,,\quad t_C\rightarrow t_C+\frac{1}{2n}D\cdot C\,,
    \end{split}
    \label{eqn:uttrafo}
\end{align}
where the contributions $\mathcal{O}(Q)$ are exponentially suppressed in the large base limit and can be neglected for our analysis.
Let us stress, that the basis~\eqref{eqn:kaehler} including the shift of $E_0^{(n)}$ by $D$ in~\eqref{eqn:shiftede0} was derived in~\cite{Schimannek:2019ijf,Cota:2019cjx} only by requiring that the K\"ahler parameters do not mix under these transformations, without any input from F-theory.

As was observed in~\cite{Schimannek:2019ijf}, the modular properties of the topological string partition function on $X_1$ under $\Gamma_1(n)$-transformations follow from the requirement that each term
\begin{align}
    Z_{C}[X_1](\tau,m,\lambda) e^{2\pi i t\cdot C}\,,
    \label{eqn:ze}
\end{align}
remains invariant under~\eqref{eqn:uttrafo}~\footnote{The $\lambda$-dependent phase that arises from the local gravitational anomaly under the $U$-transformation requires a slightly different treatment, due to its relation to the holomorphic anomaly of the topological string. However, it can be absorbed by introducing an anti-holomorphic overall factor in $Z_{C}[X_1]$~\cite{Haghighat:2013gba}. Here we assume that this has been done but keep the anti-holomorphic dependence implicit.}, such that
\begin{align}
\begin{split}
    Z_{C}[X_1]\left(\frac{\tau}{n\tau+1},m,\lambda\right)=&e^{2\pi i a}e^{2\pi i \frac{n m^2}{n\tau+1}\frac12 b\cdot C}Z_{C}[X_1](\tau,m,\lambda)\,,\\
    \quad Z_{C}[X_1](\tau+1,m,\lambda)=&\mu_n^{-\frac12 D\cdot C}Z_{C}[X_1](\tau,m,\lambda)\,,
    \end{split}
\end{align}
where again $a=\frac12 c_1(B)\cdot C$.
From this one can directly read off the geometric expressions for the local as well as the global 't~Hooft anomalies.

\section{Conclusions}
In this paper we have initiated the study of discrete anomalies in F-theory and of the corresponding discrete Green-Schwarz mechanism.
We have found new highly non-trivial consistency conditions on the spectra, that seem to imply a discrete Swampland for general 6d supergravities, and also derived a geometric expression for the discrete anomaly in F-theory on a torus fibered Calabi-Yau threefold.

By explicitly constructing the 7d anomaly theory of a self-dual chiral 2-form field we have found that this depends on a choice of a quadratic refinement of the differential cohomology pairing, as well as on an overall coupling $\ell$.
For theories with $G=\mathbb{Z}_3$ and no additional tensor multiplet, we observed that F-theory singles out a particular refinement, which imposes new non-trivial consistency conditions on the discrete charged spectrum.
On the other hand, we found that the value of $\ell$ is in general not fixed by the massless spectrum of an F-theory compactification and depends on the geometry of the Calabi-Yau.
Moreover, we have used the $\mathbb{Z}_3$-refined Gopakumar-Vafa invariants associated to a pair of geometrically inequivalent fibrations that lead to the same massless spectrum in F-Theory to demonstrate that the choice of $\ell$ affects the excitation spectrum of the non-critical strings.

To derive a geometric expression for $\ell$, we have used anomaly inflow and identified the 6d discrete Green-Schwarz coupling -- up to a sign -- with a 't~Hooft anomaly of the induced global symmetry in the worldsheet theory of the non-critical strings that couple to the 2-form fields. 
The anomalous crossing relations of topological symmetry lines then allowed us to relate this to the modular properties of the twisted twined elliptic genera of the strings. We also argued that the twisted twined elliptic genera are encoded in the A-model topological string partition function on different non-commutative resolutions of the elements of the Tate-Shafarevich groups of the Calabi-Yau.
The modular properties of the topological string partition function then allowed us to express the discrete anomaly in terms of the height-pairing of a multi-section on a smooth genus one fibration.

Our results open up many interesting avenues for future investigation.
From the supergravity perspective, we have focused on theories with only a self-dual 2-form field and $\mathbb{Z}_3$ gauge symmetry.
In future work we plan to extend our analysis to include additional tensor multiplets as well as more general discrete symmetries.
It will be very interesting to see which constraints arise in constructing the corresponding Green-Schwarz mechanisms and, eventually, to work out the boundaries of the landscape of discrete symmetries in F-theory.

An open issue that we plan to resolve is the precise relation between the global gauge and 't~Hooft anomalies under anomaly inflow, eliminating the sign ambiguity.
While the general strategy will be be analogous to the matching of the local anomalies carried out in~\cite{Shimizu:2016lbw}, extra care is necessary when treating the Bianchi identity in the presence of a string that couples to a non-trivial discrete gauge bundle.
This will likely also require us to clarify the role of the String structure.

Another aspect that we would like to investigate in future work is the relationship between local and global anomalies under Higgs transitions that break a continuous gauge symmetry to a discrete subgroup.
In general, a previously local anomaly can contribute to both the local and global anomalies of the Higgsed theory.
At the level of 2d theories, this has been studied already in~\cite{deWildPropitius:1995cf}.
From a more modern perspective, the total anomaly is encoded in the cohomology of the so-called Madsen–Tillman spectrum~\cite{Freed:2014eja,Freed:2016rqq}.
Examples of this ``anomaly interplay'' in two- and four-dimensional theories have been studied in~\cite{Davighi:2020uab}.
It will be very interesting to extend this to six-dimensional theories and to see whether the discrete anomaly cancellation conditions, that arise from the particular quadratic refinement choosen by F-theory, imply non-trivial constraints that go beyond the cancellation of local anomalies in an un-Higgsed theory.

In our analysis we have only touched upon the worldsheet theories of the self-dual strings and focused on measuring the corresponding 't~Hooft anomalies.
Working out the actual field content should be possible using the techniques developed e.g.\ in~\cite{Lawrie:2016axq}.
Consistency of these theories could potentially lead to constraints on the possible discrete symmetries themselves, analogous to the results obtained by~\cite{Kim:2019vuc,Lee:2019skh} bounding the rank of continuous gauge groups.
Moreover, the elliptic genus can be seen as an invariant of deformation families of two-dimensional superconformal field theories, which are conjecturally classified by the cohomology that is associated to the point by the spectrum of topological modular forms~\cite{stolz_teichner_2004}.
The latter is graded by the gravitational anomaly, see e.g.~\cite{Gukov:2018iiq,Tachikawa:2021mvw}.
The discrete 't~Hooft anomaly provides an additional grading and leads to a further vector valued structure, similar to the effect of the defect group on the strings that arise from compactifications of 6d SCFTs on four-manifolds, studied in~\cite{Gukov:2018iiq}.
Again, we leave a further investigation of this aspect to future work.  

\section*{Acknowledgements}
We thank Iñaki García-Etxebarria, Jie Gu, Amir-Kian Kashani-Poor, Sheldon Katz, Albrecht Klemm, Bruno Le Floch, Guglielmo Lockhart, Miguel Montero, Boris Pioline, Andrei Rotaru, Eric Sharpe and Marcus Sperling for discussions. We further thank the “Engineering in the Landscape: Geometry, Symmetries and Anomalies” workshop in Uppsala University for its hospitality while parts of this projects were advanced. T.S. also thanks the Korean Institute for Advanced Studies (KIAS) for hospitality during part of the completion of this work. The work of MD is supported by the German-Israeli Project Cooperation (DIP) on “Holography and the Swampland”. During the completion of this work P.K.O.\ has received support by
  the European Research Council (ERC) under the European Union’s Horizon 2020 research and innovation programme (grant agreement No. 851931) and  NSF CAREER grant PHY-1848089.
  The research of T.S. is supported by the Agence Nationale de la Recherche (ANR) under contract number ANR-21-CE31-0021.

\appendix
\section{Cubic fibrations from monad-bundles}
\label{app:MonadBundles}
A generic genus one fibration with a 3-section is birational to a fibration $\pi:\,X\rightarrow B$ of cubic curves in $\mathbb{P}^2$.
If the Brauer group of $B$ is trivial, the latter can always be realized as a hypersurface in the projectivization $\mathbb{P}(V)$ of an algebraic rank $3$ vector bundle on the base $B$.
However, in general $V$ will not be a toric bundle. 
To obtain a larger class of rank $3$ bundles on a base $B$ we use the so-called Monad construction.
An introduction to Monad bundles, albeit in the different context of heterotic string compactifications, can be found in~\cite{Anderson:2008uw}.

The monad bundles $V$ that we consider are defined via short exact sequences
\begin{align}
	0\rightarrow V\rightarrow W\xrightarrow{f} U\rightarrow 0\,,
\end{align}
where $W$ and $U$ are sums of line bundles on $B$ and $V=\text{ker}(f)$.
The rank $r_V$ of $V$ is then given by
\begin{align}
	r_V=r_W-r_U\,.
\end{align}
while the first and second Chern classes are
\begin{align}
	c_1(V)=c_1(W)-c_1(U)\,,\quad c_2(V)=c_2(W)+c_1(U)^2-c_2(U)-c_1(W)c_1(U)\,.
\end{align}
Using the results from~\cite{Klevers:2014bqa}, the multiplicity $n_{\pm1}$ of hypermultiplets with $\mathbb{Z}_3$ charge $\pm1$ and the Euler characteristic $\chi$ of $X$ can be expressed in terms of the Chern classes of $V$ and the first Chern class $c_1(B)$ of $B$ as
\begin{align}
	n_{\pm 1}=\frac12\left(42c_1(B)^2+3\Delta(V)\right)\,,\quad \chi=3\Delta(V)-18c_1(B)^2\,,
\end{align}
where $\Delta$ is the Bogomolov discriminant, which for a rank $3$ bundle $V$ takes the form
\begin{align}
	\Delta(V)=2(3c_2(V)-c_1(V)^2)\,.
\end{align}
The Bogomolov discriminant appears, because it is invariant under tensoring of $V$ with a line bundle and therefore an actual invariant of the projective bundle $\mathbb{P}(V)$.

We will mainly focus on fibrations over $B=\mathbb{P}^2$, such that
\begin{align}
	W=\bigoplus\limits_{i=1}^{r_W}\mathcal{O}_{\mathbb{P}^2}(w_i)\,,\quad U=\bigoplus\limits_{j=1}^{r_U}\mathcal{O}_{\mathbb{P}^2}(u_j)\,.
\end{align}
To ensure that $V$ is indeed a vector bundle it suffices to require that $u_j\ge w_i$ for all $i,j$ and that $f$ is sufficiently generic~\cite{fulton81}.
Using $\mathbb{P}(V\otimes L)=\mathbb{P}(V)$ for any line bundle $L$ we impose that $w_1=0$ and $w_i\le 0$ for $1<i\le r_W$.

The projective bundle $\mathbb{P}(V)$, and as a consequence the Calabi-Yau hypersurface, can actually be described as a complete intersection in the toric variety $\mathbb{P}(W)$.
The subspace $\mathbb{P}(V)$ is obtained as the vanishing locus of three polynomials that are linear in the homogeneous coordinates of the fiber but of respective degrees $u_i,\,i=1,\ldots,3$ with respect to the base coordinates.
The Calabi-Yau will be smooth if the complete intersection corresponds to a nef partition of the anti-canonical class of $\mathbb{P}(W)$~\cite{Batyrev:1994pg}.
By explicit calculation one finds that this is the case if and only if
\begin{align}
	\sum\limits_{i=1}^{r_W}w_i\ge-3\,.
\end{align}
On the other hand, if $u_i=w_j$ for some $i,j$, the corresponding line bundle can be removed from $W$ and $U$ without affecting $V$.
Without loss of generality we can therefore restrict to $r_W=6,\,r_U=3$ and write
\begin{align}
	(w_1,\ldots, w_6)=(-\vec{v},0,0,0)\,,
\end{align}
for some $\vec{v}\in\mathbb{N}^3$. We have to consider all choices $\vec{u},\vec{v}\in\mathbb{N}^3$ with
\begin{align}
	\sum\limits_{i=1}^3(u_i+v_i)\le 3\,.
\end{align}
To avoid redundancy we can further impose $u_i\ge u_j,\,v_i\ge v_j$ if $i>j$.

For each of the $16$ possible choices we calculate the Hodge numbers of the anti-canonical Calabi-Yau hypersurface using cohomCalg~\cite{Blumenhagen:2010pv} and find $14$ cases with $h^{1,1}=2$.
Wall's theorem implies that the homotopy type of a Calabi-Yau threefold is determined by the Hodge numbers, the intersection numbers and the second Chern class~\cite{Wall66}.
Two of the $14$ geometries turn out to be Wall equivalent and to avoid redundancy we only include one of them.
In the end we therefore obtain $13$ inequivalent Calabi-Yau threefolds that are genus one fibered over $\mathbb{P}^2$ with a $3$-section.

For each geometry we also choose a basis of divisors $D_1,D_2$ such that $D_1$ is the class of a $3$-section and $D_2=\pi^{-1}(H)$ is a vertical divisor with $H$ being the hyperplane class in $\mathbb{P}^2$, and the triple intersection numbers on the corresponding Calabi-Yau $Y$ are defined as
\begin{align}
    c_{ijk}=\int_Y D_iD_jD_k\,.
\end{align}
The topological invariants associated to the $13$ geometries, together with the resulting F-theory spectra, the invariants of a representative bundle $V$ and the corresponding values for $\vec{v},\,\vec{u}$ are listed in Table~\ref{tab:z3list}.
\begin{table}[t!]
	\centering
	\begin{align*}
		\begin{array}{c|cccc|cccc|ccc|ccc|ccc}
	\#&n_{\pm 1}&n_0&\chi&l&c_{111}&c_{112}&c_{122}&c_{222}&c_{1,V}&c_{2,V}&\Delta_V&v_1&v_2&v_3&u_1&u_2&u_3\\\hline
 1&177 & 96 & -186 & 2 & 14 & 7 & 3 & 0 & -2 & 0 & -8 & -2 & 0 & 0 &  0 & 0 & 0 \\
 2&180 & 93 & -180 & 0 & 3 & 3 & 3 & 0 & -3 & 2 & -6 & -2 & -1 & 0 &  0 & 0 & 0 \\
 3&186 & 87 & -168 & 1 & 5 & 5 & 3 & 0 & -1 & 0 & -2 & -1 & 0 & 0 & 0  & 0 & 0 \\
 4&186 & 87 & -168 & 2 & 11 & 7 & 3 & 0 & -2 & 1 & -2 & -1 & -1 & 0 &  0 & 0 & 0 \\
 5&189 & 84 & -162 & 0 & 0 & 3 & 3 & 0 & 0 & 0 & 0 & 0 & 0 & 0 & 0 & 0 & 0 \\
 6&195 & 78 & -150 & 1 & 2 & 5 & 3 & 0 & -1 & 1 & 4 & 0 & 0 & 0 & 0 & 0 & 1 \\
 7&195 & 78 & -150 & 2 & 8 & 7 & 3 & 0 & -2 & 2 & 4 & -1 & 0 & 0 & 0 &  0 & 1 \\
 8&198 & 75 & -144 & 0 & 15 & 9 & 3 & 0 & -3 & 4 & 6 & -1 & -1 & 0 & 0  & 0 & 1 \\
 9&204 & 69 & -132 & 2 & 5 & 7 & 3 & 0 & -2 & 3 & 10 & 0 & 0 & 0 & 0 &  1 & 1 \\
 10&207 & 66 & -126 & 0 & 12 & 9 & 3 & 0 & -3 & 5 & 12 & -1 & 0 & 0 & 0  & 1 & 1 \\
 11&213 & 60 & -114 & 2 & 2 & 7 & 3 & 0 & -2 & 4 & 16 & 0 & 0 & 0 & 0 &  0 & 2 \\
 12&216 & 57 & -108 & 0 & 9 & 9 & 3 & 0 & -3 & 6 & 18 & 0 & 0 & 0 &   1 & 1 & 1 \\
 13&225 & 48 &  -90 & 0 & 6 & 9 & 3 & 0 & -3 & 7 & 24 & 0 & 0 & 0 &  0 & 1 & 2
\end{array}
	\end{align*}
	\caption{Our monad construction leads to $13$ inequivalent Calabi-Yau threefolds with $h^{1,1}=2$ that are genus one fibered over $\mathbb{P}^2$ with a $3$-section. Here we list the corresponding numbers $n_{\pm 1}$ of $\mathbb{Z}_3$ charged, and $n_0$ uncharged hypermultiples in F-theory, the Euler characteristics $\chi$, the 't Hooft anomaly coefficients $l$, the triple intersection numbers $c_{ijk}$ and the data associated to the monad bundles $V$.}
	\label{tab:z3list}
\end{table}

\section{Anomaly theory for 6d supergravity without gauge group}
\label{app:gravanom}

In this appendix we summarize the anomaly theory for six-dimensional supergravity theories without gauge group. Since the anomalies consequently are pure gravitational and their global part
\begin{align}
\Omega^{\text{Spin}}_7 (\text{pt}) = 0 \,,
\end{align}
vanishes, we can directly relate our findings to the anomaly polynomial $I_8$. The absence of a gauge symmetry determines the number of vector multiplets $V = 0$, leaving us with the gravity multiplet containing a positive chirality gravitino and a self-dual tensor, $T$ tensor multiplets containing a negative chirality spinor and the anti-self-dual tensor field, as well as $H$ hypermultiplets containing negative chirality spinors.
The $\eta$-invariant for the gravitino is given by
\begin{align}
\eta^{\text{gravitino}} = - (\eta^{\text{RS}} - 2 \eta^{\text{D}}) \,,
\end{align}
with RS denoting the $\eta$-invariant for the spin-$\tfrac{3}{2}$ Rarita-Schwinger operator. The overall minus sign indicates the positive chirality of the gravitino and the subtraction of two Dirac $\eta$-invariants accounts for the additional contribution of the longitudinal polarization and extra dimension for the RS operator in seven dimensions. From the main text we have seen that the gravitational contribution of (anti-)self-dual tensor fields is given by
\begin{align}
\mathcal{A}^B_{\text{grav}} = \pm 28 \, \eta^{\text{D}} \,,
\end{align}
with $+$/$-$ sign for anti-self-dual and self-dual tensors, respectively. Giving the full contribution to the anomaly theory
\begin{align}
\mathcal{A}_{\text{grav}} = - (\eta^{\text{RS}} - 2 \, \eta^{\text{D}}) - 28 \, \eta^{\text{D}} + H \, \eta^{\text{D}} + (28 + 1) T \, \eta^{\text{D}} \,.
\end{align}
Extending the 7-manifold $N$ to an 8-manifold $W$ with $\partial W = N$ and using the index densities for Dirac and Rarita-Schwinger operator\footnote{Due to the extension by another dimension for $W$ one once more needs to subtract a Dirac contribution for $\eta^{\text{RS}}$.}
\begin{align}
\begin{split}
\eta^{\text{D}} [N] &= - \tfrac{1}{5760} \int_W (7 p_1^2 - 4 p_2) \text{ mod } \mathbb{Z} \,, \\
\eta^{\text{RS}} [N] &= - \tfrac{1}{720} \int_W (37 p_1^2 - 124 p_2) + \tfrac{1}{5760} \int_W (7 p_1^2 - 4 p_2) \text{ mod } \mathbb{Z} \,,
\end{split}
\end{align}
this can written as
\begin{align}
\begin{split}
\int_N \mathcal{A}_{\text{grav}} &= \int_W \Big( - \tfrac{1}{5760} (H + 29 T - 273) \big(7 p_1^2 - 4 p_2\big) - \tfrac{1}{4} p_1^2  \Big) \\
&= \int_W \Big( - \tfrac{1}{5760} (H + 29 T - 273) \Big( \text{tr} \mathcal{R}^4 + \tfrac{5}{4} \big( \text{tr} \mathcal{R}^2 \big)^2 \Big) - \tfrac{1}{16} \big( \text{tr} \mathcal{R}^2 \big)^2 \Big) \,.
\end{split}
\end{align}
This precisely matches the irreducible gravitational anomaly and the gravitino contribution to the reducible gravitational anomaly, see e.g.\ \cite{Park:2011ji}.

\section{Modular and Jacobi forms}
\label{app:modjacobi}
In this Appendix we collect some relevant properties of modular- and Jacobi forms.

A weak Jacobi form of weight $k$ and index $m$ on a finite index subgroup $\Gamma\subseteq SL(2,\mathbb{Z})$ is a holomorphic function $\phi_{k,m}(\tau,z)$ on $\mathbb{H}\times\mathbb{C}$
that satisfies~\cite{eichler2013theory}
\begin{align}
	\begin{split}
		\phi_{k,m}\left(\frac{a\tau+b}{c\tau+d},\frac{z}{c\tau+d}\right)=&(c\tau+d)^ke^{2\pi i \frac{mcz^2}{c\tau+d}}\phi_{k,m}(\tau,z)\,,\quad {\small\left(\begin{array}{cc}a&b\\c&d\end{array}\right)}\in\Gamma\,,\\
			\phi_{k,m}(\tau,z+\lambda\tau+\mu)=&e^{-2\pi i m(\lambda^2\tau+2\lambda z)}\phi_{k,m}(\tau,z)\,,\quad \lambda,\mu\in\mathbb{Z}\,,
	\end{split}
	\label{eqn:jacobitrafo}
\end{align} 
and admits a Fourier expansion around all the cusps, i.e.
\begin{align}
	\begin{split}
	&(c\tau+d)^{-k}e^{-2\pi i \frac{mcz^2}{c\tau+d}}\phi_{k,m}\left(\frac{a\tau+b}{c\tau+b},\frac{z}{c\tau+d}\right)\\
	=&\sum\limits_{n=0}^\infty\sum\limits_{r=-\infty}^\infty c(n,r)q^n\zeta^r\,,\quad{\small\left(\begin{array}{cc}a&b\\c&d\end{array}\right)}\in SL(2,\mathbb{Z})\,,
	\end{split}
	\label{wjacex}
\end{align}
with $q=e^{2\pi i \tau},\,\zeta=e^{2\pi i z}$.
A weak Jacobi form is called a Jacobi form if the coefficients $c(n,r)$ in~\eqref{wjacex} vanish unless $n\ge r^2/4m$.
More generally, one can consider lattice Jacobi forms $\phi_{k,M}(\tau,z_1,\ldots,z_r)$ which depend on multiple elliptic parameters $z_1,\ldots z_r$ and the index $m$ is replaced by an $r\times r$ index matrix $M$.
The generalization of~\eqref{eqn:jacobitrafo} is straightforward and we refer e.g. to~\cite{Lee:2018urn} for additional details.

The ring of even index $SL(2,\mathbb{Z})$ weak Jacobi forms $J^{\text{weak}}$ is generated over the corresponding ring of modular forms by the elements
\begin{align}
	\phi_{-2,1}(\tau,z)=\frac{\vartheta_1^2(\tau,z)}{\eta(\tau)^6}\,,\quad \phi_{0,1}(\tau,z)=4\left(\frac{\vartheta_2(\tau,z)^2}{\vartheta_2(\tau)^2}+\frac{\vartheta_3(\tau,z)^2}{\vartheta_3(\tau)^2}+\frac{\vartheta_4(\tau,z)^2}{\vartheta_4(\tau)^2}\right)\,,
\end{align}
where $\vartheta_i(\tau,z),\,i=1,\ldots,4$ are the Jacobi theta functions, with $\vartheta_i(\tau)=\vartheta_i(\tau,0)$, such that
\begin{align}
	\phi_{-2,1}(\tau,z)=&\frac{1}{\zeta}-2+\zeta-q\left(\frac{2}{\zeta^2}-\frac{8}{\zeta}+12-8\zeta+2\zeta^2\right)+\mathcal{O}(q^2)\,,\\
	\phi_{0,1}(\tau,z)=&\frac{1}{\zeta}+10+\zeta+q\left(\frac{10}{\zeta^2}-\frac{64}{\zeta}+108-64\zeta+10\zeta^2\right)+\mathcal{O}(q^2)\,.
\end{align}

We denote the ring of weight $k$ modular forms on $\Gamma_1(N)$ by $M_k(\Gamma_1(N))$.
Generators of the rings for $N\le 6$ can be found e.g. in the Appendix of~\cite{Schimannek:2021pau}.
Of particular importance to our discussion are the two weight $2N$ cusp forms on $\Gamma_1(N)$,
\begin{align}
	\begin{split}
		\Delta_{2N}(\tau)=&\frac{1}{q\cdot \phi_{-2,1}(N\tau,\tau)^N}\,,\quad \Delta'_{2N}(\tau)=\frac{1}{\phi_{-2,1}(\tau,1/N)^N}\,,
	\end{split}
\end{align}
that are related by the Fricke involution $F_N:\,\tau\rightarrow-1/N\tau$ as
\begin{align}
	F_N:\,\Delta_{2N}(\tau)\rightarrow \tau^{2N}\Delta'_{2N}(\tau)\,.
\end{align}
Under $T:\,\tau\rightarrow\tau+1$ an $N$-th root of $\Delta_{2N}(\tau)$ transforms as
\begin{align}
    T:\,\Delta_{2N}^{\frac{1}{N}}(\tau)\rightarrow \mu_N^{-1}\Delta_{2N}^{\frac{1}{N}}(\tau)\,,
\end{align}
where we defined the $N$-th root of unity $\mu_N=e^{2\pi i /N}$.

Let us also note the transformation properties of $\Delta'_{2N}(\tau)$ under
\begin{align}
	U_N:\,\tau\rightarrow\frac{\tau}{N\tau+1}\,.
\end{align}
Using the general properties of Jacobi forms it is easy to see that
\begin{align}
	\begin{split}
	U_N:\,\phi_{k,m}(\tau,1/N)\rightarrow& \mu_N^{-m}(N\tau+1)^k\phi_{k,m}(\tau,1/N)\,,
	\end{split}
\end{align}
and therefore ${\Delta'_{2N}}^{\frac{1}{N}}$ transforms under the action~\eqref{eqn:utrafo} as
\begin{align}
	U_N:\,{\Delta'_{2N}}^{\frac{1}{N}}\rightarrow \mu_N(N\tau+1)^2{\Delta'_{2N}}^{\frac{1}{N}}\,.
	\label{eqn:dptrafo}
\end{align}
On the other hand, the Dedekind eta function transforms as
\begin{align}
	U_N:\,\eta(\tau)\rightarrow e^{-N\pi i/12}\sqrt{N\tau+1}\eta(\tau)\,.
	\label{eqn:etatrafo}
\end{align}

\section{Anomalies in supergravity theories with Abelian symmetries}
\label{app:AbelianAnomalies}
Six-dimensional $\mathcal{N}=1$ supergravities are highly constrained by perturbative anomalies.
Nevertheless, those anomalies still allow for a wide range of solutions.
It is expected that the majority of these can not be realized in string theory and are in the Swampland.

For a  $U(1)^r$ theory coupled to gravity, conditions that the pure gauge and gauge-gravitational anomaly can be cancelled by the generalized Green-Schwarz mechanism are given by
\begin{align} 
\label{eq:U1Anomalies}
\begin{array}{rrl}
U(1)_i  \cdot    U(1)_j\cdot U(1)_m \cdot U(1)_n  :&   b_{i,j} \cdot b_{m,n}   &= \sum_k q_i q_j q_m q_n  n_{(q_i, q_j,q_m,q_n)} \, ,  \\ 
U(1)_i \cdot U(1)_j \cdot \text{Grav}^2:& \qquad    a \cdot   b_{i,j}  &= \sum_k q_i  q_j  n_{(q_i, q_j)} \, ,
\end{array}
\end{align}
with $a , b_{i,j}$ being vectors in the $SO(1,T)$ anomaly lattice and $n_{(q_i,q_j)}$ the multiplicity of hypermultiplets with charges $(q_i,q_j)$ under $U(1)_i, U(1)_j$. Together with the corresponding conditions for the pure gravitational anomalies
\begin{align}
\label{eq:GravAnomalies}
a \cdot a = 9-T \, , \qquad   H-r+29T = 273 \, ,
\end{align}
these still allow for an infinite set of seemingly consistent theories, as has systematically been analyzed in~\cite{Park:2011wv,Taylor:2018khc}. For example, in theories with $T=9$ and no massless charged hypermultiplets, the rank $r$ is unbounded~\cite{Park:2011wv}. However, imposing consistency of the worldsheet theories of the non-critical strings, an upper bound $r\le 16$ for theories with $T>0$ was recently derived in~\cite{Lee:2019skh}.
  
Similarly, the anomaly conditions themselves are not strong enough to bound the maximal $U(1)$ charge of hypermultiplets. When setting  $T=0$ one may consider a massless spectrum~\cite{Taylor:2018khc} which is given as 
\begin{align}
H: \{ n_{q}, n_p, n_{p+q}, n_0 \} = \{  54,54, 54, 112   \} \, , \quad \text{ with   } a=3  \, , b_{11}=6 (p^2 + pq + q^2) \, .
\end{align}
This solves \eqref{eq:U1Anomalies} and \eqref{eq:GravAnomalies} for any pair $q,p$.
At this point, there is also no general argument from string theory that imposes an upper bound on the largest possible charge.
The only constraints so far, rely on various concrete string constructions: Fully resolved compact Calabi-Yau threefolds that lead to F-theory vacua with hypermultiplets of $U(1)$ charge $q=5$, that can also be used to Higgs to a $\mathbb{Z}_5$, have been constructed in~\cite{Knapp:2021vkm}. Semi-local as well as perturbative IIB constructions yield charges up to $q=6$, see for example~\cite{Collinucci:2019fnh,Cianci:2018vwv}, while an indirect construction, based on a Higgs chain starting with a non-Abelian group and exotic matter, might even yield charges up to $q=21$~\cite{Raghuram:2018hjn}. 

The maximal possible $U(1)$ charges for massless hypermultiplets is, via Higgs transitions, closely related to the question about possible discrete gauge symmetries.
Currently there is also no bound on the degree of the (minimal) multi-sections in genus-one fibrations that could indicate a maximal $\mathbb{Z}_n$ symmetry.
Attempts in classifying genus one fibrations have been put forward in \cite{Hajouji:2020ddi} and indirect arguments for such bounds have been given in~\cite{Oehlmann:2016wsb, Anderson:2019kmx,Hajouji:2019vxs,Schimannek:2021pau}, indicating that $\mathbb{Z}_6$ might be the maximum.

Anomalies of $\mathbb{Z}_n$ discrete gauge theories have also been considered in~\cite{Monnier:2018nfs}, using similar techniques to the ones we apply in this work.
There it was found that the differences $\Delta n_s$ of the multiplicities of discrete charged singlets of two theories that  are free of  anomalies have to satisfy
\begin{align}
\label{eq:DiscAnomaly}
    \sum_{s=1}^{n-1} \frac{\Delta n_s }{24n} \left(   2 s^2 - 2ns -n^2 s^2 +2n s^3 -s^4 \right) = 0 \text{ mod } 1 \,.
\end{align}
Including 2-form fields and using the general structure of the 7d Wu-Chern-Simons theory and its quadratic refinements it was argued~\cite{Monnier:2018nfs} that this condition is a priori weakened and, specialized to $\mathbb{Z}_3$, the generic constraint on the difference of spectra is
\begin{align}
\label{eq:AnomalyRef}
\begin{array}{cl}
     \mathbb{Z}_3:& \frac13 (\Delta n_1 + \Delta n_2) =  0 \text{ mod } 1\,.
     \end{array} 
\end{align}
Note that this does not constrain the absolute multiplicities $n_1$ and $n_2$ in the theories themselves.

\section{Gopakumar-Vafa invariants}
\label{sec:gvinvariants}

\begin{table}[t!]
\centering
\begin{align*}\tiny
\begin{array}{|c|ccccc|}\hline
n^{(d_1,d_2)}_{g=0}&d_2=0&1&2&3&4\\\hline
d_1=0& 0 & 18 & -2 & 0 & 0 \\
 1&186 & 2439 & 442 & -512 & 768 \\
 2&186 & 58032 & 480662 & 109488 & -119142 \\
 3&168 & 709281 & 39212788 & 232354413 & 60915700 \\
 4&186 & 5931612 & 1402803900 & 35519961168 & 165368736700\\\hline
\end{array}
\end{align*}
\caption{Genus 0 GV-invariants associated to $X_1^{(1)}$.}
\label{tab:gvx11}
\end{table}

\begin{table}[t!]
\centering
\begin{align*}\tiny
\begin{array}{|c|cccc|}\hline
n^{(d_1,d_2),q=0}_{g=0}&d_2=0&1&2&3\\\hline
 d_1=0&0 & 3 & -6 & 27 \\
 1&168 & -336 & 840 & -5376 \\
 2&168 & 47952 & -192144 & 1688904 \\
 3&168 & 68025204 & 24932016 & -304414536 \\
 4&168 & 7257658029 & -16644397488 & 74703349650 \\\hline\hline
n^{(d_1,d_2),q=\pm 1}_{g=0}&d_2=0&1&2&3\\\hline
d_1=0&0 & 0 & 0 & 0 \\
 1&186 & -372 & 930 & -5952 \\
 2&186 & 47709 & -191208 & 1681533 \\
 3&186 & 68022990 & 24939252 & -304484232 \\
 4&186 & 7257644763 & -16644331086 & 74702754525 \\\hline
\end{array}
\end{align*}
\caption{$\mathbb{Z}_3$-refined genus 0 GV-invariants with $\mathbb{Z}_3$ charge $q\in\{0,\pm1\}$ associated to $X_{0}^{(1)}$.}
\label{tab:gvx01}
\end{table}

\begin{table}[t!]
\centering
\begin{align*}\tiny
\begin{array}{|c|ccccc|}\hline
n^{(d_1,d_2)}_{g=0}&d_2=0&1&2&3&4\\\hline
d_1=0& 0 & 1 & 0 & 0 & 0 \\
 1&186 & 640 & 0 & 0 & 0 \\
 2&186 & 22295 & 10032 & 0 & 0 \\
 3&168 & 324012 & 2367222 & 288384 & 0 \\
 4&186 & 3017570 & 138486166 & 239874519 & 10979984 \\\hline
\end{array}
\end{align*}
\caption{Genus 0 GV-invariants associated to $X_1^{(2)}$.}
\label{tab:gvx12}
\end{table}

\begin{table}[t!]
\centering
\begin{align*}\tiny
\begin{array}{|c|cccc|}\hline
n^{(d_1,d_2),q=0}_{g=0}&d_2=0&1&2&3\\\hline
 d_1=0&0 & 3 & -6 & 27 \\
 1&168 & -336 & 840 & -5376 \\
 2&168 & 47952 & -192144 & 1688904 \\
 3&168 & 68024232 & 24932016 & -304414536 \\
 4&168 & 7257653655 & -16644388740 & 74703327780 \\\hline\hline
n^{(d_1,d_2),q=\pm 1}_{g=0}&d_2=0&1&2&3\\\hline
d_1=0&0 & 0 & 0 & 0 \\
 1&186 & -372 & 930 & -5952 \\
 2&186 & 47709 & -191208 & 1681533 \\
 3&186 & 68023476 & 24939252 & -304484232 \\
 4&186 & 7257646950 & -16644335460 & 74702765460 \\\hline
\end{array}
\end{align*}
\caption{$\mathbb{Z}_3$-refined genus 0 GV-invariants with $\mathbb{Z}_3$ charge $q\in\{0,\pm1\}$ associated to $X_{0}^{(2)}$.}
\label{tab:gvx02}
\end{table}
Here we compare the Gopakumar-Vafa invariants associated to a pair $(X_1^{(1)},X_1^{(2)})$ of inequivalent genus one fibered Calabi-Yau threefolds that lead to a $G=\mathbb{Z}_3$ gauge symmetry and identical massless spectra in F-theory.
We also compare the $\mathbb{Z}_3$-refined invariants associated to the corresponding Jacobian fibrations $(X_0^{(1)},X_0^{(2)})$ following~\cite{Schimannek:2021pau}.
We focus on the geometries 3, 4 in Tables~\ref{tab:z3listred} and~\ref{tab:z3list}.
They can be realized as anti-canonical hypersurfaces in the respective toric ambient spaces
\begin{align}
    M^{(1)}=\mathbb{P}(\mathcal{O}(-1)\oplus\mathcal{O}^{\oplus 2}\rightarrow\mathbb{P}^2)\,,\quad M^{(2)}=\mathbb{P}(\mathcal{O}(-1)^{\oplus 2}\oplus\mathcal{O}\rightarrow\mathbb{P}^2)\,.
\end{align}

Gopakumar-Vafa invariants are physically defined as certain traces over multipicities of BPS states in the five-dimensional M-theory compactification on a Calabi-Yau and are encoded in the A-model topological string partition function~\cite{Gopakumar:1998ii,Gopakumar:1998jq}.
In the case of a genus one fibration $X_1$ with an $n$-section, it was recently found that there are, in addition to the GV-invariants associated to $X_1$ itself, also $\mathbb{Z}_n$-refined GV invariants that can be extracted by combining information from the topological string partition function on the non-commutative resolutions $X_{0,\text{n.c.}s}$ of the singular Jacobian fibration $X_0$ and its smooth deformation $X_{0,\text{def.}}$~\cite{Schimannek:2021pau}.
The refined invariant are associated to M-theory on $X_0$ and take into account the charge of the BPS particles under the $\mathbb{Z}_n$ gauge symmetry.
We will just state the result of the calculation at genus zero and refer to~\cite{Schimannek:2021pau} for any  of the technical details.

Some genus zero Gopakumar-Vafa invariants $n^{(d_1,d_2)}_g$ associated to $X_1^{(1)}$ and $X_1^{(2)}$ are listed in Tables~\ref{tab:gvx11} and~\ref{tab:gvx12}.
The corresponding $\mathbb{Z}_3$-refined GV invariants $n^{(d_1,d_2),q}_g$ associated to the Jacobian fibrations $X_{0}^{(1)}$ and $X_{0}^{(2)}$ are provided in Tables~\ref{tab:gvx01} and~\ref{tab:gvx02}.
In each case $d_1$ refers to the degree with respect to the fiber or, in the case of the genus one fibrations, a component of a reducible $I_2$ such that the generic fiber has degree $d_1=3$.
The degree with respect to the hyperplane class of the base $\mathbb{P}^2$ is denoted by $d_2$.
The refined invariants are also labelled by the $\mathbb{Z}_3$ charge $q$.
The genus $g$ is related to the representations under the $SO(4)$ little group.

Comparing the GV-invariants associated to $X_1^{(1)}$ and $X_1^{(2)}$ in Tables~\ref{tab:gvx11} and~\ref{tab:gvx12}, it is immediately clear that $X_1^{(1)}$ are $X_1^{(2)}$ lead to different M-theory vacua.
The $\mathbb{Z}_3$-refined invariants in the Tables~\ref{tab:gvx01} and~\ref{tab:gvx02} also differ but only for degrees $d_1\ge 3$, corresponding to higher excitations of the self-dual strings.
Nevertheless, this demonstrates that the excitation spectra of the strings, and therefore the F-theory vacua associated to $X_0^{(1)}$ and $X_0^{(2)}$, are also physically distinct.
It would be very interesting to develop a better understanding of the corresponding worldsheet theories and to study the effect of the anomaly further in future work.

\addcontentsline{toc}{section}{References}
\bibliographystyle{JHEP}
\bibliography{names}

\providecommand{\href}[2]{#2}\begingroup\raggedright\begin{thebibliography}{100}

\bibitem{Green:1984bx}
M.B.~Green, J.H.~Schwarz and P.C.~West, \emph{{Anomaly Free Chiral Theories in
  Six-Dimensions}},
  \href{https://doi.org/10.1016/0550-3213(85)90222-6}{\emph{Nucl. Phys. B}
  {\bfseries 254} (1985) 327}.

\bibitem{GREEN1984117}
M.B.~Green and J.H.~Schwarz, \emph{Anomaly cancellations in supersymmetric d =
  10 gauge theory and superstring theory},
  \href{https://doi.org/https://doi.org/10.1016/0370-2693(84)91565-X}{\emph{Physics
  Letters B} {\bfseries 149} (1984) 117}.

\bibitem{Sagnotti:1992qw}
A.~Sagnotti, \emph{{A Note on the Green-Schwarz mechanism in open string
  theories}}, \href{https://doi.org/10.1016/0370-2693(92)90682-T}{\emph{Phys.
  Lett. B} {\bfseries 294} (1992) 196}
  [\href{https://arxiv.org/abs/hep-th/9210127}{{\ttfamily hep-th/9210127}}].

\bibitem{Vafa:2005ui}
C.~Vafa, \emph{{The String landscape and the swampland}},
  \href{https://arxiv.org/abs/hep-th/0509212}{{\ttfamily hep-th/0509212}}.

\bibitem{Lee:2019skh}
S.-J.~Lee and T.~Weigand, \emph{{Swampland Bounds on the Abelian Gauge
  Sector}}, \href{https://doi.org/10.1103/PhysRevD.100.026015}{\emph{Phys. Rev.
  D} {\bfseries 100} (2019) 026015}
  [\href{https://arxiv.org/abs/1905.13213}{{\ttfamily 1905.13213}}].

\bibitem{Brennan:2017rbf}
T.D.~Brennan, F.~Carta and C.~Vafa, \emph{{The String Landscape, the Swampland,
  and the Missing Corner}},
  \href{https://doi.org/10.22323/1.305.0015}{\emph{PoS} {\bfseries TASI2017}
  (2017) 015} [\href{https://arxiv.org/abs/1711.00864}{{\ttfamily
  1711.00864}}].

\bibitem{Palti:2019pca}
E.~Palti, \emph{{The Swampland: Introduction and Review}},
  \href{https://doi.org/10.1002/prop.201900037}{\emph{Fortsch. Phys.}
  {\bfseries 67} (2019) 1900037}
  [\href{https://arxiv.org/abs/1903.06239}{{\ttfamily 1903.06239}}].

\bibitem{vanBeest:2021lhn}
M.~van Beest, J.~Calder\'on-Infante, D.~Mirfendereski and I.~Valenzuela,
  \emph{{Lectures on the Swampland Program in String Compactifications}},
  \href{https://doi.org/10.1016/j.physrep.2022.09.002}{\emph{Phys. Rept.}
  {\bfseries 989} (2022) 1} [\href{https://arxiv.org/abs/2102.01111}{{\ttfamily
  2102.01111}}].

\bibitem{Vafa:1996xn}
C.~Vafa, \emph{{Evidence for F theory}},
  \href{https://doi.org/10.1016/0550-3213(96)00172-1}{\emph{Nucl. Phys. B}
  {\bfseries 469} (1996) 403}
  [\href{https://arxiv.org/abs/hep-th/9602022}{{\ttfamily hep-th/9602022}}].

\bibitem{Morrison:1996na}
D.R.~Morrison and C.~Vafa, \emph{{Compactifications of F theory on Calabi-Yau
  threefolds. 1}},
  \href{https://doi.org/10.1016/0550-3213(96)00242-8}{\emph{Nucl. Phys. B}
  {\bfseries 473} (1996) 74}
  [\href{https://arxiv.org/abs/hep-th/9602114}{{\ttfamily hep-th/9602114}}].

\bibitem{Morrison:1996pp}
D.R.~Morrison and C.~Vafa, \emph{{Compactifications of F theory on Calabi-Yau
  threefolds. 2.}},
  \href{https://doi.org/10.1016/0550-3213(96)00369-0}{\emph{Nucl. Phys. B}
  {\bfseries 476} (1996) 437}
  [\href{https://arxiv.org/abs/hep-th/9603161}{{\ttfamily hep-th/9603161}}].

\bibitem{Bershadsky:1996nh}
M.~Bershadsky, K.A.~Intriligator, S.~Kachru, D.R.~Morrison, V.~Sadov and
  C.~Vafa, \emph{{Geometric singularities and enhanced gauge symmetries}},
  \href{https://doi.org/10.1016/S0550-3213(96)90131-5}{\emph{Nucl. Phys. B}
  {\bfseries 481} (1996) 215}
  [\href{https://arxiv.org/abs/hep-th/9605200}{{\ttfamily hep-th/9605200}}].

\bibitem{Sadov:1996zm}
V.~Sadov, \emph{{Generalized Green-Schwarz mechanism in F theory}},
  \href{https://doi.org/10.1016/0370-2693(96)01134-3}{\emph{Phys. Lett. B}
  {\bfseries 388} (1996) 45}
  [\href{https://arxiv.org/abs/hep-th/9606008}{{\ttfamily hep-th/9606008}}].

\bibitem{Aspinwall:2000kf}
P.S.~Aspinwall, S.H.~Katz and D.R.~Morrison, \emph{{Lie groups, Calabi-Yau
  threefolds, and F theory}},
  \href{https://doi.org/10.4310/ATMP.2000.v4.n1.a2}{\emph{Adv. Theor. Math.
  Phys.} {\bfseries 4} (2000) 95}
  [\href{https://arxiv.org/abs/hep-th/0002012}{{\ttfamily hep-th/0002012}}].

\bibitem{Grassi:2000we}
A.~Grassi and D.R.~Morrison, \emph{{Group representations and the Euler
  characteristic of elliptically fibered Calabi-Yau threefolds}},
  \href{https://arxiv.org/abs/math/0005196}{{\ttfamily math/0005196}}.

\bibitem{Grassi:2011hq}
A.~Grassi and D.R.~Morrison, \emph{{Anomalies and the Euler characteristic of
  elliptic Calabi-Yau threefolds}},
  \href{https://doi.org/10.4310/CNTP.2012.v6.n1.a2}{\emph{Commun. Num. Theor.
  Phys.} {\bfseries 6} (2012) 51}
  [\href{https://arxiv.org/abs/1109.0042}{{\ttfamily 1109.0042}}].

\bibitem{Kumar:2009ac}
V.~Kumar, D.R.~Morrison and W.~Taylor, \emph{{Mapping 6D N = 1 supergravities
  to F-theory}}, \href{https://doi.org/10.1007/JHEP02(2010)099}{\emph{JHEP}
  {\bfseries 02} (2010) 099} [\href{https://arxiv.org/abs/0911.3393}{{\ttfamily
  0911.3393}}].

\bibitem{Park:2011ji}
D.S.~Park, \emph{{Anomaly Equations and Intersection Theory}},
  \href{https://doi.org/10.1007/JHEP01(2012)093}{\emph{JHEP} {\bfseries 01}
  (2012) 093} [\href{https://arxiv.org/abs/1111.2351}{{\ttfamily 1111.2351}}].

\bibitem{Arras:2016evy}
P.~Arras, A.~Grassi and T.~Weigand, \emph{{Terminal Singularities, Milnor
  Numbers, and Matter in F-theory}},
  \href{https://doi.org/10.1016/j.geomphys.2017.09.001}{\emph{J. Geom. Phys.}
  {\bfseries 123} (2018) 71}
  [\href{https://arxiv.org/abs/1612.05646}{{\ttfamily 1612.05646}}].

\bibitem{Grassi:2018rva}
A.~Grassi and T.~Weigand, \emph{{On topological invariants of algebraic
  threefolds with ($\mathbb Q$-factorial) singularities}},
  \href{https://arxiv.org/abs/1804.02424}{{\ttfamily 1804.02424}}.

\bibitem{Grassi1991}
A.~Grassi, \emph{On minimal models of elliptic threefolds.},
  {\emph{Mathematische Annalen} {\bfseries 290} (1991) 287}.

\bibitem{gross1994finiteness}
M.~Gross, \emph{A finiteness theorem for elliptic calabi-yau threefolds},
  {\emph{Duke Mathematical Journal} {\bfseries 74} (1994) 271}
  [\href{https://arxiv.org/abs/alg-geom/9305002}{{\ttfamily
  alg-geom/9305002}}].

\bibitem{Kumar:2010ru}
V.~Kumar, D.R.~Morrison and W.~Taylor, \emph{{Global aspects of the space of 6D
  N = 1 supergravities}},
  \href{https://doi.org/10.1007/JHEP11(2010)118}{\emph{JHEP} {\bfseries 11}
  (2010) 118} [\href{https://arxiv.org/abs/1008.1062}{{\ttfamily 1008.1062}}].

\bibitem{Taylor:2018khc}
W.~Taylor and A.P.~Turner, \emph{{An infinite swampland of U(1) charge spectra
  in 6D supergravity theories}},
  \href{https://doi.org/10.1007/JHEP06(2018)010}{\emph{JHEP} {\bfseries 06}
  (2018) 010} [\href{https://arxiv.org/abs/1803.04447}{{\ttfamily
  1803.04447}}].

\bibitem{Kim:2019vuc}
H.-C.~Kim, G.~Shiu and C.~Vafa, \emph{{Branes and the Swampland}},
  \href{https://doi.org/10.1103/PhysRevD.100.066006}{\emph{Phys. Rev. D}
  {\bfseries 100} (2019) 066006}
  [\href{https://arxiv.org/abs/1905.08261}{{\ttfamily 1905.08261}}].

\bibitem{Krause:2012yh}
S.~Krause, C.~Mayrhofer and T.~Weigand, \emph{{Gauge Fluxes in F-theory and
  Type IIB Orientifolds}},
  \href{https://doi.org/10.1007/JHEP08(2012)119}{\emph{JHEP} {\bfseries 08}
  (2012) 119} [\href{https://arxiv.org/abs/1202.3138}{{\ttfamily 1202.3138}}].

\bibitem{Morrison:2012ei}
D.R.~Morrison and D.S.~Park, \emph{{F-Theory and the Mordell-Weil Group of
  Elliptically-Fibered Calabi-Yau Threefolds}},
  \href{https://doi.org/10.1007/JHEP10(2012)128}{\emph{JHEP} {\bfseries 10}
  (2012) 128} [\href{https://arxiv.org/abs/1208.2695}{{\ttfamily 1208.2695}}].

\bibitem{Grimm:2013oga}
T.W.~Grimm, A.~Kapfer and J.~Keitel, \emph{{Effective action of 6D F-Theory
  with U(1) factors: Rational sections make Chern-Simons terms jump}},
  \href{https://doi.org/10.1007/JHEP07(2013)115}{\emph{JHEP} {\bfseries 07}
  (2013) 115} [\href{https://arxiv.org/abs/1305.1929}{{\ttfamily 1305.1929}}].

\bibitem{Braun:2014oya}
V.~Braun and D.R.~Morrison, \emph{{F-theory on Genus-One Fibrations}},
  \href{https://doi.org/10.1007/JHEP08(2014)132}{\emph{JHEP} {\bfseries 08}
  (2014) 132} [\href{https://arxiv.org/abs/1401.7844}{{\ttfamily 1401.7844}}].

\bibitem{Morrison:2014era}
D.R.~Morrison and W.~Taylor, \emph{{Sections, multisections, and U(1) fields in
  F-theory}},  \href{https://arxiv.org/abs/1404.1527}{{\ttfamily 1404.1527}}.

\bibitem{Anderson:2014yva}
L.B.~Anderson, I.n.~Garc\'\i{}a-Etxebarria, T.W.~Grimm and J.~Keitel,
  \emph{{Physics of F-theory compactifications without section}},
  \href{https://doi.org/10.1007/JHEP12(2014)156}{\emph{JHEP} {\bfseries 12}
  (2014) 156} [\href{https://arxiv.org/abs/1406.5180}{{\ttfamily 1406.5180}}].

\bibitem{Klevers:2014bqa}
D.~Klevers, D.K.~Mayorga~Pena, P.-K.~Oehlmann, H.~Piragua and J.~Reuter,
  \emph{{F-Theory on all Toric Hypersurface Fibrations and its Higgs
  Branches}}, \href{https://doi.org/10.1007/JHEP01(2015)142}{\emph{JHEP}
  {\bfseries 01} (2015) 142} [\href{https://arxiv.org/abs/1408.4808}{{\ttfamily
  1408.4808}}].

\bibitem{Mayrhofer:2014laa}
C.~Mayrhofer, E.~Palti, O.~Till and T.~Weigand, \emph{{On Discrete Symmetries
  and Torsion Homology in F-Theory}},
  \href{https://doi.org/10.1007/JHEP06(2015)029}{\emph{JHEP} {\bfseries 06}
  (2015) 029} [\href{https://arxiv.org/abs/1410.7814}{{\ttfamily 1410.7814}}].

\bibitem{Cvetic:2015moa}
M.~Cveti\v{c}, R.~Donagi, D.~Klevers, H.~Piragua and M.~Poretschkin,
  \emph{{F-theory vacua with $\mathbb Z_3$ gauge symmetry}},
  \href{https://doi.org/10.1016/j.nuclphysb.2015.07.011}{\emph{Nucl. Phys. B}
  {\bfseries 898} (2015) 736}
  [\href{https://arxiv.org/abs/1502.06953}{{\ttfamily 1502.06953}}].

\bibitem{Garcia-Etxebarria:2014qua}
I.n.~Garc\'\i{}a-Etxebarria, T.W.~Grimm and J.~Keitel, \emph{{Yukawas and
  discrete symmetries in F-theory compactifications without section}},
  \href{https://doi.org/10.1007/JHEP11(2014)125}{\emph{JHEP} {\bfseries 11}
  (2014) 125} [\href{https://arxiv.org/abs/1408.6448}{{\ttfamily 1408.6448}}].

\bibitem{Lin:2015qsa}
L.~Lin, C.~Mayrhofer, O.~Till and T.~Weigand, \emph{{Fluxes in F-theory
  Compactifications on Genus-One Fibrations}},
  \href{https://doi.org/10.1007/JHEP01(2016)098}{\emph{JHEP} {\bfseries 01}
  (2016) 098} [\href{https://arxiv.org/abs/1508.00162}{{\ttfamily
  1508.00162}}].

\bibitem{Oehlmann:2016wsb}
P.-K.~Oehlmann, J.~Reuter and T.~Schimannek, \emph{{Mordell-Weil Torsion in the
  Mirror of Multi-Sections}},
  \href{https://doi.org/10.1007/JHEP12(2016)031}{\emph{JHEP} {\bfseries 12}
  (2016) 031} [\href{https://arxiv.org/abs/1604.00011}{{\ttfamily
  1604.00011}}].

\bibitem{Kimura:2016crs}
Y.~Kimura, \emph{{Discrete Gauge Groups in F-theory Models on Genus-One Fibered
  Calabi-Yau 4-folds without Section}},
  \href{https://doi.org/10.1007/JHEP04(2017)168}{\emph{JHEP} {\bfseries 04}
  (2017) 168} [\href{https://arxiv.org/abs/1608.07219}{{\ttfamily
  1608.07219}}].

\bibitem{Cvetic:2018bni}
M.~Cveti\v{c} and L.~Lin, \emph{{TASI Lectures on Abelian and Discrete
  Symmetries in F-theory}},
  \href{https://doi.org/10.22323/1.305.0020}{\emph{PoS} {\bfseries TASI2017}
  (2018) 020} [\href{https://arxiv.org/abs/1809.00012}{{\ttfamily
  1809.00012}}].

\bibitem{Anderson:2018heq}
L.B.~Anderson, A.~Grassi, J.~Gray and P.-K.~Oehlmann, \emph{{F-theory on
  Quotient Threefolds with (2,0) Discrete Superconformal Matter}},
  \href{https://doi.org/10.1007/JHEP06(2018)098}{\emph{JHEP} {\bfseries 06}
  (2018) 098} [\href{https://arxiv.org/abs/1801.08658}{{\ttfamily
  1801.08658}}].

\bibitem{Anderson:2019kmx}
L.B.~Anderson, J.~Gray and P.-K.~Oehlmann, \emph{{F-Theory on Quotients of
  Elliptic Calabi-Yau Threefolds}},
  \href{https://doi.org/10.1007/JHEP12(2019)131}{\emph{JHEP} {\bfseries 12}
  (2019) 131} [\href{https://arxiv.org/abs/1906.11955}{{\ttfamily
  1906.11955}}].

\bibitem{Cota:2019cjx}
C.F.~Cota, A.~Klemm and T.~Schimannek, \emph{{Topological strings on genus one
  fibered Calabi-Yau 3-folds and string dualities}},
  \href{https://doi.org/10.1007/JHEP11(2019)170}{\emph{JHEP} {\bfseries 11}
  (2019) 170} [\href{https://arxiv.org/abs/1910.01988}{{\ttfamily
  1910.01988}}].

\bibitem{Oehlmann:2019ohh}
P.-K.~Oehlmann and T.~Schimannek, \emph{{GV-Spectroscopy for F-theory on
  genus-one fibrations}},
  \href{https://doi.org/10.1007/JHEP09(2020)066}{\emph{JHEP} {\bfseries 09}
  (2020) 066} [\href{https://arxiv.org/abs/1912.09493}{{\ttfamily
  1912.09493}}].

\bibitem{Knapp:2021vkm}
J.~Knapp, E.~Scheidegger and T.~Schimannek, \emph{{On genus one fibered
  Calabi-Yau threefolds with 5-sections}},
  \href{https://arxiv.org/abs/2107.05647}{{\ttfamily 2107.05647}}.

\bibitem{Schimannek:2021pau}
T.~Schimannek, \emph{{Modular curves, the Tate-Shafarevich group and
  Gopakumar-Vafa invariants with discrete charges}},
  \href{https://doi.org/10.1007/JHEP02(2022)007}{\emph{JHEP} {\bfseries 02}
  (2022) 007} [\href{https://arxiv.org/abs/2108.09311}{{\ttfamily
  2108.09311}}].

\bibitem{Garcia-Etxebarria:2018ajm}
I.n.~Garc\'\i{}a-Etxebarria and M.~Montero, \emph{{Dai-Freed anomalies in
  particle physics}},
  \href{https://doi.org/10.1007/JHEP08(2019)003}{\emph{JHEP} {\bfseries 08}
  (2019) 003} [\href{https://arxiv.org/abs/1808.00009}{{\ttfamily
  1808.00009}}].

\bibitem{Monnier:2019ytc}
S.~Monnier, \emph{{A Modern Point of View on Anomalies}},
  \href{https://doi.org/10.1002/prop.201910012}{\emph{Fortsch. Phys.}
  {\bfseries 67} (2019) 1910012}
  [\href{https://arxiv.org/abs/1903.02828}{{\ttfamily 1903.02828}}].

\bibitem{Dai:1994kq}
X.-z.~Dai and D.S.~Freed, \emph{{eta invariants and determinant lines}},
  \href{https://doi.org/10.1063/1.530747}{\emph{J. Math. Phys.} {\bfseries 35}
  (1994) 5155} [\href{https://arxiv.org/abs/hep-th/9405012}{{\ttfamily
  hep-th/9405012}}].

\bibitem{Witten:1996hc}
E.~Witten, \emph{{Five-brane effective action in M theory}},
  \href{https://doi.org/10.1016/S0393-0440(97)80160-X}{\emph{J. Geom. Phys.}
  {\bfseries 22} (1997) 103}
  [\href{https://arxiv.org/abs/hep-th/9610234}{{\ttfamily hep-th/9610234}}].

\bibitem{Belov:2006jd}
D.~Belov and G.W.~Moore, \emph{{Holographic Action for the Self-Dual Field}},
  \href{https://arxiv.org/abs/hep-th/0605038}{{\ttfamily hep-th/0605038}}.

\bibitem{Hsieh:2020jpj}
C.-T.~Hsieh, Y.~Tachikawa and K.~Yonekura, \emph{{Anomaly Inflow and p-Form
  Gauge Theories}},
  \href{https://doi.org/10.1007/s00220-022-04333-w}{\emph{Commun. Math. Phys.}
  {\bfseries 391} (2022) 495}
  [\href{https://arxiv.org/abs/2003.11550}{{\ttfamily 2003.11550}}].

\bibitem{cmp/1104180750}
R.~Dijkgraaf and E.~Witten, \emph{{Topological gauge theories and group
  cohomology}}, \href{https://doi.org/cmp/1104180750}{\emph{Communications in
  Mathematical Physics} {\bfseries 129} (1990) 393 }.

\bibitem{Gaiotto:2014kfa}
D.~Gaiotto, A.~Kapustin, N.~Seiberg and B.~Willett, \emph{{Generalized Global
  Symmetries}}, \href{https://doi.org/10.1007/JHEP02(2015)172}{\emph{JHEP}
  {\bfseries 02} (2015) 172} [\href{https://arxiv.org/abs/1412.5148}{{\ttfamily
  1412.5148}}].

\bibitem{Bhardwaj:2017xup}
L.~Bhardwaj and Y.~Tachikawa, \emph{{On finite symmetries and their gauging in
  two dimensions}}, \href{https://doi.org/10.1007/JHEP03(2018)189}{\emph{JHEP}
  {\bfseries 03} (2018) 189}
  [\href{https://arxiv.org/abs/1704.02330}{{\ttfamily 1704.02330}}].

\bibitem{Chang:2018iay}
C.-M.~Chang, Y.-H.~Lin, S.-H.~Shao, Y.~Wang and X.~Yin, \emph{{Topological
  Defect Lines and Renormalization Group Flows in Two Dimensions}},
  \href{https://doi.org/10.1007/JHEP01(2019)026}{\emph{JHEP} {\bfseries 01}
  (2019) 026} [\href{https://arxiv.org/abs/1802.04445}{{\ttfamily
  1802.04445}}].

\bibitem{Schimannek:2019ijf}
T.~Schimannek, \emph{{Modularity from Monodromy}},
  \href{https://doi.org/10.1007/JHEP05(2019)024}{\emph{JHEP} {\bfseries 05}
  (2019) 024} [\href{https://arxiv.org/abs/1902.08215}{{\ttfamily
  1902.08215}}].

\bibitem{Weigand:2018rez}
T.~Weigand, \emph{{F-theory}}, {\emph{PoS} {\bfseries TASI2017} (2018) 016}
  [\href{https://arxiv.org/abs/1806.01854}{{\ttfamily 1806.01854}}].

\bibitem{Mayrhofer:2014haa}
C.~Mayrhofer, E.~Palti, O.~Till and T.~Weigand, \emph{{Discrete Gauge
  Symmetries by Higgsing in four-dimensional F-Theory Compactifications}},
  \href{https://doi.org/10.1007/JHEP12(2014)068}{\emph{JHEP} {\bfseries 12}
  (2014) 068} [\href{https://arxiv.org/abs/1408.6831}{{\ttfamily 1408.6831}}].

\bibitem{Bhardwaj:2015oru}
L.~Bhardwaj, M.~Del~Zotto, J.J.~Heckman, D.R.~Morrison, T.~Rudelius and
  C.~Vafa, \emph{{F-theory and the Classification of Little Strings}},
  \href{https://doi.org/10.1103/PhysRevD.93.086002}{\emph{Phys. Rev. D}
  {\bfseries 93} (2016) 086002}
  [\href{https://arxiv.org/abs/1511.05565}{{\ttfamily 1511.05565}}].

\bibitem{silverman2009arithmetic}
J.H.~Silverman, \emph{The Arithmetic of Elliptic Curves}, Graduate Texts in
  Mathematics, Springer New York (2009).

\bibitem{joseph1994advanced}
J.H.~Silverman, \emph{Advanced Topics in the Arithmetic of Elliptic Curves},
  Graduate texts in mathematics, Springer-Verlag (1994).

\bibitem{deBoer:2001wca}
J.~de~Boer, R.~Dijkgraaf, K.~Hori, A.~Keurentjes, J.~Morgan, D.R.~Morrison
  et~al., \emph{{Triples, fluxes, and strings}},
  \href{https://doi.org/10.4310/ATMP.2000.v4.n5.a1}{\emph{Adv. Theor. Math.
  Phys.} {\bfseries 4} (2002) 995}
  [\href{https://arxiv.org/abs/hep-th/0103170}{{\ttfamily hep-th/0103170}}].

\bibitem{Kohl:2021rxy}
F.B.~Kohl, M.~Larfors and P.-K.~Oehlmann, \emph{{F-theory on 6D symmetric
  toroidal orbifolds}},
  \href{https://doi.org/10.1007/JHEP05(2022)064}{\emph{JHEP} {\bfseries 05}
  (2022) 064} [\href{https://arxiv.org/abs/2111.07998}{{\ttfamily
  2111.07998}}].

\bibitem{oggshafa}
I.~Dolgachev and M.~Gross, \emph{Elliptic three-folds i: Ogg-shafarevich
  theory},  \href{https://arxiv.org/abs/alg-geom/9210009}{{\ttfamily
  alg-geom/9210009}}.

\bibitem{wip}
S.~Katz, A.~Klemm, T.~Schimannek and E.~Sharpe, \emph{{Topological Strings on
  Non-Commutative Resolutions (work in progress)}}, .

\bibitem{caldararu2001derived}
A.~C\u{a}ld\u{a}raru, \emph{Derived categories of twisted sheaves on elliptic
  threefolds}, \href{https://doi.org/10.1515/crll.2002.022}{\emph{J. Reine
  Angew. Math.} {\bfseries 544} (2002) 161}
  [\href{https://arxiv.org/abs/math/0012083}{{\ttfamily math/0012083}}].

\bibitem{Batyrev:1994pg}
V.V.~Batyrev and L.A.~Borisov, \emph{{On Calabi-Yau complete intersections in
  toric varieties}},  \href{https://arxiv.org/abs/alg-geom/9412017}{{\ttfamily
  alg-geom/9412017}}.

\bibitem{Monnier:2018nfs}
S.~Monnier and G.W.~Moore, \emph{{Remarks on the Green--Schwarz Terms of
  Six-Dimensional Supergravity Theories}},
  \href{https://doi.org/10.1007/s00220-019-03341-7}{\emph{Commun. Math. Phys.}
  {\bfseries 372} (2019) 963}
  [\href{https://arxiv.org/abs/1808.01334}{{\ttfamily 1808.01334}}].

\bibitem{Witten:2015aba}
E.~Witten, \emph{{Fermion Path Integrals And Topological Phases}},
  \href{https://doi.org/10.1103/RevModPhys.88.035001}{\emph{Rev. Mod. Phys.}
  {\bfseries 88} (2016) 035001}
  [\href{https://arxiv.org/abs/1508.04715}{{\ttfamily 1508.04715}}].

\bibitem{Yonekura:2016wuc}
K.~Yonekura, \emph{{Dai-Freed theorem and topological phases of matter}},
  \href{https://doi.org/10.1007/JHEP09(2016)022}{\emph{JHEP} {\bfseries 09}
  (2016) 022} [\href{https://arxiv.org/abs/1607.01873}{{\ttfamily
  1607.01873}}].

\bibitem{Hsieh:2018ifc}
C.-T.~Hsieh, \emph{{Discrete gauge anomalies revisited}},
  \href{https://arxiv.org/abs/1808.02881}{{\ttfamily 1808.02881}}.

\bibitem{Freed:2014iua}
D.S.~Freed, \emph{{Anomalies and Invertible Field Theories}},
  \href{https://doi.org/10.1090/pspum/088/01462}{\emph{Proc. Symp. Pure Math.}
  {\bfseries 88} (2014) 25} [\href{https://arxiv.org/abs/1404.7224}{{\ttfamily
  1404.7224}}].

\bibitem{Debray:2022wcd}
A.~Debray and M.~Yu, \emph{{What bordism-theoretic anomaly cancellation can do
  for U}},  \href{https://arxiv.org/abs/2210.04911}{{\ttfamily 2210.04911}}.

\bibitem{Ibanez:1991hv}
L.E.~Ibanez and G.G.~Ross, \emph{{Discrete gauge symmetry anomalies}},
  \href{https://doi.org/10.1016/0370-2693(91)91614-2}{\emph{Phys. Lett. B}
  {\bfseries 260} (1991) 291}.

\bibitem{Garcia-Etxebarria:2017crf}
I.n.~Garc\'\i{}a-Etxebarria, H.~Hayashi, K.~Ohmori, Y.~Tachikawa and
  K.~Yonekura, \emph{{8d gauge anomalies and the topological Green-Schwarz
  mechanism}}, \href{https://doi.org/10.1007/JHEP11(2017)177}{\emph{JHEP}
  {\bfseries 11} (2017) 177}
  [\href{https://arxiv.org/abs/1710.04218}{{\ttfamily 1710.04218}}].

\bibitem{Debray:2021vob}
A.~Debray, M.~Dierigl, J.J.~Heckman and M.~Montero, \emph{{The anomaly that was
  not meant IIB}},  \href{https://arxiv.org/abs/2107.14227}{{\ttfamily
  2107.14227}}.

\bibitem{10.1007/BFb0075216}
J.~Cheeger and J.~Simons, \emph{Differential characters and geometric
  invariants},  in \emph{Geometry and Topology}, (Berlin, Heidelberg),
  pp.~50--80, Springer Berlin Heidelberg, 1985.

\bibitem{Freed:2006yc}
D.S.~Freed, G.W.~Moore and G.~Segal, \emph{{Heisenberg Groups and
  Noncommutative Fluxes}},
  \href{https://doi.org/10.1016/j.aop.2006.07.014}{\emph{Annals Phys.}
  {\bfseries 322} (2007) 236}
  [\href{https://arxiv.org/abs/hep-th/0605200}{{\ttfamily hep-th/0605200}}].

\bibitem{Lee:2022spd}
Y.~Lee and K.~Yonekura, \emph{{Global anomalies in 8d supergravity}},
  \href{https://doi.org/10.1007/JHEP07(2022)125}{\emph{JHEP} {\bfseries 07}
  (2022) 125} [\href{https://arxiv.org/abs/2203.12631}{{\ttfamily
  2203.12631}}].

\bibitem{Shimizu:2016lbw}
H.~Shimizu and Y.~Tachikawa, \emph{{Anomaly of strings of 6d $
  \mathcal{N}=\left(1,0\right) $ theories}},
  \href{https://doi.org/10.1007/JHEP11(2016)165}{\emph{JHEP} {\bfseries 11}
  (2016) 165} [\href{https://arxiv.org/abs/1608.05894}{{\ttfamily
  1608.05894}}].

\bibitem{Sati:2009ic}
H.~Sati, U.~Schreiber and J.~Stasheff, \emph{{Differential twisted String and
  Fivebrane structures}},
  \href{https://doi.org/10.1007/s00220-012-1510-3}{\emph{Commun. Math. Phys.}
  {\bfseries 315} (2012) 169}
  [\href{https://arxiv.org/abs/0910.4001}{{\ttfamily 0910.4001}}].

\bibitem{Monnier:2017oqd}
S.~Monnier, G.W.~Moore and D.S.~Park, \emph{{Quantization of anomaly
  coefficients in 6D $\mathcal{N}=(1,0)$ supergravity}},
  \href{https://doi.org/10.1007/JHEP02(2018)020}{\emph{JHEP} {\bfseries 02}
  (2018) 020} [\href{https://arxiv.org/abs/1711.04777}{{\ttfamily
  1711.04777}}].

\bibitem{Numasawa:2017crf}
T.~Numasawa and S.~Yamaguchi, \emph{{Mixed global anomalies and boundary
  conformal field theories}},
  \href{https://doi.org/10.1007/JHEP11(2018)202}{\emph{JHEP} {\bfseries 11}
  (2018) 202} [\href{https://arxiv.org/abs/1712.09361}{{\ttfamily
  1712.09361}}].

\bibitem{Kikuchi:2019ytf}
K.~Kikuchi and Y.~Zhou, \emph{{Two-dimensional Anomaly, Orbifolding, and
  Boundary States}},  \href{https://arxiv.org/abs/1908.02918}{{\ttfamily
  1908.02918}}.

\bibitem{ZUBER1986127}
J.-B.~Zuber, \emph{Discrete symmetries of conformal theories},
  \href{https://doi.org/https://doi.org/10.1016/0370-2693(86)90936-6}{\emph{Physics
  Letters B} {\bfseries 176} (1986) 127}.

\bibitem{etingof2016tensor}
P.~Etingof, S.~Gelaki, D.~Nikshych and V.~Ostrik, \emph{Tensor Categories},
  Mathematical Surveys and Monographs, American Mathematical Society (2016).

\bibitem{deWildPropitius:1995cf}
M.D.F.~de~Wild~Propitius, \emph{{Topological interactions in broken gauge
  theories}}, Ph.D. thesis, Amsterdam U., 1995.
\newblock \href{https://arxiv.org/abs/hep-th/9511195}{{\ttfamily
  hep-th/9511195}}.

\bibitem{Haghighat:2013gba}
B.~Haghighat, A.~Iqbal, C.~Koz\c{c}az, G.~Lockhart and C.~Vafa,
  \emph{{M-Strings}},
  \href{https://doi.org/10.1007/s00220-014-2139-1}{\emph{Commun. Math. Phys.}
  {\bfseries 334} (2015) 779}
  [\href{https://arxiv.org/abs/1305.6322}{{\ttfamily 1305.6322}}].

\bibitem{Haghighat:2015ega}
B.~Haghighat, S.~Murthy, C.~Vafa and S.~Vandoren, \emph{{F-Theory, Spinning
  Black Holes and Multi-string Branches}},
  \href{https://doi.org/10.1007/JHEP01(2016)009}{\emph{JHEP} {\bfseries 01}
  (2016) 009} [\href{https://arxiv.org/abs/1509.00455}{{\ttfamily
  1509.00455}}].

\bibitem{Lawrie:2016axq}
C.~Lawrie, S.~Schafer-Nameki and T.~Weigand, \emph{{Chiral 2d theories from N =
  4 SYM with varying coupling}},
  \href{https://doi.org/10.1007/JHEP04(2017)111}{\emph{JHEP} {\bfseries 04}
  (2017) 111} [\href{https://arxiv.org/abs/1612.05640}{{\ttfamily
  1612.05640}}].

\bibitem{DelZotto:2018tcj}
M.~Del~Zotto and G.~Lockhart, \emph{{Universal Features of BPS Strings in
  Six-dimensional SCFTs}},
  \href{https://doi.org/10.1007/JHEP08(2018)173}{\emph{JHEP} {\bfseries 08}
  (2018) 173} [\href{https://arxiv.org/abs/1804.09694}{{\ttfamily
  1804.09694}}].

\bibitem{Bobev:2015kza}
N.~Bobev, M.~Bullimore and H.-C.~Kim, \emph{{Supersymmetric Casimir Energy and
  the Anomaly Polynomial}},
  \href{https://doi.org/10.1007/JHEP09(2015)142}{\emph{JHEP} {\bfseries 09}
  (2015) 142} [\href{https://arxiv.org/abs/1507.08553}{{\ttfamily
  1507.08553}}].

\bibitem{Benjamin:2016fhe}
N.~Benjamin, E.~Dyer, A.L.~Fitzpatrick and S.~Kachru, \emph{{Universal Bounds
  on Charged States in 2d CFT and 3d Gravity}},
  \href{https://doi.org/10.1007/JHEP08(2016)041}{\emph{JHEP} {\bfseries 08}
  (2016) 041} [\href{https://arxiv.org/abs/1603.09745}{{\ttfamily
  1603.09745}}].

\bibitem{DelZotto:2016pvm}
M.~Del~Zotto and G.~Lockhart, \emph{{On Exceptional Instanton Strings}},
  \href{https://doi.org/10.1007/JHEP09(2017)081}{\emph{JHEP} {\bfseries 09}
  (2017) 081} [\href{https://arxiv.org/abs/1609.00310}{{\ttfamily
  1609.00310}}].

\bibitem{DelZotto:2017mee}
M.~Del~Zotto, J.~Gu, M.-X.~Huang, A.-K.~Kashani-Poor, A.~Klemm and G.~Lockhart,
  \emph{{Topological Strings on Singular Elliptic Calabi-Yau 3-folds and
  Minimal 6d SCFTs}},
  \href{https://doi.org/10.1007/JHEP03(2018)156}{\emph{JHEP} {\bfseries 03}
  (2018) 156} [\href{https://arxiv.org/abs/1712.07017}{{\ttfamily
  1712.07017}}].

\bibitem{Weigand:2017gwb}
T.~Weigand and F.~Xu, \emph{{The Green-Schwarz Mechanism and Geometric Anomaly
  Relations in 2d (0,2) F-theory Vacua}},
  \href{https://doi.org/10.1007/JHEP04(2018)107}{\emph{JHEP} {\bfseries 04}
  (2018) 107} [\href{https://arxiv.org/abs/1712.04456}{{\ttfamily
  1712.04456}}].

\bibitem{Lee:2018urn}
S.-J.~Lee, W.~Lerche and T.~Weigand, \emph{{Tensionless Strings and the Weak
  Gravity Conjecture}},
  \href{https://doi.org/10.1007/JHEP10(2018)164}{\emph{JHEP} {\bfseries 10}
  (2018) 164} [\href{https://arxiv.org/abs/1808.05958}{{\ttfamily
  1808.05958}}].

\bibitem{Kim:2016foj}
H.-C.~Kim, S.~Kim and J.~Park, \emph{{6d strings from new chiral gauge
  theories}},  \href{https://arxiv.org/abs/1608.03919}{{\ttfamily 1608.03919}}.

\bibitem{Klemm:1996hh}
A.~Klemm, P.~Mayr and C.~Vafa, \emph{{BPS states of exceptional noncritical
  strings}}, \href{https://doi.org/10.1016/S0920-5632(97)00422-2}{\emph{Nucl.
  Phys. B Proc. Suppl.} {\bfseries 58} (1997) 177}
  [\href{https://arxiv.org/abs/hep-th/9607139}{{\ttfamily hep-th/9607139}}].

\bibitem{Haghighat:2014vxa}
B.~Haghighat, A.~Klemm, G.~Lockhart and C.~Vafa, \emph{{Strings of Minimal 6d
  SCFTs}}, \href{https://doi.org/10.1002/prop.201500014}{\emph{Fortsch. Phys.}
  {\bfseries 63} (2015) 294} [\href{https://arxiv.org/abs/1412.3152}{{\ttfamily
  1412.3152}}].

\bibitem{Huang:2015sta}
M.-x.~Huang, S.~Katz and A.~Klemm, \emph{{Topological String on elliptic CY
  3-folds and the ring of Jacobi forms}},
  \href{https://doi.org/10.1007/JHEP10(2015)125}{\emph{JHEP} {\bfseries 10}
  (2015) 125} [\href{https://arxiv.org/abs/1501.04891}{{\ttfamily
  1501.04891}}].

\bibitem{Lee:2018spm}
S.-J.~Lee, W.~Lerche and T.~Weigand, \emph{{A Stringy Test of the Scalar Weak
  Gravity Conjecture}},
  \href{https://doi.org/10.1016/j.nuclphysb.2018.11.001}{\emph{Nucl. Phys. B}
  {\bfseries 938} (2019) 321}
  [\href{https://arxiv.org/abs/1810.05169}{{\ttfamily 1810.05169}}].

\bibitem{Candelas:1994hw}
P.~Candelas, A.~Font, S.H.~Katz and D.R.~Morrison, \emph{{Mirror symmetry for
  two parameter models. 2.}},
  \href{https://doi.org/10.1016/0550-3213(94)90155-4}{\emph{Nucl. Phys. B}
  {\bfseries 429} (1994) 626}
  [\href{https://arxiv.org/abs/hep-th/9403187}{{\ttfamily hep-th/9403187}}].

\bibitem{haghighat2016}
B.~Haghighat, S.~Murthy, C.~Vafa and S.~Vandoren, \emph{F-theory, spinning
  black holes and multi-string branches},
  \href{https://doi.org/10.1007/jhep01(2016)009}{\emph{Journal of High Energy
  Physics} {\bfseries 2016} (2016) }.

\bibitem{Ferrara:1996wv}
S.~Ferrara, R.~Minasian and A.~Sagnotti, \emph{{Low-energy analysis of M and F
  theories on Calabi-Yau threefolds}},
  \href{https://doi.org/10.1016/0550-3213(96)00268-4}{\emph{Nucl. Phys. B}
  {\bfseries 474} (1996) 323}
  [\href{https://arxiv.org/abs/hep-th/9604097}{{\ttfamily hep-th/9604097}}].

\bibitem{Grimm:2010ks}
T.W.~Grimm, \emph{{The N=1 effective action of F-theory compactifications}},
  \href{https://doi.org/10.1016/j.nuclphysb.2010.11.018}{\emph{Nucl. Phys. B}
  {\bfseries 845} (2011) 48} [\href{https://arxiv.org/abs/1008.4133}{{\ttfamily
  1008.4133}}].

\bibitem{Bonetti:2011mw}
F.~Bonetti and T.W.~Grimm, \emph{{Six-dimensional (1,0) effective action of
  F-theory via M-theory on Calabi-Yau threefolds}},
  \href{https://doi.org/10.1007/JHEP05(2012)019}{\emph{JHEP} {\bfseries 05}
  (2012) 019} [\href{https://arxiv.org/abs/1112.1082}{{\ttfamily 1112.1082}}].

\bibitem{Freed:2014eja}
D.S.~Freed, \emph{{Short-range entanglement and invertible field theories}},
  \href{https://arxiv.org/abs/1406.7278}{{\ttfamily 1406.7278}}.

\bibitem{Freed:2016rqq}
D.S.~Freed and M.J.~Hopkins, \emph{{Reflection positivity and invertible
  topological phases}},
  \href{https://doi.org/10.2140/gt.2021.25.1165}{\emph{Geom. Topol.} {\bfseries
  25} (2021) 1165} [\href{https://arxiv.org/abs/1604.06527}{{\ttfamily
  1604.06527}}].

\bibitem{Davighi:2020uab}
J.~Davighi and N.~Lohitsiri, \emph{{The algebra of anomaly interplay}},
  \href{https://doi.org/10.21468/SciPostPhys.10.3.074}{\emph{SciPost Phys.}
  {\bfseries 10} (2021) 074}
  [\href{https://arxiv.org/abs/2011.10102}{{\ttfamily 2011.10102}}].

\bibitem{stolz_teichner_2004}
S.~Stolz and P.~Teichner, \emph{What is an elliptic object?},  in
  \emph{Topology, Geometry and Quantum Field Theory: Proceedings of the 2002
  Oxford Symposium in Honour of the 60th Birthday of Graeme Segal},
  U.~Tillmann, ed., London Mathematical Society Lecture Note Series,
  p.~247–343, Cambridge University Press (2004),
  \href{https://doi.org/10.1017/CBO9780511526398.013}{DOI}.

\bibitem{Gukov:2018iiq}
S.~Gukov, D.~Pei, P.~Putrov and C.~Vafa, \emph{{4-manifolds and topological
  modular forms}}, \href{https://doi.org/10.1007/JHEP05(2021)084}{\emph{JHEP}
  {\bfseries 05} (2021) 084}
  [\href{https://arxiv.org/abs/1811.07884}{{\ttfamily 1811.07884}}].

\bibitem{Tachikawa:2021mvw}
Y.~Tachikawa, \emph{{Topological modular forms and the absence of a heterotic
  global anomaly}}, \href{https://doi.org/10.1093/ptep/ptab060}{\emph{PTEP}
  {\bfseries 2022} (2022) 04A107}
  [\href{https://arxiv.org/abs/2103.12211}{{\ttfamily 2103.12211}}].

\bibitem{Anderson:2008uw}
L.B.~Anderson, Y.-H.~He and A.~Lukas, \emph{{Monad Bundles in Heterotic String
  Compactifications}},
  \href{https://doi.org/10.1088/1126-6708/2008/07/104}{\emph{JHEP} {\bfseries
  07} (2008) 104} [\href{https://arxiv.org/abs/0805.2875}{{\ttfamily
  0805.2875}}].

\bibitem{fulton81}
W.~Fulton and R.~Lazarsfeld, \emph{On the connectedness of degeneracy loci and
  special divisors}, \href{https://doi.org/10.1007/BF02392466}{\emph{Acta
  Mathematica} {\bfseries 146} (1981) 271}.

\bibitem{Blumenhagen:2010pv}
R.~Blumenhagen, B.~Jurke, T.~Rahn and H.~Roschy, \emph{{Cohomology of Line
  Bundles: A Computational Algorithm}},
  \href{https://doi.org/10.1063/1.3501132}{\emph{J. Math. Phys.} {\bfseries 51}
  (2010) 103525} [\href{https://arxiv.org/abs/1003.5217}{{\ttfamily
  1003.5217}}].

\bibitem{Wall66}
C.T.C.~Wall, \emph{Classification problems in differential topology. v},
  \href{https://doi.org/10.1007/BF01389738}{\emph{Inventiones mathematicae}
  {\bfseries 1} (1966) 355}.

\bibitem{eichler2013theory}
M.~Eichler and D.~Zagier, \emph{The Theory of Jacobi Forms}, Progress in
  Mathematics, Birkh{\"a}user Boston (2013).

\bibitem{Park:2011wv}
D.S.~Park and W.~Taylor, \emph{{Constraints on 6D Supergravity Theories with
  Abelian Gauge Symmetry}},
  \href{https://doi.org/10.1007/JHEP01(2012)141}{\emph{JHEP} {\bfseries 01}
  (2012) 141} [\href{https://arxiv.org/abs/1110.5916}{{\ttfamily 1110.5916}}].

\bibitem{Collinucci:2019fnh}
A.~Collinucci, M.~Fazzi, D.R.~Morrison and R.~Valandro, \emph{{High electric
  charges in M-theory from quiver varieties}},
  \href{https://doi.org/10.1007/JHEP11(2019)111}{\emph{JHEP} {\bfseries 11}
  (2019) 111} [\href{https://arxiv.org/abs/1906.02202}{{\ttfamily
  1906.02202}}].

\bibitem{Cianci:2018vwv}
F.M.~Cianci, D.K.~Mayorga Pe\~na and R.~Valandro, \emph{{High U(1) charges in
  type IIB models and their F-theory lift}},
  \href{https://doi.org/10.1007/JHEP04(2019)012}{\emph{JHEP} {\bfseries 04}
  (2019) 012} [\href{https://arxiv.org/abs/1811.11777}{{\ttfamily
  1811.11777}}].

\bibitem{Raghuram:2018hjn}
N.~Raghuram and W.~Taylor, \emph{{Large U(1) charges in F-theory}},
  \href{https://doi.org/10.1007/JHEP10(2018)182}{\emph{JHEP} {\bfseries 10}
  (2018) 182} [\href{https://arxiv.org/abs/1809.01666}{{\ttfamily
  1809.01666}}].

\bibitem{Hajouji:2020ddi}
N.~Hajouji, \emph{{Trace Zero Points on Elliptic Fibrations}}, Ph.D. thesis,
  UC, Santa Barbara (main), 2020.

\bibitem{Hajouji:2019vxs}
N.~Hajouji and P.-K.~Oehlmann, \emph{{Modular Curves and Mordell-Weil Torsion
  in F-theory}}, \href{https://doi.org/10.1007/JHEP04(2020)103}{\emph{JHEP}
  {\bfseries 04} (2020) 103}
  [\href{https://arxiv.org/abs/1910.04095}{{\ttfamily 1910.04095}}].

\bibitem{Gopakumar:1998ii}
R.~Gopakumar and C.~Vafa, \emph{{M theory and topological strings. 1.}},
  \href{https://arxiv.org/abs/hep-th/9809187}{{\ttfamily hep-th/9809187}}.

\bibitem{Gopakumar:1998jq}
R.~Gopakumar and C.~Vafa, \emph{{M theory and topological strings. 2.}},
  \href{https://arxiv.org/abs/hep-th/9812127}{{\ttfamily hep-th/9812127}}.

\end{thebibliography}\endgroup
\end{document}